\documentclass[draft]{agujournal2019}
\usepackage{url}
\usepackage{lineno}
\usepackage{amsmath,amssymb,gensymb}
\usepackage{rotating}

\usepackage{subcaption}
\usepackage[inline]{trackchanges}
\usepackage{soul}

\draftfalse

\journalname{JGR: Space Physics}

\begin{document}

\title{Energetic proton losses reveal Io's extended and longitudinally asymmetrical atmosphere}

\authors{H.L.F. Huybrighs\affil{1,2,3,4,*}, C.P.A. van Buchem\affil{4,5,*}, A. Bl\"{o}cker\affil{6,7}, V. Dols\affil{8}, C.F. Bowers\affil{1}, C.M. Jackman\affil{1}}

\affiliation{1}{School of Cosmic Physics, DIAS Dunsink Observatory, Dublin Institute for Advanced Studies, Dublin 15, Ireland}
\affiliation{2}{Space and Planetary Science Center, Khalifa University, Abu Dhabi, UAE}
\affiliation{3}{Department of Mathematics, Khalifa University, Abu Dhabi, UAE}
\affiliation{4}{European Space Agency (ESA), European Space Research and Technology Centre (ESTEC), Keplerlaan 1, 2201 AZ Noordwijk, The Netherlands}
\affiliation{5}{Leiden University, Leiden, the Netherlands}
\affiliation{6}{KTH, Royal Institute of Technology, Stockholm, Sweden}
\affiliation{7}{Department of Earth and Environmental Sciences, Ludwig Maximilian University of Munich, Munich, Germany}
\affiliation{8}{Laboratory for Atmospheric \& Space Physics, University of Colorado, Boulder, CO, USA}
\affiliation{*}{HLF Huybrighs and CPA van Buchem contributed equally to this work and are considered as joint first authors.}

\correspondingauthor{Hans Huybrighs}{hans@cp.dias.ie}

\begin{keypoints}
\item Atmospheric charge exchange is a major or dominant loss process of energetic protons (155-224 keV) during three close Galileo flybys of Io
\item Proton losses suggest an expanded atmosphere on the day/downstream side and a lack of atmospheric collapse on the night/upstream side. 
\item Discrepancies between the data and model hint at a global large scale height atmosphere
\end{keypoints}

\begin{abstract}
Along the I24, I27 and I31 flybys of Io (1999-2001), the Energetic Particle Detector (EPD) onboard the Galileo spacecraft observed localised regions of energetic protons losses (155 keV-1250 keV). Using back-tracking particle simulations combined with a prescribed atmospheric distribution and a magnetohydrodynamics (MHD) model of the plasma/atmosphere interaction, we investigate the possible causes of these depletions. We focus on a limited region within two Io radii, which is dominated by Io's SO$_2$ atmosphere. Our results show that  charge exchange of protons with the SO$_2$ atmosphere, absorption by the surface and the configuration of the electromagnetic field contribute to the observed proton depletion along the Galileo flybys. In the 155-240 keV energy range, charge exchange is either a major or the dominant loss process, depending on the flyby altitude. In the 540-1250 keV range, as the charge exchange cross sections are small, the observed decrease of the proton flux is attributed to absorption by the surface and the perturbed electromagnetic fields, which divert the protons away from the detector. From a comparison between the modelled losses and the data we find indications of an extended atmosphere on the day/downstream side of Io, a lack of atmospheric collapse on the night/upstream side as well as a more global extended atmospheric component ($> 1$ Io radius). Our results demonstrate that observations and modeling of proton depletion around the moon constitute an important tool to constrain the electromagnetic field configuration around Io and the radial and longitudinal atmospheric distribution, which is still poorly understood.
\end{abstract}

\section*{Plain Language Summary}
Io is a moon of Jupiter with active volcanoes and an atmosphere of which the structure and variability is poorly understood. A fascinating object on its own, neutral gas from Io also serves as the main source of plasma in Jupiter's magnetosphere. Improving our understanding of Io's atmosphere will allow us to better understand the precise link between Io's neutral environment and the plasma torus surrounding Jupiter. In this work we analyse data from the historic Galileo spacecraft that encountered Io. Specifically, we analyse regions close to Io where (normally abundant) energetic protons are disappearing. We find that charge exchange between these particles and Io's atmosphere along with the effect of the electric and magnetic fields on the paths of these particles can cause such decreases. Charge exchange with the atmosphere is either a major or the dominant source of losses, depending on the flyby altitude. The losses of the protons are related to Io's atmosphere and hint at an atmosphere that is more extended on Io's day side (also the downstream side from the plasma flow's point of view), doesn't collapse fully on the night (upstream) side, and appears to be more extended than we assumed.

\section{Introduction}\label{sec:introduction}
Io is the most volcanically active body in our solar system due to the periodic deformation of its interior by Jupiter's tidal forces. A fascinating object in its own right, Io is also considered to be the main source of plasma in Jupiter's magnetosphere. However, it is not understood how changes in the Io plasma torus' density exceeding 300\% are related to Io \cite{Roth2024}. This is in part due to our limited understanding of Io's atmospheric structure and variability. Dropouts of energetic ions could provide us with a new diagnostic of Io's neutral environment. In this introduction we review features of Io's environment that are relevant to the high energy protons around Io. A more detailed review of Io's atmosphere and its plasma environment can be found in \citeA{Bagenal2020TheEuropa}. 

\subsection{Io's atmosphere and role in Jupiter's magnetosphere}

The SO$_2$ frost sublimation (e.g. \citeA{Ingersoll1985,Moreno1991,Doute2001}), volcanic plumes (e.g. \citeA{Morabito1979,Spencer2000,Walker2010}), atmospheric sputtering  (e.g. \citeA{McGrath1987MagnetosphericAtmosphere,Pospieszalska1996})  and physical chemistry \cite{Dols2008ATorus,Dols2012AsymmetryFlybys} contribute to the formation of an extended atmosphere of SO$_2$, S and O around Io. Minor species are also present such as SO \cite{Lellouch1996DetectionAtmosphere} and NaCl \cite{Lellouch2003VolcanicallyTorus}. The atmosphere extends into Hill's sphere ($\sim$6 Io radii R$_{Io}$, 1821 km) as a corona, which is mainly composed of S and O \cite{Wolven2001EmissionCorona} and is thought to be produced by atmospheric sputtering \cite{McGrath1987MagnetosphericAtmosphere} and physical chemistry processes \cite{Dols2008ATorus}. 

The atmosphere of Io is thought to be asymmetrical along different axes (e.g. \citeA{Blocker2018MHDAtmosphere}). Being sublimation dominated, the dayside is more dense than the nightside (e.g. \citeA{Tsang2016TheEclipse}). It is thought that the atmosphere could partially collapse on the nightside due to the condensation of atmospheric SO$_2$ in the cold night temperature. The atmospheric density at the poles is thought to be only a few percent of the atmosphere at the equator, due to the equatorial distribution of frost, the preferential heating of the sub solar point and the location of some main volcanoes closer to the equator (e.g. \citeA{Strobel2001TheHydrogen,Spencer2005Mid-infraredAtmosphere}). The anti‐Jovian atmosphere is denser and latitudinally more extended than the sub‐Jovian side (e.g. \citeA{Feaga2009IosAtmosphere,Spencer2005Mid-infraredAtmosphere,Moullet2010SimultaneousArray}). Lastly the atmospheric density varies longitudinally with a denser upstream than downstream atmosphere (e.g. \citeA{Saur2002InterpretationPasses}). Major uncertainties remain in our understanding of the radial profile, longitudinal asymmetries, minor species, and how much the atmosphere collapses at night or in eclipse.

Some of the neutrals escape Io's gravity field to form giant neutral clouds of S and O (several Jupiter radii in size, 71492 km) that extend along Io's orbit. The neutral clouds at Io's orbital velocity are continuously passed by the plasma flow  of the Io torus at a relative velocity $\sim$57km/s \cite{Kivelson2004MagnetosphericSatellites}. They experience charge exchange and electron-impact ionization, which supply the torus with fresh plasma. The flux of plasma directly from Io is about $\sim300$ kg/s (e.g. \citeA{Saur2003TheIo,Dols2008ATorus}). The plasma torus has an average electron density $\sim$2000 cm$^{-3}$ and an average thermal temperature $\sim$60-100 eV \cite{Kivelson2004MagnetosphericSatellites,Bagenal2020TheEuropa}. Centrifugal forces lead the plasma to flow radially outwards along Jupiter's magnetic equator thereby forming a plasma sheet \cite{Bagenal1989Torus-MagnetosphereCoupling}. 

While circulating through the magnetosphere, charged particles are accelerated to suprathermal energies \cite{Paranicas2000EnergeticEuropa} and are able to diffuse back radially inwards \cite{Khurana2004TheMagnetosphere} towards Jupiter's moons.
These particles are referred to as energetic particles in the rest of the text. 
In summary, Io is exposed to a flow of corotating low energy plasma ($\sim$100 eV) and to a population of energetic charged particles, including the energetic protons ($>100$ keV) on which this study focuses.

\subsection{Energetic ions near Io}

The Energetic Particle Detector (EPD) onboard Galileo \cite{Williams1992TheDetector} detected dropouts of energetic ions during close flybys of Io. In this work we investigate specifically the cause of energetic proton losses. Several causes of energetic ion depletion have previously been proposed for various moons: impact of the energetic protons on the moon's surface, deflection of the protons by inhomogeneous electromagnetic fields, charge exchange of the protons with atmospheric neutrals and wave particle interactions. 

Energetic ions that impact the surface are absorbed and therefore result in a depletion, such losses have been widely reported at Io \cite{Selesnick2009ChargeIo}
and various moons of Jupiter and Saturn e.g. \citeA{Paranicas2000EnergeticEuropa,Kivelson2004MagnetosphericSatellites,Roussos2022ABelts,Kotova2015ModelingDione,Krupp2020Magnetospheric20052015}. Furthermore, a magnetic field gradient can restrict access of energetic ions in certain regions and cause apparent losses. Such features have been reported at Io  \cite{Selesnick2009ChargeIo} and other moons of Jupiter and Saturn e.g. \citeA{Paranicas2000EnergeticEuropa,Roussos2012EnergeticInteraction,Huybrighs2023}. We will refer to losses due to perturbed fields as ``forbidden regions".

When energetic protons charge exchange with atmospheric neutrals they turn into energetic neutral atoms (ENA) which will result in a depletion of the energetic protons. Atmospheric charge exchange has been investigated as a loss process of energetic ions at moons of Jupiter and Saturn e.g. \citeA{Mitchell2005EnergeticMagnetosphere,Wulms2010EnergeticFields,Kotova2015ModelingDione,Addison2021InfluenceWeathering}. Specifically at Europa \citeA{Huybrighs2020AnDepletions} demonstrated, using a Monte Carlo particle tracing code, that proton depletions (115-244 keV) measured during the E26 Galileo flyby were affected, in addition to absorption by the surface, by atmospheric charge exchange and inhomogeneous fields. This study builds on the modelling efforts of \citeA{Huybrighs2020AnDepletions}. 

Specifically at Io, ion dropouts during three close Galileo flybys (I27, I31 and I32) have been analysed by \citeA{Selesnick2009ChargeIo}. Using ion back-tracing simulations they demonstrated that losses of energetic oxygen and sulfur ions ($>$5 MeV/nucleon) measured by Galileo's Heavy Ion Counter (HIC) could be attributed to absorption by the surface and the effect Io's Alfv\'en wings. \citeA{Selesnick2009ChargeIo} accounted for the Alfv\'en wings by using an idealized analytical description, but did not account for other perturbations such as magnetic field pile-up. The addition of the Alfv\'en wings provides a better reproduction of the measured ion dropouts during polar flybys I31 and I32. At certain segments of the flyby, the Alfv\'en wings deflect ions that would otherwise collide with Io's surface, while enhancing losses in other segments. For flyby I27 the effect of the Alfv\'en wing appeared to be small.

Ion dropouts have also been reported away from Io, but near the Io orbit, using Voyager 1, Voyager 2 and Galileo data \cite{Armstrong1981Low-energyMagnetosphere,Cheng1983EnergeticOrbitb,Williams1996,Mauk1998Galileo-measuredEpoch,Lagg1998DeterminationMeasurements}. Specifically, \citeA{Cheng1983EnergeticOrbit} and \citeA{Lagg1998DeterminationMeasurements} proposed charge exchange with neutrals as a loss process for the energetic ions, 
More recently, \citeA{Mauk2022} argued, based on \cite{Smith2019}, that near Io's orbit charge exchange with low energy ions would dominate over charge exchange with neutrals. \citeA{Mauk2022} also argues that near Io's orbit charge exchange losses dominate for protons (roughly below 200 keV) over losses from scattering. Furthermore, they suggest that at energies above 300 keV the scattering dominates over charge exchange.

Lastly, ion dropouts have also been measured along field lines connected to Io, but not in a close flyby of Io. The JEDI instrument on Juno measured losses of energetic protons with energies in the 10-800 keV range over a region larger than Io along field lines connected with Io \cite{Paranicas2019IosData}. They attribute these losses to wave-particle interaction, building on \citeA{Nenon2018AOrbit} which proposes that near Io ion cyclotron waves could resonate with energetic protons and cause proton losses. \citeA{Nenon2018AOrbit} suggest that protons at energies between about 0.3 and 4 MeV are affected by electromagnetic ion cyclotron waves near Io's L shells. However, \citeA{Paranicas2019IosData} also suggests that below 100 keV, proton losses to charge exchange may become significant due to the large charge exchange cross section. 

\subsection{Scope of this study}

In this study we investigate the cause of energetic proton losses measured close to Io ($<2 R_{Io}$) by the Energetic Particle Detector (EPD) on the Galileo spacecraft. While the effect of charge exchange has been considered as a loss process for Io's orbital environment, previous studies investigating particle losses near Io \cite{Selesnick2009ChargeIo} have not yet considered the effect of charge exchange with the atmosphere, and they employed a highly simplified model of the perturbed fields. Therefore, we opt to include charge exchange with Io's asymmetrical atmosphere as a loss process as well as a detailed MHD model of the perturbed fields. To investigate the cause of the losses we  employ a particle back-tracing model which we use to reproduce the flux of energetic protons as measured by Galileo during three Io flybys (I24, I27 and I31). 

We consider this study in the first place as a feasibility study to demonstrate the sensitivity of the proton dropouts to the atmospheric structure and magnetic field configuration. Using previously validated models of fields and atmosphere we will show that the models are not able to fully reproduce the energetic proton dropouts and that the proton data can thus provide new observational constraints to modelers. Our aim is to provide hypotheses and directions to encourage further modelling studies that exploit the additional constraint offered by the energetic ions to the fullest.

\section{Method: Energetic Particle Detector (EPD) and data selection}\label{sec:epd}

The EPD telescope contains the Composition Measurement System (CMS) that uses a time of flight mass spectrometer to separate protons from other ions. A triple coincidence detection system is used in order to limit the effect of noise by penetrating radiation \cite{Williams1992TheDetector}. The EPD telescope can scan the whole sky in $\sim$140 seconds using the combined motion of a rotating platform that moves through 7 motor positions and the $\sim$20 seconds period spacecraft spin. The EPD CMS channels have an opening angle of 18 degrees. Further details are documented in the EPD user manual \cite{Kollmann2022}.

We focus specifically on EPD measurements of energetic protons. Though Oxygen and Sulfur ions fluxes have also been measured by EPD, there are large uncertainties in the charge exchange cross section for those species. Furthermore, protons have a single charge state, while (part of the) oxygen and sulfur ions could have higher charge states \cite{Clark2016ChargeMagnetosphere,Nenon2019EvidenceMeasurements}, which would introduce more free parameters in the analysis.

Building on \citeA{Huybrighs2021ReplyDepletions} we focus specifically on two energy channels, one channel (155-224 keV, TP1) that is sensitive to charge exchange with Io's atmosphere due to the relatively large charge exchange cross section (see Figure \ref{fig_cross}) and a second channel (540-1250 keV, TP3) that is not sensitive to charge exchange due to the strongly reduced charge exchange cross section. The energy range for TP1 is a different energy range than originally defined in \citeA{Williams1992TheDetector} due to 'dead' layer formation \cite{Lee-Payne2020CorrectionCorrection}. The effect of dead layer formation is not as pronounced for TP3, hence the original range of 540-1250 keV is used for that channel.

The data from TP1 is especially well suited to the study of atmospheric charge exchange, due to the relatively large charge exchange cross section of protons on the atmosphere. SO$_2$ is the dominant atmospheric component. To our knowledge, there are no published values of charge exchange cross sections of H$^+$ on SO$_2$. Consequently we use the cross section of H$^+$ on O$_2$ as a reasonable approximation, with experimental data from \cite{Basu1987LinearObservations} extrapolated to the higher energies of TP1 and TP3 (Figure \ref{fig_cross}) as in \cite{Huybrighs2020AnDepletions}. It is well known that charge exchange cross sections drop dramatically for very high energies \cite{Lo1971}. We contrast the TP1 channel with the TP3 channel, rather than the intermediate TP2 channel for which data is also available, because the charge exchange cross section in TP3 is so small that it is not sensitive to atmospheric charge exchange. Therefore, if losses occur in TP1 but not in TP3, this would argue in favour of the losses being caused by charge exchange. 
In addition, due to the low sensitivity to charge exchange, the TP3 channel could allow us to better isolate apparent losses associated with forbidden regions caused by perturbations in the electromagnetic fields surrounding Io. A last argument for the use of the channels is that these energies are small enough for gyration scales ($<$30 km for TP1 and $<$100 km for TP3, see Figure \ref{fig_gyro}) to be sensitive to inhomogeneous fields. 

\begin{figure}
\centering
\noindent\includegraphics[width=0.9\textwidth]{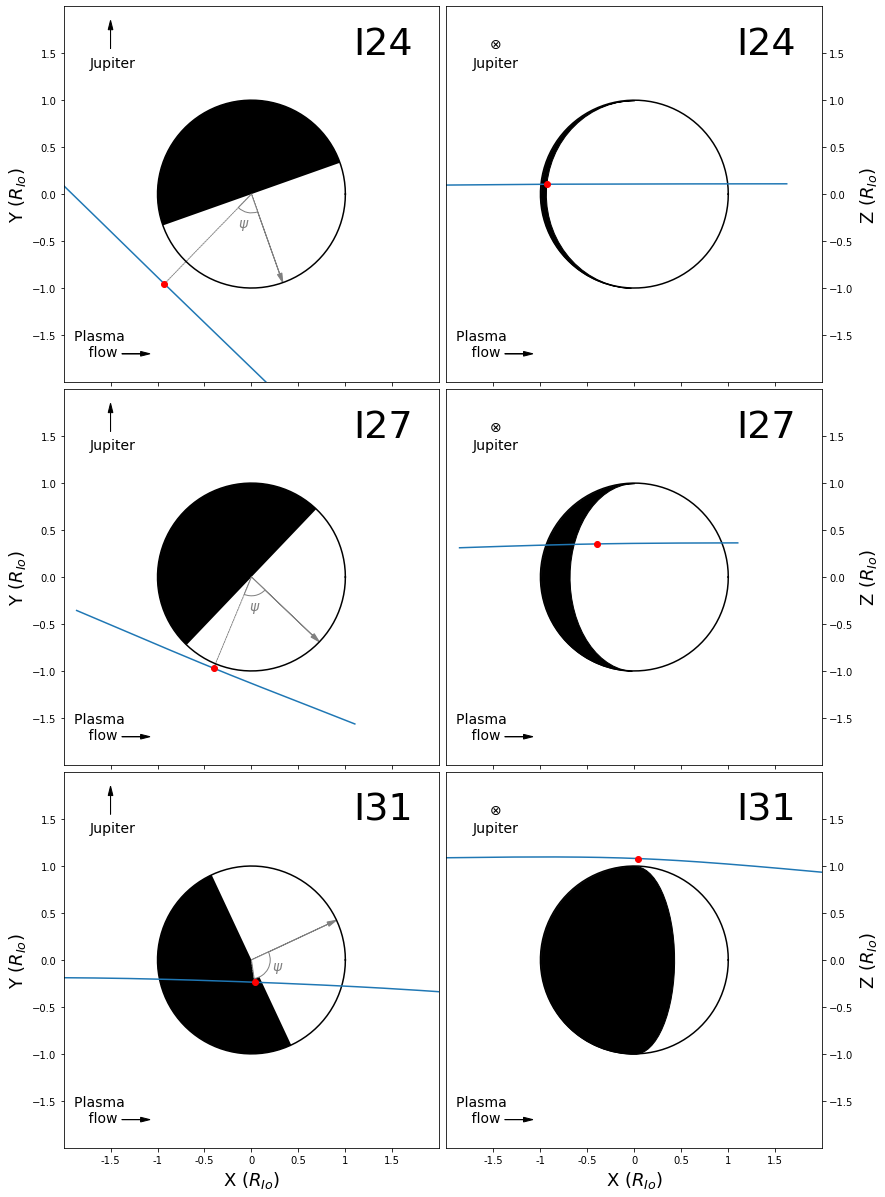}
\caption{Overview of the different Galileo flybys. The red dot marks the point of closest approach, the blue line the flight path and $\psi$ indicates the zenith angle. The black areas indicate Io's night side. Note that on the left we have a `top-down' view, with the vertical axis along the y-axis (towards Jupiter), while on the right we have an `edge-on' view of Io with the vertical direction along the z-axis (pointing towards the north of the system). In all panels the horizontal axis points along the direction of the plasma flow. For all of these flybys Galileo flew from the left to the right of the panels.}
\label{fig:flight_overview}
\end{figure}

The three flybys that we focus on are I24, I27, and I31. See Table \ref{tab:Flyby parameters} for the parameters of the closest approach and Figure \ref{fig:flight_overview} for a visualisation of the geometry of the flybys. The different flybys will allows us to demonstrate that the causes of the proton losses depend on the flyby altitude and geometry. Furthermore,
different flybys will reveal different asymmetries (or lack thereof) in the atmosphere. Two of the notable differences between the three flybys are the fact that I24 passes Io at a distance which is about three times greater than I27 and I31, and that I24 and I27 are equatorial flybys while I31 passes over Io's north pole. Note that we do not include flybys J0 and I32 in this study. For flyby J0 we were not able to reconstruct the EPD pointing using the data in the PDS, for flyby I32 no high resolution EPD data is available. 

Although the EPD signal shows proton depletion far from Io (e.g. \citeA{Lagg1998DeterminationMeasurements}), we focus here on the depletion features that are unique to the environment close to Io within $\sim2$R$_{Io}$ from the surface, a region dominated by the SO$_2$ atmosphere. In the chosen region the interaction with Io's surface, atmosphere and perturbed fields is most apparent. 

\begin{table}[]
\caption{Parameters for the Galileo flybys of Io studied in this paper. All parameters are given for the respective points of closest approach.}
\centering
\begin{tabular}{c c c c c c}
\textbf{Flyby name} & \textbf{\begin{tabular}[c]{@{}c@{}}Date time \\ (UT)\end{tabular}} & \textbf{\begin{tabular}[c]{@{}c@{}}Local time\\ (h)\end{tabular}} & \textbf{\begin{tabular}[c]{@{}c@{}}Altitude \\ (km)\end{tabular}} & \textbf{\begin{tabular}[c]{@{}c@{}}Longitude ($\phi$) \\ ($\degree$)\end{tabular}} & \textbf{\begin{tabular}[c]{@{}c@{}}Latitude ($\theta$) \\ ($\degree$)\end{tabular}} \\ \hline
I24 & 1999-10-11 04:33:03 & 10.7 & 611 & -134.13 & 4.51 \\ \hline
I27 & 2000-02-22 13:46:42 & 8.91 & 198 & -112.54 & 18.54 \\ \hline
I31 & 2001-08-06 04:59:21 & 4.33 & 193 & -81.71 & 77.49 \\ \hline
\end{tabular}
\label{tab:Flyby parameters}
\end{table}

\section{Method: modelling}\label{sec:method}

We simulate the flux of energetic protons using a back-tracing approach under different cases, to determine the relative contribution of surface impact, charge exchange with the atmosphere and the electromagnetic fields to the measured proton losses. We will consider cases with charge exchange, without charge exchange, without perturbed fields and with perturbed fields and compare them with each other. We will also consider two different atmospheric models, one model with asymmetries (day-night, upstream-downstream, Jupiter facing) and one without, which will allow us to determine the sensitivity of the proton losses to different atmospheric configurations.

\subsection{Atmospheric models}\label{sec:method_atmospheres}

\begin{figure}
\center
\noindent\includegraphics[width=1.0\textwidth]{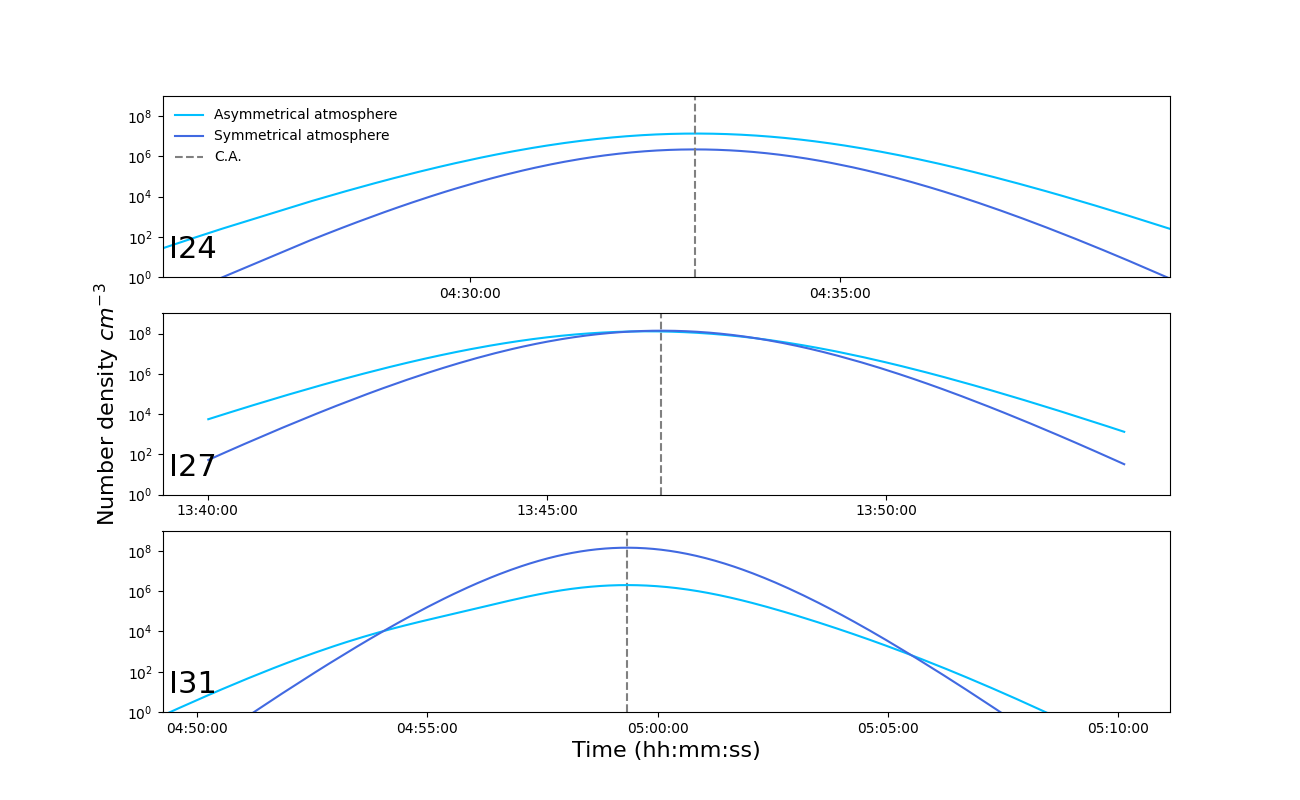}
\caption{Number density of the two atmospheric models along the trajectories of the three flybys (from top to bottom) I24, I27, and I31.}
\label{fig:atmos_density_fp}
\end{figure}

\begin{figure}
    \centering
    \makebox[\textwidth][c]{
    \includegraphics[width=1.3\textwidth]{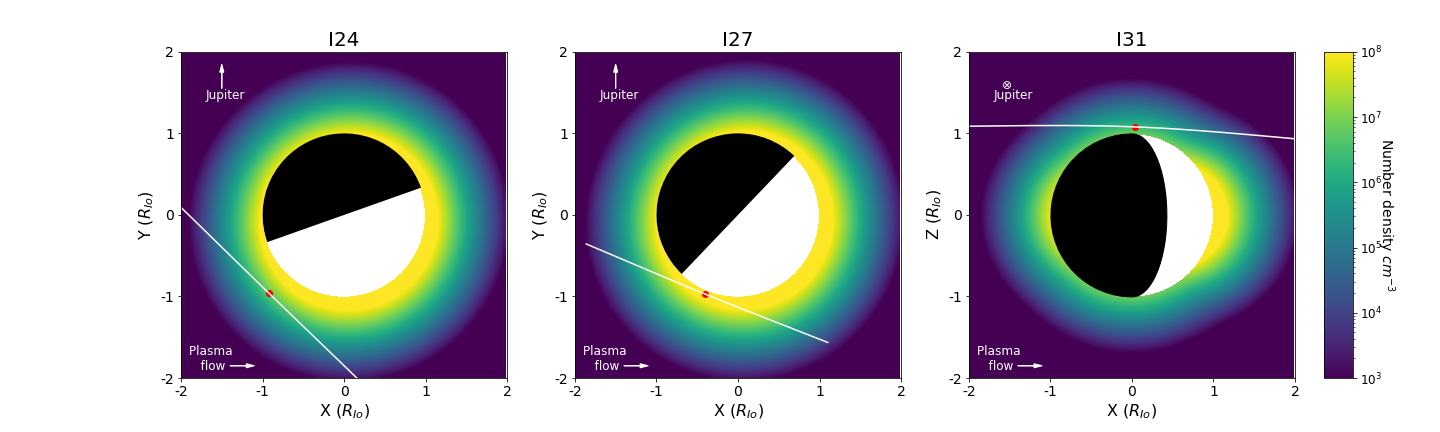}}
\caption{Visualisation of the longitude and latitude dependent atmospheric model used in this study and presented in \protect\citeA{Blocker2018MHDAtmosphere}. The I24, I27, and I31 flybys are shown respectively from left to right. The day/night hemispheres are indicated with the white and black sides respectively. The closest approach is indicated with a red point. In section \ref{sec:method_atmospheres} we give a description of the model. Note that for I31 an 'edge-on' view of the system is shown in order to better visualise the polar flyby and show the latitudinal variation of the atmosphere model.}
\label{fig:atmos_vis}
\end{figure}

In this work,  we will focus on two models of the SO$_2$ distribution around Io: one radially symmetrical model (referred to as the 'symmetrical atmosphere' in the rest of the text) and one asymmetrical model that takes into account longitudinal and latitudinal asymmetries (referred to as 'asymmetrical atmosphere' in the rest of the text), leaving to future work a more comprehensive study of atmospheric and field perturbation models. 
A comparison between the two models will reveal the sensitivity of the proton losses to different atmospheric configurations. Profiles of the two atmospheric models for the different flybys are plotted in Figure \ref{fig:atmos_density_fp}. The global distribution of the asymmetrical atmosphere is visualised for the three flybys in Figure \ref{fig:atmos_vis}.

\subsubsection{Symmetrical atmosphere}

The first model that we consider is a one-dimensional, spherically symmetric exponential model based on \citeA{Saur2002InterpretationPasses}. This model is described by:

\begin{equation}\label{eq:hom_atm}
    n(r) = n_0 \exp \frac{-z}{H},
\end{equation}

Where $n_0$ is the number density at the surface, $z$ is the altitude (given by $r-R_{Io}$), and $H$ is the scale height. We used $n_0=10^9$ cm\textsuperscript{-3} and $H = 100$ km, giving a column density of $1\times10^{16}$ cm\textsuperscript{-2}.
The SO$_2$ vertical column density is not well determined. It varies between $1\times10^{16}$ cm\textsuperscript{-2} and $15\times10^{16}$ cm\textsuperscript{-2} depending on the location of the observation and the method of observation \cite{Bagenal2020TheEuropa}. 
For our symmetrical atmosphere we use an SO$_2$ column density of $1\times10^{16}$ cm\textsuperscript{-2} as usually inferred from plasma/atmosphere interaction modeling (e.g. \citeA{Blocker2018MHDAtmosphere}).

\subsubsection{Atmosphere with longitudinal and latitudinal asymmetries}
\label{ss_asymmetrical_atmosphere}

Our second model is an SO$_2$ atmosphere that presents latitudinal and longitudinal variations based on \citeA{Blocker2018MHDAtmosphere}, which is visualised in Figure \ref{fig:atmos_vis}. The latitudinal variation in this model is based on the analysis of reflected solar Lyman $\alpha$ intensity data from the Hubble Space Telescope (HST) by \citeA{Strobel2001TheHydrogen}. It is thought that this latitude dependency coincides with the regions of active volcanism and stronger sublimation of frost layers \cite<e.g.>[]{Spencer2005Mid-infraredAtmosphere}. Described by:

\begin{equation}\label{eq:blocker1}
    n(r,\theta,\phi) = (n_{pol} + n_s(\theta,\phi)) \exp\left(\frac{-z}{H}\right)
\end{equation}

\noindent where $n_{pol} = 0.02n_0$ and $n_0$ is the surface density at the equator. We set $n_0 = 4 \times 10^8$ cm\textsuperscript{-3} and $H=140$ km. These values are the same as in atmosphere model 2 in \citeA{Blocker2018MHDAtmosphere}. The latitude and longitude dependencies are included as three factors $\beta_1$, $\beta_2$, and $\beta_3$:

\begin{equation}\label{eq:blocker2}
    n_s(\theta,\phi) = (n_0-n_{pol})\exp\left(-\left(\frac{0.5\pi-\theta}{0.625rad}\right)^2\right)\beta_1\beta_2\beta_3
\end{equation}

\noindent $\beta_1$ is the day/night asymmetry. This asymmetry is thought to exist due to the strong temperature dependence of sublimation and its role in supporting the atmosphere. Using high-resolution spectra gathered by the Gemini telescope, \citeA{Tsang2016TheEclipse} found a decrease in atmospheric column density from the day-side to the night-side of factor five. This work is the basis upon which the day-night asymmetry is modeled using:

\begin{equation}
    \beta_1(\theta,\phi) = \left(1+\frac{2}{3} \cos\left(\psi(\theta,\phi,\theta_s,\phi_s)\right)\right),
\end{equation}

\noindent where $\psi$ is the solar zenith angle. The local times used in order to calculate the subsolar point, given by $\theta_s$ and $\phi_s$, can be found in Table \ref{tab:Flyby parameters}. $\beta_2$ adds the sub-/anti-Jovian asymmetry. An asymmetry in the atmospheric structure of Io is supported by both HI Lyman-$\alpha$ images from HST \cite{Feaga2009IosAtmosphere} and data from the Submillimeter Array \cite{Moullet2010SimultaneousArray}. This is described using:

\begin{equation}
        \beta_2(\phi) = \left(1+\frac{3}{5}\sin\left(\phi-\frac{1}{2}\pi\right)\right).
\end{equation}

\noindent $\beta_3$ represents the downstream/upstream asymmetry. According to \citeA{Saur2002InterpretationPasses} this asymmetry exists due to the drag force of the flowing plasma on Io's atmosphere. It is described using:

\begin{equation}\label{eq:last_blocker}
    \beta_3(\phi)=\left(1+\frac{1}{3}\sin(\phi-\pi)\right).
\end{equation}

The properties of the asymmetrical atmosphere assumed for each flyby are summarized in Table \ref{tab:atm_io}.

\subsection{Electromagnetic fields around Io}

\begin{table}
\caption{Atmospheric Properties of Io's Simulation Runs}\label{tab:atm_io}
\begin{tabular}{l l l l   }
Model Scenario & $n_{0}$ (cm$^{-3}$)& $H_0$ (km) & Asymmetry\textsuperscript{a}\\
\hline
atm. model I24&4.0~$\times$~10$^{8}$ & 140&  up-/downstream, anti-/sub-Jovian, day/night\\
atm. model I27&4.0~$\times$~10$^{8}$ & 140 &  up-/downstream, anti-/sub-Jovian, day/night\\
atm. model I31&2.0~$\times$~10$^{8}$ & 100 &  up-/downstream, anti-/sub-Jovian, day/night\\
\hline
\multicolumn{4}{l}{\textsuperscript{a}\footnotesize{Applied longitudinal asymmetry in the atmospheric configuration.}}\\
\end{tabular}
\end{table}

In our simulations we will consider both homogeneous and inhomogeneous electromagnetic fields. In the homogeneous electromagnetic fields, no perturbations due to the interaction of Io and the plasma that is corotating with Jupiter will occur. In the inhomogeneous fields the perturbations are simulated using an MHD approach. By simulating the particle flux for both cases we can determine the contribution of the perturbed fields to the measured flux, e.g. by determining if forbidden fields cause any of the measured proton losses. We will demonstrate in the next section that the inhomogeneous fields (as opposed to the homogeneous fields) can have a strong influence on the energetic charged particle depletion.
 
The homogeneous magnetic field $\textbf{B}$ is obtained from the Galileo MAG data from Io and does not take into account the tilt of the background field. The electric field $\textbf{E}$ is calculated using the relation $\textbf{E}=-\textbf{v}\times\textbf{B}$, in which $\textbf{v}$ is the velocity of the corotational plasma. The values for $\textbf{B}$ and $\textbf{v}$ in the homogeneous case are the same as the initial values used for the inhomogeneous case (see Table \ref{tab:initial_io}). 

The inhomogeneous magnetic field is obtained using the MHD simulations of \citeA{Blocker2018MHDAtmosphere}, consistent with the asymmetrical atmospheric description proposed in Section \ref{sec:method_atmospheres}. The unique feature of the \citeA{Blocker2018MHDAtmosphere} MHD simulations is the absence of a prescribed magnetic field induced in Io’s putative magma ocean, as initially proposed by \citeA{Khurana2011EvidenceInterior}.  \citeA{Blocker2018MHDAtmosphere} propose a convincing match to the Galileo Magnetometer data with the simple assumption of an asymmetrical atmosphere and no induced dipole. Quoting their paper, they claim that “we  find that the measured perturbations can be primarily caused by the plasma interaction with the longitudinally asymmetric atmosphere. This implies that a significant magnetic induction signal from a partially molten magma ocean is not necessarily required to explain the Galileo magnetometer data." These simulated magnetic perturbations, consistent with the Galileo observations, are used in our study of the proton losses. For further details on the MHD simulation we refer the reader to the paper. We will demonstrate in the next section that the inhomogeneous fields (as opposed to the homogeneous fields) can have a strong influence on the energetic charged particle depletion.

\begin{figure}
    \centering
    \makebox[\textwidth][c]{
    \includegraphics[width=1.3\textwidth]{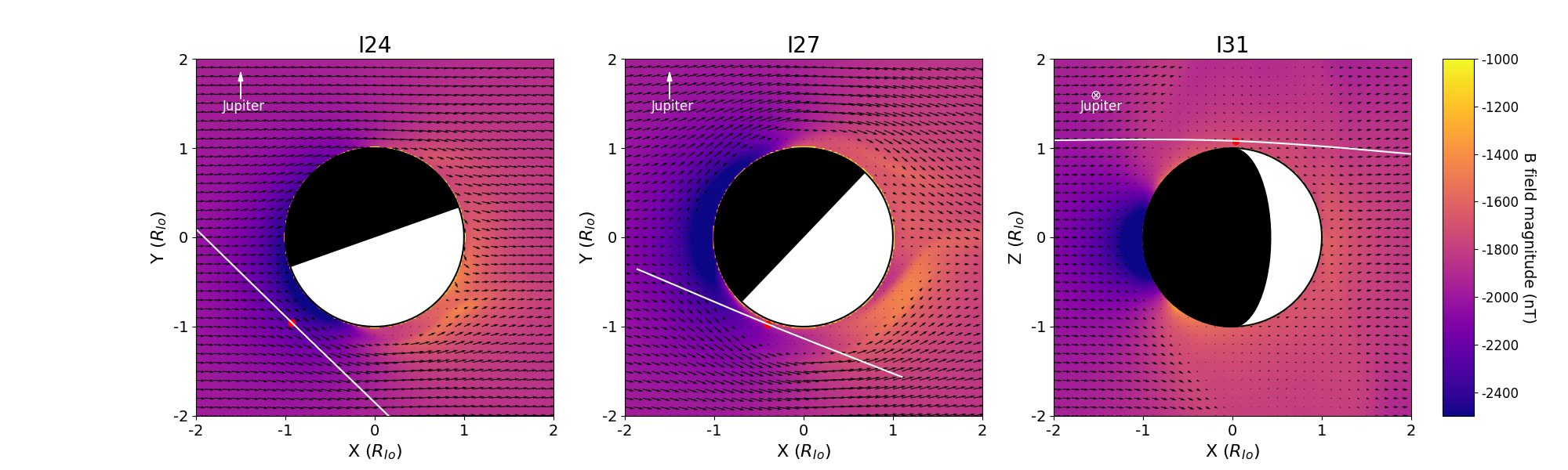}}
    \caption{Visualisation of the magnetic field magnitude (color map) and the plasma flow (arrows) produced by the MHD model as run by \protect\citeA{Blocker2018MHDAtmosphere} for flybys I24, I27, and I31. The first two panels represent cuts at z=0 and the third panel is a cut at y=0. The day/night side are indicated with the white and black sides respectively. The point of closest approach is indicated with the red point. Note that for I31 an `edge-on' view of the system is displayed in order to better visualise the polar flyby.}
    \label{fig:MHD_vis}
\end{figure}

The MHD simulations are used to self-consistently model the plasma interaction of Io's atmosphere with the corotating plasma from Jupiter's magnetosphere. The applied MHD model has already been used to study Io's plasma interaction by \citeA{Blocker2018MHDAtmosphere}. The authors adjusted the model to Io's plasma interaction from the model which was developed for Ganymede's plasma interaction by \citeA{Duling2014}. 
A detailed description of the model is given in \citeA{Blocker2018MHDAtmosphere}.
The model solves a set of three-dimensional equations consisting of the continuity equation, the momentum equation, the induction equation, and an equation for the internal energy on a spherical grid. 
It takes into account collisions between ions and neutrals, plasma production and loss due to electron impact ionization and dissociative recombination, and the ionospheric Hall effect. 
Io's atmosphere is implemented as a globally asymmetric static SO$_2$ neutral atmosphere (see Equation \ref{eq:blocker2}) which is described in detail in Section \ref{sec:method_atmospheres}.  
The atmosphere description is introduced in the MHD model in the terms of ion-neutral collisions, electron impact ionization, and the ionospheric Hall effect.
An overview of the parameter settings for the initial and boundary conditions in this model is given in Table \ref{tab:initial_io}.\\ 
Figure \ref{fig:MHD_vis} shows the magnetic perturbations around Io for each Galileo flyby addressed here. In the three panels we see a strong increase of the magnetic field magnitude upstream of the moon  due to the pile-up of the magnetic field lines. The formation of the Alfv\'en wings is clearly visible in the reduced plasma flow velocity (arrows in third plot) north and south of the moon. They are bent back with respect to the unperturbed background magnetic field.  
The plots in the equatorial plane  (I24 and I27) show that the plasma flow is diverted around the moon and accelerated at the flanks. 
Most of the plasma flow into the ionosphere is reduced and is swept around the moon and the wings. 
Upstream of the moon the plasma flow velocity is decreased and downstream of the moon it increases again to the background value due to the Lorentz force. The asymmetry of the plasma flow and the magnetic field between the Jupiter-facing side and anti-Jupiter facing side shown in the first two plots is due to the plasma interaction with an asymmetric atmosphere. The comparison between the plasma flow between the I24 and I27 case shows that the plasma flow upstream of the moon is slowed down more effectively for the I27 case than for the I24 case. 
This is an effect of the asymmetric atmosphere (see Figure \ref{fig:atmos_vis}). The Galileo flyby I31 provided a close pass of the north polar region as presented in the trajectory in the third panel of Figure \ref{fig:MHD_vis}.
The spacecraft crossed the northern Alfv\'en wing. Due to its  close flyby to Io the electromagnetic perturbations are not only affected by the Alfv\'enic interaction but also by the ionospheric current system.

In \citeA{Blocker2018MHDAtmosphere} a comparison between the MHD model and the measured components of the magnetic field are shown in Figure 7 for I24 (atm. Model 2) and Figure 10 for I27. Atm. model 2 is not shown in the plots for I27, but the results are similar as to atm. Model 3. For those flybys a comparison is also made to electron density derived from Galileo PWS and plasma velocity components from Galileo PLS (PWS data is shown in Figures 9 and 12). In \ref{app_I31} we provide the MHD simulations for I31,  and a comparison with the magnetic field data which were not provided in \citeA{Blocker2018MHDAtmosphere}.

\begin{table}
 \caption{Initial and Boundary Condition Values and Calculated Parameters in the IPhiO System for Io's Simulations}
 \label{tab:initial_io}
 \begin{tabular}{lcccc}
 Flyby & $\underline{B}_0$& $\underline{v}_0$ & $\rho_0$  & $\epsilon_0$\textsuperscript{a} \\
 & (nT)& (km/s) & (amu/m$^{-3}$) & (nPa) \\
 \hline
 I24& (310, 530, $-$1910)&(57, 0, 0)      & 3.06~$\times$~10$^{10}$ & 21.3\\
 I27 & (300, 510, $-$1890)&(57, 0, 0)      & 2.05~$\times$~10$^{10}$ & 24.6\\
 I31 & (250, $-$675, $-$1930)&(57, 0, 0)& 2.95~$\times$~10$^{10}$ & 29.9\\
 \hline
 \multicolumn{5}{l}{\textsuperscript{a}\footnotesize{$\epsilon_0=3/2 n k_B(T_e+T_{i,0})$ with $k_B T_{i,0}=50$eV for I24 \cite{Frank2000}, $k_B T_{i,0}=90$eV for I27 \cite{Frank2001},}}\\
\multicolumn{5}{l}{\textsuperscript{}\footnotesize{and $k_B T_{i,0}=75$eV otherwise \cite{Kivelson2004MagnetosphericSatellites}.}}
 \end{tabular}
\end{table}

\subsection{Particle tracing}
\label{s_tracing}

\begin{table}
\caption{Parameter values used in the particle tracing simulations.}
\centering
\begin{tabular}{l c}
\hline
 Technical parameters particle tracing simulation  & Value  \\
\hline
  Integration time step particle trajectory [s]  & 0.001  \\
  Number of integration time steps per gyration  & $\sim32$ \\
  Maximum number of time steps per particle &  90000\\
  Number of particles per spacecraft trajectory step &  100  \\
  Number of energy bins & 10 \\
  Spacecraft trajectory time step [s] & 0.6s \\
  \hline
 Physical parameters particle tracing simulation  & Value  \\
   \hline
  Surface density symmetrical atmosphere [cm$^{-3}$] & 1~$\times$~10$^{9}$ \\
  Scale height symmetrical atmosphere [km] & 100 \\ 
  Surface density asymmetrical atmosphere\textsuperscript{a} [cm$^{-3}$] & 4.0~$\times$~10$^{8}$ \\
  Scale height asymmetrical atmosphere\textsuperscript{a} [km] & 140 \\
  $\underline{B}_0$ [nT] & (0, 0, $-$2000) \\
  $\underline{v}_0$ [km/s] & (60, 0, 0)  \\
  \hline
   \multicolumn{1}{l}{\textsuperscript{a}\footnotesize{Includes following asymmetries: up-/downstream, anti-/sub-Jovian, day/night, see Section \ref{ss_asymmetrical_atmosphere}}.}
   \\
\hline
\label{tab_particle}
\end{tabular}
\end{table}

We simulate the measured flux of energetic protons using a Monte Carlo particle back-tracing code, developed from \citeA{Huybrighs2017OnMission,Huybrighs2020AnDepletions,Huybrighs2023}. Particle tracing has been employed previously at various Galilean moons to investigate the ion dynamics as affected by field perturbations (e.g. \citeA{Selesnick2009ChargeIo,Poppe2018ThermalMagnetosphere,Liuzzo2019EnergeticCallistob,Breer2019EnergeticEuropa,Carnielli2019FirstIonosphere,Plainaki2020,Addison2022EffectIce,Nordheim2022MagnetosphericSurface,Huybrighs2023}) and atmospheric charge exchange (e.g. \citeA{Huybrighs2020AnDepletions,Addison2021InfluenceWeathering}).

Because the energetic protons have no influence on the electromagnetic fields they can be treated as test particles. In the simulation protons are traced back in time from the EPD detector. For each particle it is determined if it impacts and/or charge exchanges. If, at any point along the particle trajectory, a surface impact ($r\leq R_{\text{Io}}$) or charge exchange with the atmosphere occurs, the particle is considered lost. This analysis is performed for many particles to determine the depletion as a fraction of the (normalized) undisturbed flux. All the steps of this process, from initiating the particle to the possible loss, are described in the following paragraphs. The relevant input parameters are summarized in Table \ref{tab_particle}.

We take into account the energy distribution of the proton flux in the simulation. Particles are initiated at a random energy within the energy range of the TP1 or TP3 channel, using a Monte Carlo approach. Once the particles are initiated they are binned according to their initial energy (10 bins) and pitch angle (64 bins). The energetic proton flux decreases with increasing proton energy, though less than an order of magnitude over both the TP1 or TP3 channels \cite{Paranicas2003}. We account for this decrease in flux with increasing energy by weighting the particles according to the energy distribution of the flux. Since we found in  \citeA{Huybrighs2020AnDepletions} that the simulation is not very sensitive to the specific drop off profile in one energy channel, we have used the same relative change in flux ($\sim$40 percent decrease in flux from the beginning to the end of the energy channel) as used in \citeA{Huybrighs2020AnDepletions,Huybrighs2021ReplyDepletions,Huybrighs2023} for Europa in this study.

We use a leap frog numerical integration scheme of the Lorentz force to trace the protons back in time \cite{Huybrighs2017OnMission,Huybrighs2020AnDepletions,Huybrighs2021ReplyDepletions,Huybrighs2023}. We simulate the trajectories of energetic protons under a homogeneous and an inhomogeneous electromagnetic field (the latter is discussed in the next section). In the homogeneous case we neglect any perturbations in the electromagnetic field resulting from the interaction of the corotational plasma with Io. Despite that the situation with homogeneous fields is not expected to occur at Io in reality, we consider including this case relevant so that we can determine relative the contribution of the inhomogeneous fields to the proton losses.

We neglect the pitch angle distribution of the protons, thus our pitch angle distribution is uniform. The consequence of this simplification is that our simulations will not be able to reproduce losses that are the result from variations in the background pitch angle distribution. However, we consider this approach suitable because our simulations will show that the deviations from the background distribution are much more pronounced features than the measured variations in the data upstream of Io.

In the simulation back-traced particles are assumed to contribute to a flux decrease, if they impact on Io's surface or charge exchange with neutrals from its atmosphere. This is because when a back-traced particle that impacts Io or charge exchanges, is traced forward it would have to originate from the atmosphere or the surface, and neither of those are sources of energetic protons. It is then assumed that the corresponding particle does not exist and should thus count towards a loss of flux. Perturbed fields can form 'forbidden regions' which will lead simulated particles to impact on Io's surface, resulting in an apparent loss as measured by the instrument. In reality the particle has effectively been deflected away from the 'forbidden region'. In the simulation a particle is considered to have impacted Io's surface when its trajectory intersects the moon's surface. If the energetic proton charge exchanges, it is assumed to turn into an energetic neutral atom (ENA). ENAs are not bound by the electromagnetic fields, and will escape from the system at a high velocity, thereby contributing to the proton loss. The probability of charge exchange with the atmospheric neutrals is determined using Equation (\ref{eq:prob_charge_ex}) from \citeA{Birdsall1991}:

\begin{equation}\label{eq:prob_charge_ex}
    P = 1 - \exp(-n\sigma g \Delta t)
\end{equation}

\noindent Where $n$ is the local number density (of the atmosphere), $\sigma$ is the energy dependent charge exchange cross-section, and $g$ is the velocity of the ion relative to Io. For each time-step along the trajectory of the back-traced particle, $P$ is calculated \cite{Brieda2011ChargeCEX}. Charge exchange is said to occur if at any point along the trajectory a randomly generated number from a uniform distribution between 0 and 1 is less than the calculated probability. The collision cross section ($\sigma$) has to be determined empirically for different particles. Note that because we use the charge exchange cross section of protons on O$_2$ as an approximation of protons on SO$_2$, any neutral SO$_2$ densities derived from our particle tracing analysis are also an approximation.

We neglect the following loss processes: charge exchange with the low energy ions, energy loss due to Coulomb collisions with ions and electrons, and wave particle interaction. \citeA{Mauk2022} suggest that charge exchange with the low energy ions could dominate over charge exchange with neutrals near Io's orbit. However, in this study we are dealing with the environment very close to Io ($\sim2 $R$_{Io}$) where the neutral density of the atmosphere is orders of magnitude larger than the ion density ($\sim$2000 cm$^{-3}$; \citeA{Bagenal2020TheEuropa}). Furthermore, the charge exchange cross section of the energetic protons on the low energy ions is two orders of magnitude smaller than the cross section of energetic protons and the neutrals (see Table 9 in \cite{Mauk2022}). We therefore expect charge exchange with the low energy plasma to be only a minor effect compared to charge exchange with the neutral environment close to Io. 

We also assume that the effect of Coulomb collisions as a loss process is negligible. Coulomb collisions could result in energy losses of energetic protons, which would result in apparent losses of the protons. However, \citeA{Nenon2018AOrbit} has shown that for energies of 0.1 MeV the loss rates due to charge exchange with neutrals of the Io and Europa torus are orders of magnitude larger than the losses due to Coulomb collisions. They specifically modelled Coulomb collisions with bound electrons of the neutral particles of the Jovian hydrogen corona and Io and Europa gas torus, with bound electrons of the cold plasma ions and with free electrons of the cold plasma of the inner magnetosphere of Jupiter. We also neglect Coulomb collisions as a loss process for the TP3 channel, if Coulomb losses with the particle environment (which at the distances of this work is dominated by the atmospheric density) were a major loss process in this channel for the data segments we considered, then we would expect to see a correlation between the density of the atmosphere and the dropout, which is not what we observe for TP3.

We also neglect wave particle interaction as a loss process that plays a major role in the proton loss features we observe within $\sim2 $R$_{Io}$. We build on \citeA{Mauk2022} which argues that near Io's orbit charge exchange losses dominate for protons (roughly below 200 keV). As we will demonstrate later, the depletion features at 540-1250 keV can be explained by the field configuration, indicating that also at these higher energies wave particle interactions do not play a major role in proton depletion within $\sim2 $R$_{Io}$.

We ignore the possibility of protons re-entering the simulation box after exiting, because the bounce periods are larger than the time it takes the corotation to cover the spatial scales of the relevant proton loss features. Therefore, we consider it unlikely that losses occurring closer to upstream would create the measured losses further downstream by modifying the proton distributions. See \ref{appendix_bounce_period} for a more detailed discussion.

Considering the complexity of the processes leading to proton depletion in Io's atmosphere (electric and magnetic field perturbation, neutral density etc.) the atmospheric density cannot be directly inferred from the measurements. Thus, we analyze the data with a forward modeling approach, prescribing an atmospheric distribution and electric and magnetic field perturbations around Io to compute the proton depletion and compare it to the Galileo data.

\section{Results}\label{sec:results}

\subsection{Measured energetic proton depletions during flybys I24, I27, and I31}\label{sec:epd_data}

\begin{sidewaysfigure}
    \includegraphics[width=\textwidth]{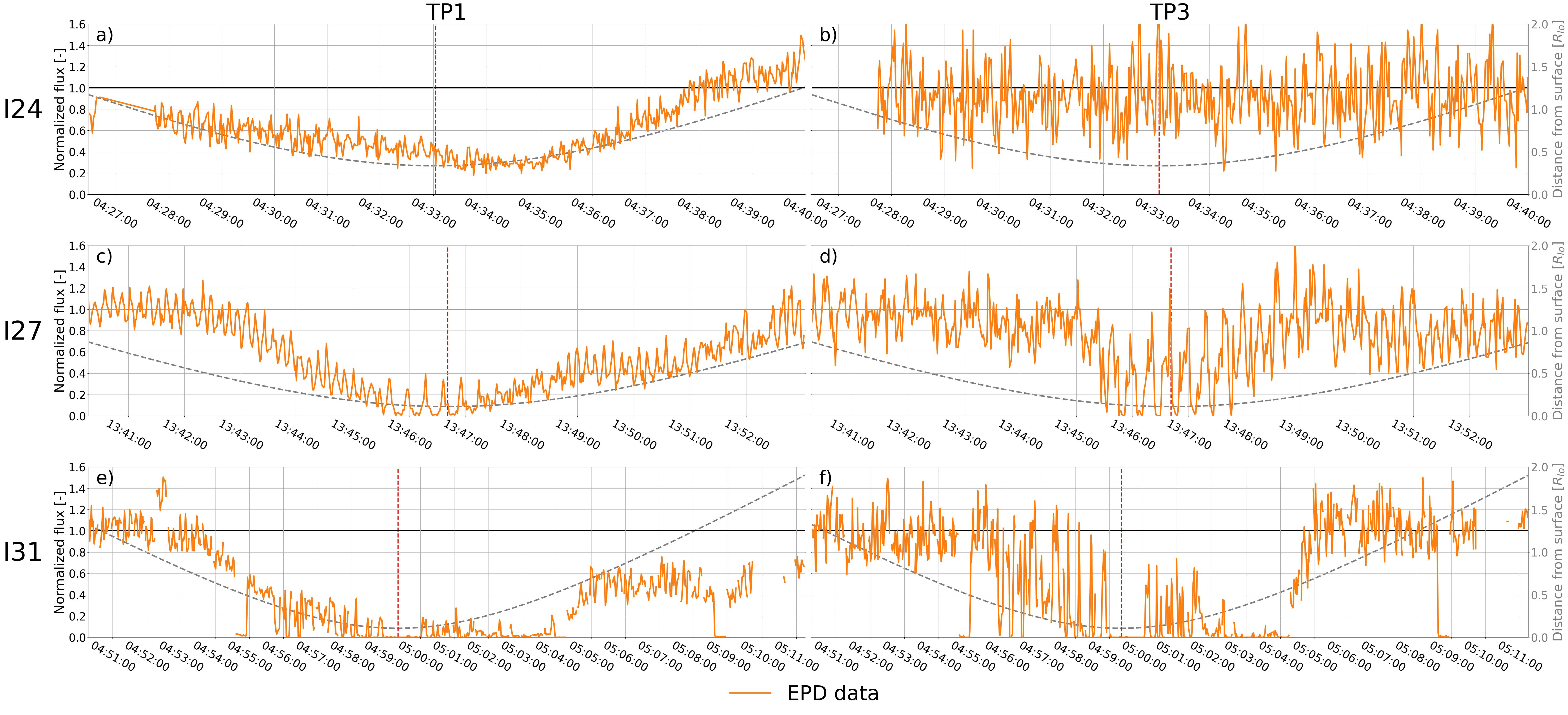}
    \caption{The normalized energetic proton flux as detected by the EPD is shown using the solid orange line for the three Io flybys I24 (a and b), I27 (c and d), and I31 (e and f). The TP1 (155-224 keV) flux is shown in the left side panels (a, c, and e) and the TP3 flux (540-1250 keV) in the right side panels (b, d, and f). The point of closest approach is marked with a vertical red dashed line. Note the periodic variation in the flux due to the rotation and scanning motion of the spacecraft. The dotted grey line shows the distance to Io in R$_{Io}$ (axis on the right side).}
    \label{fig:epd_data}
\end{sidewaysfigure}

In Figure \ref{fig:epd_data} we show the measured energetic proton flux (solid orange line) for the three flybys in the TP1 and the TP3 channel. The vertical red dashed line indicates the point of closest approach for each flyby and the grey dashed line indicates the radial distance to Io. The flux in these plots has been normalized to the mean of two cycles of the instrument from the start of the flyby segment. This allows for comparison with the particle tracing code which expresses depletion numbers in percentages. The periodic variation of the flux measurements is due to the rotation and scanning motion of the spacecraft (as mentioned above, this takes about 140 s). 

We observe that the region over which depletions occur along the spacecraft trajectory is larger in the TP1 channel than the TP3 channel for I27 and I31. Furthermore, for flyby I24 no depletion is observed in the TP3 channel (b) at all. This decrease in size or disappearance of the depletion region in TP3 is suggestive of charge exchange playing a major role in the depletion because the charge exchange cross section decreases significantly from the TP1 to the TP3 channel (see Figure \ref{fig_cross}). 
For I24 in TP1 the depletion region is symmetrical with the distance to Io, while in I27 and I31 the symmetry breaks down. During both I27 and I31 (TP1) the depletion region is more extended on the day/downstream side (right side in panels a c and e). For TP3 I24 does not show a trend of depletion. The extent of the depletion region in I27 and I31 (TP3) is roughly symmetrical with respect to the distance to Io. However, during I31 there is a complicated structure, with a  near complete depletion occurring after the closest approach from 05:02 to 05:04 that does not occur post closest approach. 

Note that we do not attribute the measured depletions to the gyroradius effect. That is, the effect where particles impact on Io's surface because they encounter it during their gyromotion along the field line. Figure \ref{fig_gyro} shows that the gyroradius of the respective protons is smaller than the closest approach altitude, even during the closer flyby I27. Furthermore, the gyroradii shown in Figure \ref{fig_gyro} are for the upstream conditions, and represent a lower estimate. Near Io the magnitude of the field will increase due to magnetic pile up, therefore the gyroradius will be even smaller near Io.

An artifact that introduces nonphysical dropouts can occur in the EPD data, when measurements are taken on motor position 7 \cite{Jia2021CommentEtal.,Huybrighs2021ReplyDepletions}. For flybys I24 and I27 this artifact does not occur because the EPD was kept in a fixed motor position, which is different from motor 7. For the data segment of I24 and I27, considered in this study, the motor position is fixed to position number 4. During I31 motor position 7 is one of the motor positions that is scanned, however we find that the EPD data gathered, when the motor is in position 7, does not appear to fall outside of the general trend. We conclude that the data artifact will not affect the conclusions of this study, because we focus on the general trend and structure of the depletion rather than depletion features measured at single motor positions. 

\subsection{Data-simulation comparison }\label{sec:results_tp1}

\begin{sidewaysfigure}
    \includegraphics[width=\textwidth]{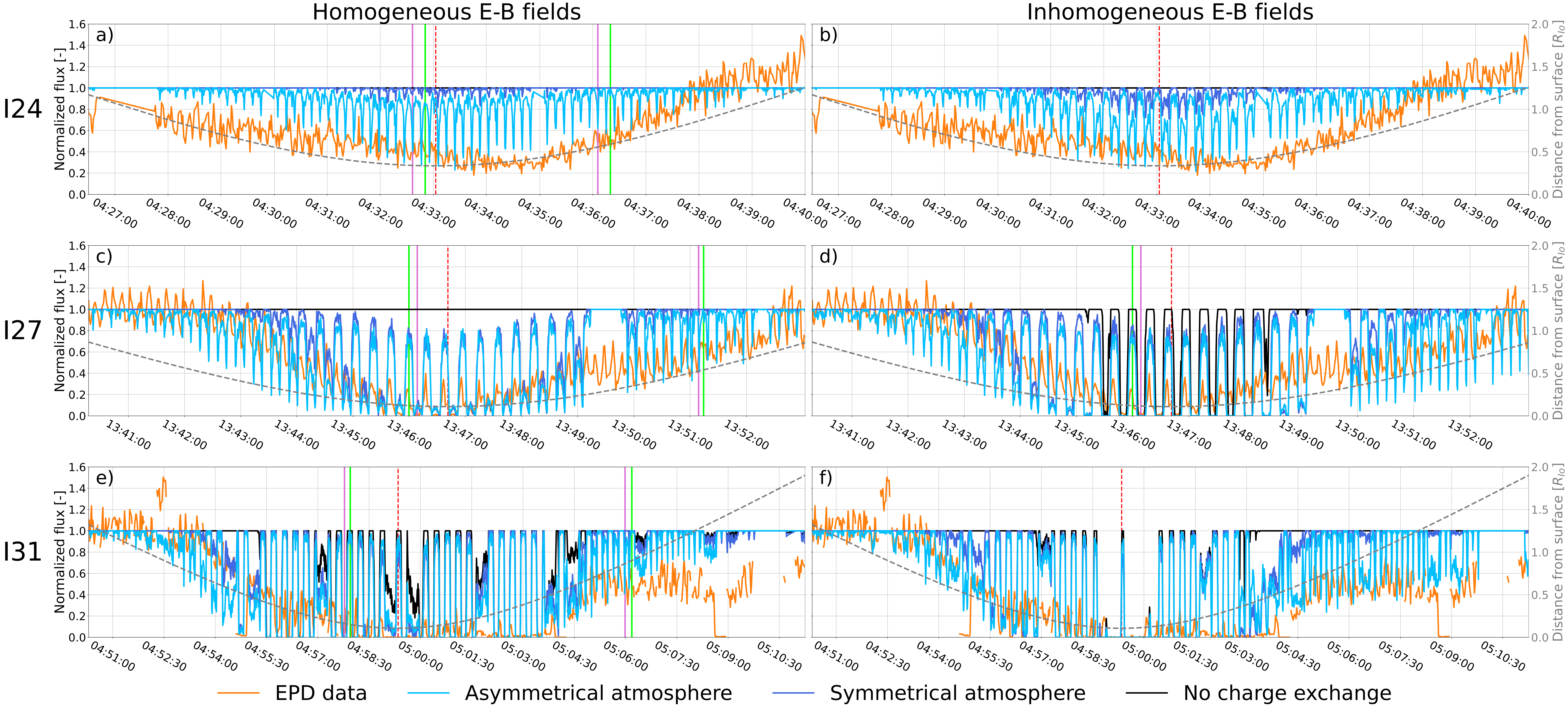}
    \caption{Normalized energetic proton flux as detected by EPD in the TP1 channel (155-224 keV) (orange solid line) compared to the depletion as predicted in the particle back-tracing simulations for the three Io flybys I24 (a and b), I27 (c and d), and I31 (e and f). The simulations results where no atmospheric charge exchange was taken into account are shown in black. Simulations that consider charge exchange with the symmetrical atmospheric model are shown in dark blue and those considering the asymmetrical model are shown in light blue. The left side panels (a, c, and e) show the results when assuming homogeneous electromagnetic fields, and the right side panels (b, d, and f) show the results when perturbed electromagnetic fields are taken into account. The dotted grey line shows the distance to Io in R$_{Io}$ (axis on the right side). Vertical green and pink lines in the right column correspond to the starting point of particle trajectories shown in Figure \ref{fig:I27_traj_perturbed}. Vertical green and pink lines in the left column correspond to individual particles trajectories plotted in Figures \ref{fig:I24_traj}, \ref{fig:I27_traj} and \ref{fig:I31_traj}. }
    \label{fig:results1}
\end{sidewaysfigure}

\begin{sidewaysfigure}
    \includegraphics[width=\textwidth]{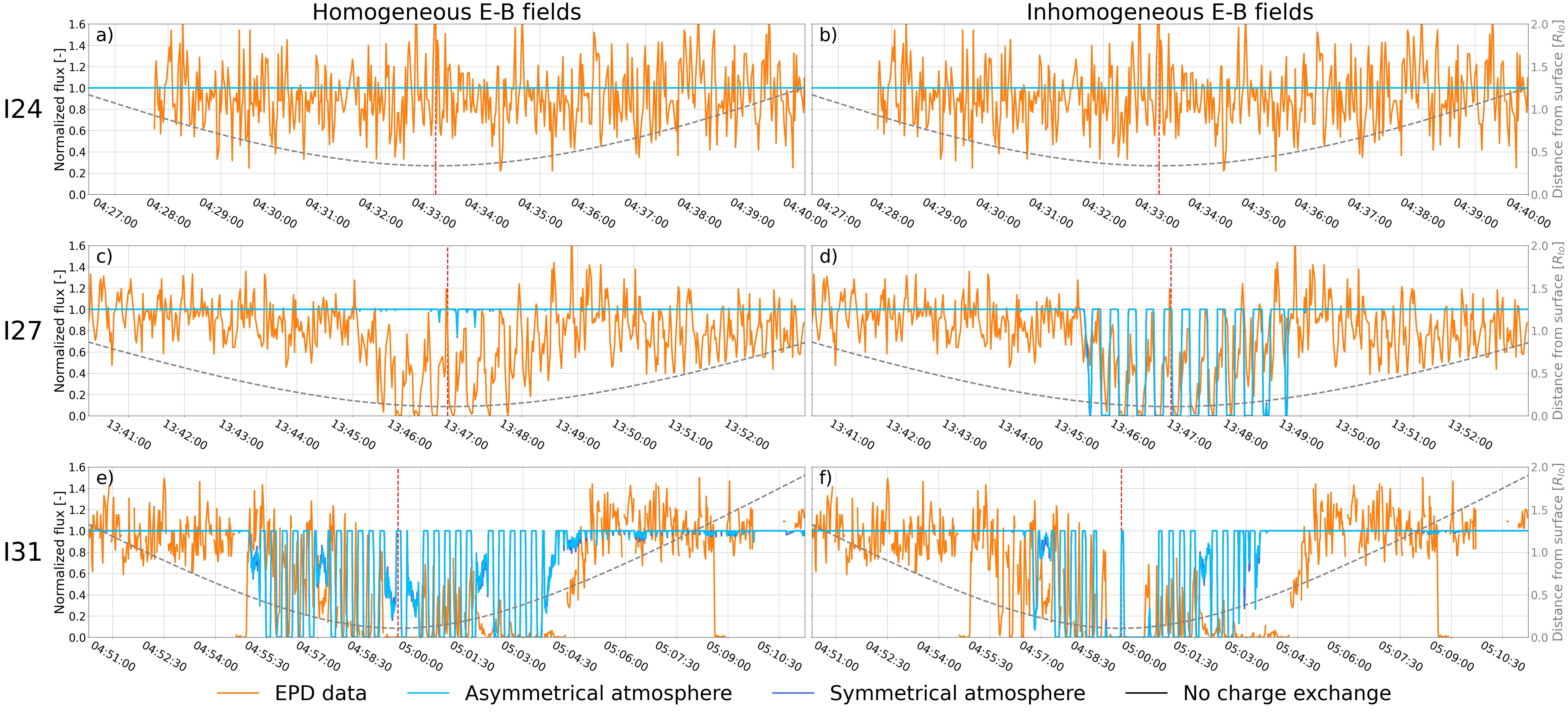}
    \caption{Normalized energetic proton flux as detected by the EPD in the TP3 channel (540-1250 keV) (orange solid line) compared to the depletion as predicted in the particle back-tracing simulations for the three Io flybys I24 (a and b), I27 (c and d), and I31 (e and f). Note that the simulations without charge exchange and with charge exchange fall together exactly, because the effect of charge exchange is negligible. The left side panels (a, c, and e) show the results when assuming homogeneous electromagnetic fields, and the right side panels (b, d, and f) show the results when assume perturbed electromagnetic fields. The dotted grey line shows the distance to Io in R$_{Io}$ (axis on the right side).}
    \label{fig:results2}
\end{sidewaysfigure}

In Figure \ref{fig:results1} the results are shown for the back-tracing simulations for the TP1 energy channel (155-224 keV). In the plots on the left of the figure (a, c, and e) the results are shown for the homogeneous fields, while the right side (b, d, and f) shows the results for the inhomogeneous fields. The first row (a and b) shows the results for I24, the second row (c and d) for I27, and the third row (e and f) for I31. In each plot, the EPD measurement is shown in orange. The depletion predicted while assuming no atmospheric interactions is shown in black, and the simulations considering charge exchange with the atmosphere are shown in dark blue for the symmetrical atmosphere and light blue for the asymmetrical atmosphere. In Figure \ref{fig:results2} simulation results for the TP3 energy channel (540-1250 keV) are shown. The same colors and labels are used as in Figure \ref{fig:results1}. 

We will argue that at 155-224 keV charge exchange is the dominant loss process (I24) or a main loss process combined with  surface absorption (I27 and I31). The measured losses are also sensitive to the description of the electromagnetic field (I27 and I31). Due to the collision cross section decreasing exponentially with energy, the likelihood of charge exchange occurring with high energy particles is smaller than for lower energy particles. Therefore the 540-1250 keV (TP3) simulation results do not show any significant difference when atmospheric charge exchange in included, all the solutions are overlapping each other (black, dark blue en light blue line). We will argue that 540-1250 keV protons during I24 are not affected by Io and that inhomogeneous electromagnetic fields are responsible for the proton depletion observed during the two closer flybys I27 and I31. Due to I31's polar trajectory surface absorption is also an important loss process during that flyby.

\subsubsection{Flyby I24}

Panels a and b in Figure \ref{fig:results1} show the results for flyby I24 (155-224 keV, TP1). If atmospheric charge exchange is not taken into account (black line), no depletion at all is predicted for I24 for either the homogeneous fields (a) or the MHD fields (b). This indicates that neither impact on the surface or the configuration of the electromagnetic fields can explain the proton losses during flyby I24.

The asymmetrical atmospheric model (light blue) partially reaches the approximate depth of the depletion. We interpret the partial agreement as an indication that charge exchange with the atmosphere is likely the main cause of the observed depletion region and therefore the dominant loss process at 155-224 keV during this flyby. However, the simulated depletion does not reach the depth of the depletion consistently. Instead we observe that the simulated flux periodically returns to the unperturbed level, whereas the data does not. This periodic discrepancy will be discussed further in Section \ref{s_discrepancy}. 

For the symmetrical atmospheric model (dark blue line), we see depletion of no more than about 10\% around the point of closest approach (red dashed line) when assuming homogeneous electromagnetic fields (a). The difference with the symmetrical atmosphere is that the density at the altitude of the I24 flyby is higher due to a higher scale height (Figure \ref{fig:atmos_density_fp}). When also including the MHD fields (b), we see a slight increase in depletion up to about 20\%. Though the MHD simulations have not been run self-consistently with the symmetrical atmosphere, it also appears unlikely that a combination of this particular symmetrical atmosphere and perturbed fields will explain the depth of the proton loss. Overall, we conclude that it is unlikely that charge exchange with this symmetrical atmosphere can explain the observed proton losses. 
While perturbed fields (panel b) slightly modify the depletion pattern for both atmospheric models, we conclude that the overall effect of the perturbed fields on the proton losses for this flyby is minor.

At 540-1250 keV (TP3), Figure \ref{fig:results2}, even though there is a variability of the measured proton flux, there appears to be no correlation between the measured proton flux and the distance to Io. We consider it likely that the proton flux at 540-1250 keV during this flyby is not affected by Io. Indeed, the simulations demonstrate that neither surface impact, charge exchange or perturbed fields are expected to have any effect during this flyby. The lack of depletions in the TP3 channel, and the partial agreement between the simulated and measured depletions in the lower energy TP1 channel argues in favour of charge exchange being the dominant loss process at the TP1 energies (155-224 keV).

\subsubsection{Flyby I27}

\begin{figure}
    \centering
    \includegraphics[width=1.0\textwidth]{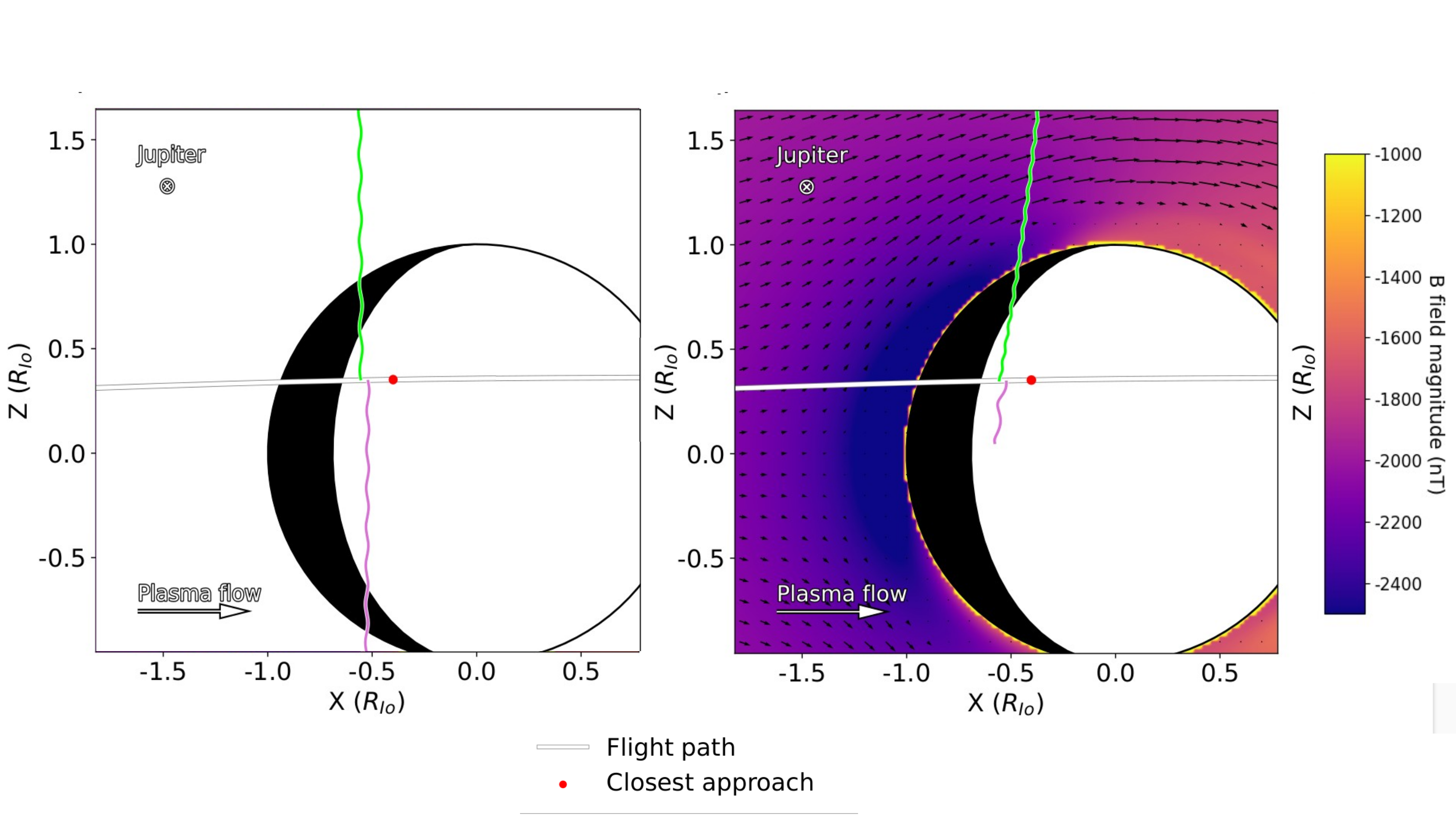}
    \caption{Particle trajectories during flyby I27 with homogeneous fields (left), and perturbed and tilted fields (right). The green particle is not considered lost, it neither impacts on Io's surface or charge exchanges. On the left the pink particle is lost because it charge exchanges with the atmosphere, however on the right its trajectory is guided towards the surface and impacts. In the background of the right panel the magnitude of the perturbed electromagnetic fields has been plotted. Note also that the gyromotion of the particle has been plotted. The panels on the left and right correspond, respectively, to the lines with the same color on the left of the closest approach in panels 'c' and 'd' in Figure \ref{fig:results1}.}
    \label{fig:I27_traj_perturbed}
\end{figure}

In the simulation results for flyby I27 (155-224 keV, TP1), no depletion is seen in the homogeneous fields plot (black line in panel c in Figure \ref{fig:results1}) when neglecting atmospheric charge exchange. However, with the inhomogeneous fields and no charge exchange (black line, panel d) we do see a strong periodic depletion between the 13:45:00 and 13:49:00, but no depletion predicted outside of this area. We attribute these depletions to the electromagnetic fields used in case d, which guide the particles to Io's surface in the back-tracing sense due to a tilt in the magnetic field that is neglected in the homogeneous case. An example of such a particle trajectory is shown in Figure \ref{fig:I27_traj_perturbed}. In forward sense this particle does not exist, because Io's surface does not emit energetic protons. The periodic losses in the black line in panel d can thus be considered as a 'shadowing' by Io's surface which absorbs the ions and that occurs specifically under this field configuration.

With atmospheric charge exchange taken into account the simulations show deeper depletions compared to I24 in the case with atmospheric charge exchange. We attribute this difference to the lower altitude of the closest approach (see Table \ref{tab:Flyby parameters}) and the corresponding higher atmospheric densities for both atmospheric models. We also observe that the difference in depth of depletion between the two models is smaller than for I24. Towards the extremities of the flyby the asymmetrical model still exceeds the simulated depletion depth for the symmetrical model, due to higher densities at higher altitudes in the asymmetrical model (see Figure \ref{fig:atmos_density_fp}). Near closest approach the density of the symmetrical model is comparable to that in the asymmetrical model, and the depth of depletion in the symmetrical case is also comparable to the asymmetrical case.
Just as for flyby I24 we observe that the simulated flux periodically returns to the unperturbed level, whereas the data does not. This periodic discrepancy will be discussed further in Section \ref{s_discrepancy}.

The extent of the loss region in the case with only the perturbed fields is smaller than the loss region in the data. For both atmospheric models the loss region with atmospheric charge exchange extends beyond the region in which losses occur associated with the choice of fields (black line, d), indicating that both charge exchange and an appropriate description of the fields are needed to explain the loss region.

At 540-1250 keV (TP3) proton depletion is visible in the EPD data around the point of closest approach (13:47:00). The predicted depletion for the homogeneous fields (c) is negligible. However, for the case with inhomogeneous and tilted fields (d) a series of depletions are predicted. We find that the simulation reproduces the general structure of depletion features well, with the depth of the depletion being reproduced, as well as the increase in flux between each depletion. From this general agreement we conclude that an appropriate choice of electromagnetic field can explain the majority of the depletion features during I27 at 540-1250 keV.

\subsubsection{Flyby I31}

I31 results are shown in panels e and f in Figure \ref{fig:results1}. Without taking atmospheric charge exchange into account (black lines) strong and periodic depletions are predicted for the homogeneous fields. This is due to the fact that I31 is a polar flyby (see Figure \ref{fig:flight_overview}). Charged particles move along the magnetic field lines, which are approximately parallel to Io's rotation axis. Therefore, when Galileo (along with the EPD) flies over the northern pole of Io, Io is placed directly in the path of charged particles moving (in forward sense) from south to north which blocks their path to the EPD (see Figure \ref{fig:I31_traj}, left column).
When assuming inhomogeneous fields (f), the depletion is localised closer around the point of closest approach between 04:57:00 and 05:03:00. We attribute this difference to deflection of energetic ions away from the surface by the inhomogeneous fields. However the simulations with the inhomogeneous fields are not able to explain the extent of the depletion region along the flyby.

When including the symmetrical (in dark blue) and asymmetrical (in light blue) atmospheric models we see that the extent of the depletion region is better reproduced. However, just as for flyby I24 and I27 we observe that the simulated flux periodically returns to the unperturbed level, whereas the data does not. This periodic discrepancy will be discussed further in Section  \ref{s_discrepancy}. 

Whereas for flyby I24 the depletion region correlates roughly with the radial distance to Io, for the downstream side of the I31 this correlation breaks down as an extended depletion region is visible, both in the data and the simulation with the asymmetrical atmosphere. It appears that the simulated dropout is sensitive to the enhanced atmosphere on the dayside/downstream side of Io. Figure \ref{fig:atmos_density_fp} shows that the density along the trajectory is symmetrical with respect to the closest approach, however the simulated proton loss is not. The proton loss is thus not fully determined by the density at the trajectory, but also by density structures elsewhere. We emphasize that charge exchange simulation is not exclusively sensitive to the atmospheric density along the spacecraft trajectory, but the density along the entire particle trajectory.

At 540-1250 keV (TP3) we observe that, just as for the TP1 channel, depletions are expected to occur even in the absence of perturbed fields (e). The periodicity is due to the spin of Galileo and EPD looking into directions for which Io blocks the field of view and prevents particles from reaching the detector.

We observe that by introducing perturbed fields the complete depletion $\sim$2 mins around the closest approach is reproduced. This implies that in the back-tracking simulation all ions impact Io's surface in this region. We consider this a 'forbidden field' region, wherein the specific configuration of the field prevents access of protons. However, large discrepancies between the simulated and measured proton loss occur elsewhere along the flyby. The predicted extent of the depletion regions is reduced in the inhomogeneous case as compared to the homogeneous case. For the homogeneous fields we see depletion between 04:55:30 and 05:04:30, while for the inhomogeneous fields the depletion is concentrated between 04:57:00 and 05:03:00. Which is shorter than the time during which depletions are observed in the measurement. A periodic depletion is seen in the measurement, around 04:56:00, but is not reproduced. A near complete region of depletion occurs from 05:01:00 to 05:04:30, which is poorly reproduced by the simulation. Our simulations for I27 show that at these high energies the bulk of the depletion feature can be explained solely by an appropriate choice of electromagnetic field. Therefore, we attribute the discrepancies between the simulated proton flux and the measurement to discrepancies between the simulated magnetic field and the real configuration.

\subsection{Periodic discrepancy between data and model at 155-224 keV}
\label{s_discrepancy}

As described in the previous sections a periodic discrepancy occurs between the simulated losses and the measured loss at 155-224 keV (TP1). While the simulated flux periodically returns to the unperturbed level, the data does not. We attribute this periodicity to the approximately 20 second spin of the Galileo spacecraft, which makes EPD scan different directions. Meaning that in certain directions the losses are predicted better than in others. We further interpret this periodic discrepancy as an indication that our atmospheric model can be improved in the directions where we underpredict the loss, in particular by increasing the scale height of the atmosphere.

Firstly, we point out that the periodic matching and mismatching of the data by the model does not occur at 540-1250 keV (TP3) and is therefore an energy dependent effect. In fact, the overall agreement between the simulation with perturbed fields and the data is notably better at 540-1250 keV (where charge exchange is negligible) than for the simulations at lower energies (155-224 keV) without charge exchange. Considering that atmospheric charge exchange is more likely to occur in the lower energy range than the higher energy range. The fact that the model more accurately reproduces the data in the higher energy regime compared to the lower energy regime suggests that the model is not sufficiently capturing the entirety of the charge exchange occurring at Io. This implies that if the atmospheric model is improved, charge exchange at lower energies would be enhanced and the depletion features at 155-224keV will be better reproduced.

\begin{figure}
    \centering
    \makebox[\textwidth][c]{
    \includegraphics[width=0.8\textwidth]{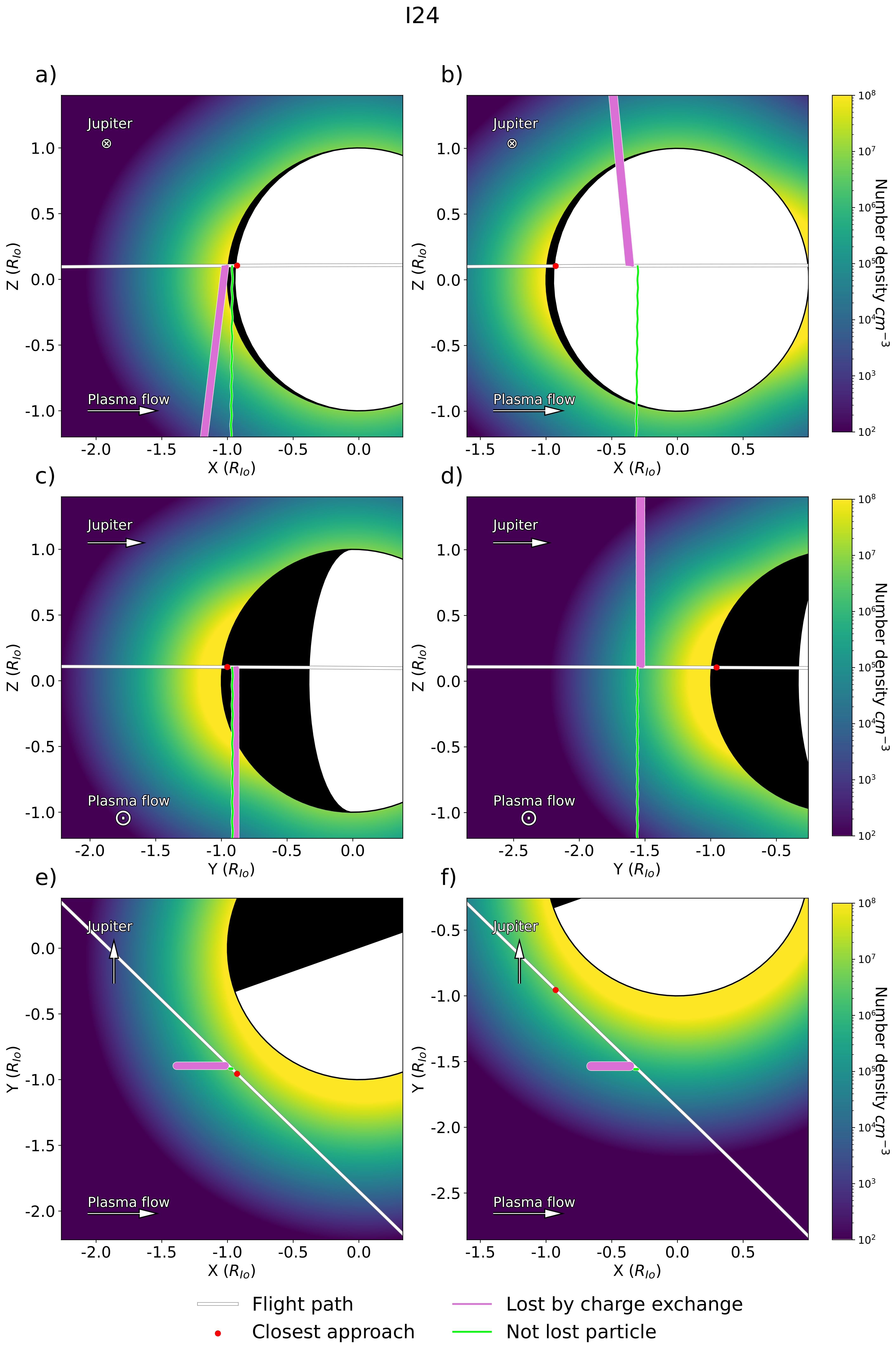}}
    \caption{Particle trajectories during flyby I24. Green particles are particles that are not lost in the simulation, pink ones are lost by charge exchange.
    Note also that the gyromotion of the particle has been plotted. Green particles correspond to smaller gyroradii, while pink particles correspond to larger gyroradii.
    Furthermore, pink particles corresponding to times along the trajectory where the simulation matches the depth of the dropout in the data approximately, green ones correspond to points where the loss is underpredicted. The corresponding points along the trajectory are indicated in Figure \ref{fig:results1} panel 'a', left column on the left side of the closest approach, right column on the right side.}
    \label{fig:I24_traj}
\end{figure}

Next we will discuss the discrepancy for flyby I24 as an example. Because charge exchange is the dominant loss process during this flyby, the interpretation of the discrepancy is simpler. During the entire duration of I24 a periodic matching and mismatching of the depth of the depletion associated with the asymmetrical model occurs. In Figure \ref{fig:I24_traj} we show particle trajectories for four points along the trajectory where the depletion reaches approximately the depth of the data (pink trajectory), and where it does not (green trajectory). The trajectories corresponding to points where the depth of depletion is underestimated have a smaller gyroradius, or equivalently, higher speed parallel to the magnetic field. The trajectories corresponding to those points where the depth of the depletion is approximately attained, have larger gyroradius, or equivalently lower speed parallel to the magnetic field.

The periodicity in the simulation (light blue line, Figure \ref{fig:results1}, panel a) has three dips within 20 seconds (the spin of the spacecraft). This is because during a full rotation of 20 seconds the detector viewing direction changes from perpendicular to the field line (larger gyroradius, large dip \#1), to parallel (smaller gyroradius, no or small dip), perpendicular again but in opposite direction (larger gyroradius, large dip \#2), parallel again but in opposite direction (smaller gyroradius, no or small dip) and back to the initial viewing angle (larger gyroradius, equivalent to large dip \#1).

\begin{figure}
    \centering
    \makebox[\textwidth][c]{
    \includegraphics[width=0.8\textwidth]{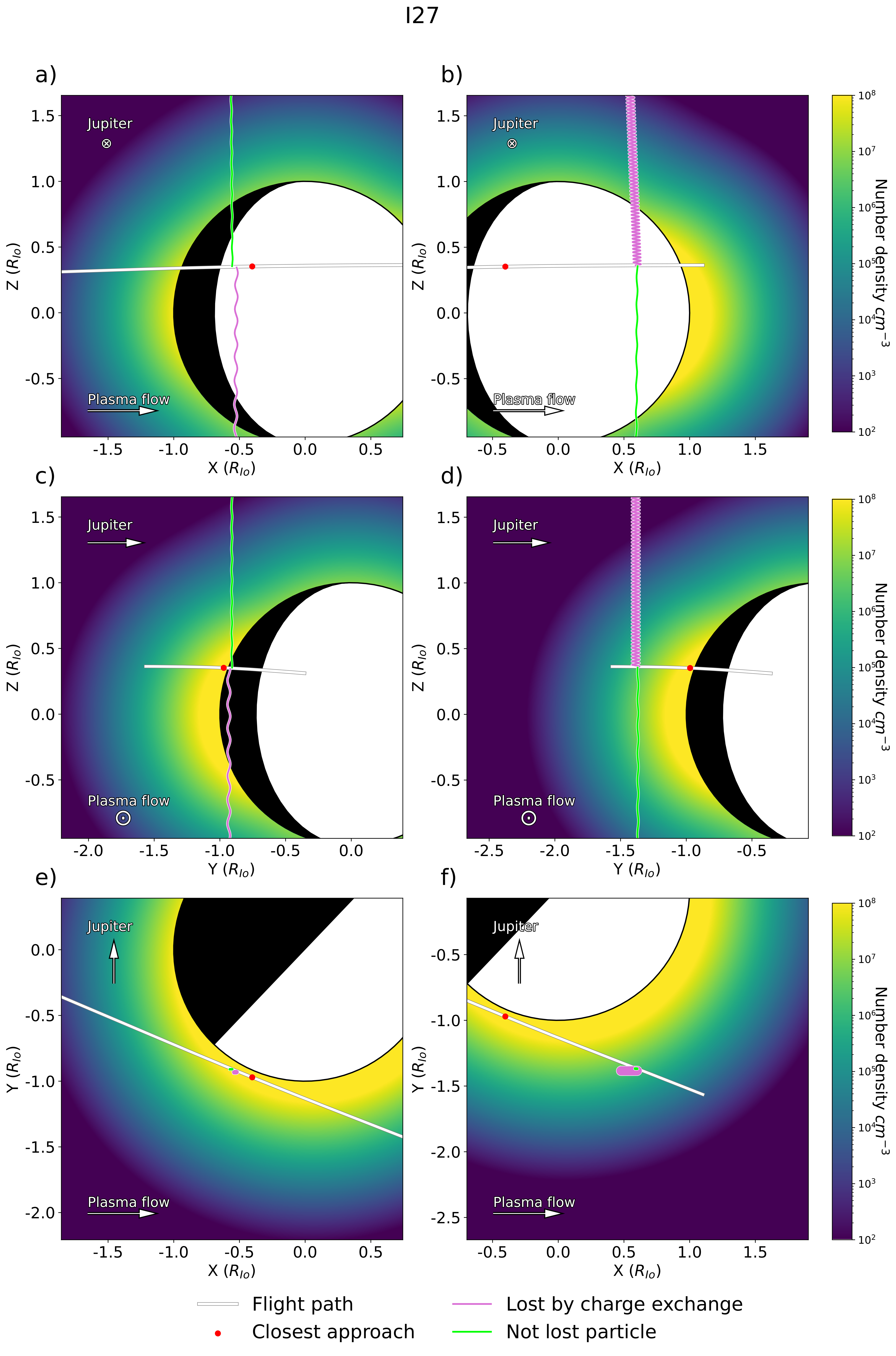}}
    \caption{Particle trajectories during flyby I27. The green trajectories are particles that are not lost, while the pink trajectories are particles that are lost by charge exchange. Note also that the gyromotion of the particle has been plotted. Furthermore, pink particles correspond to points along the trajectory for which the depth of the depletion is approximately recreated, while it is underestimated for the points corresponding to the green trajectories. The points along the trajectory corresponding to these particles is indicated in Figure \ref{fig:results1} panel 'c', left column on the left side of the closest approach, right column on the right side.}
    \label{fig:I27_traj}
\end{figure}

Next we discuss flyby I27, during which a different pattern of agreement between the data and the model than for I24 occurs. At the extremities of the I27 the depletion pattern is similar to I24: i.e. three approximately equal dips are visible every 20 seconds, with an underprediction of the measured depletion in between the dips (e.g. near the pink and green line on the right side of panel c in Figure \ref{fig:results1}). However, near the closest approach the pattern of depletion becomes different. Within the 20 second spin we see an underprediction of the depletion (green line on the left of the closest approach), a deep depletion that matches the data (purple line) and an undeprediction again. Thus, compared to I24, or the extremities of I27, we see only one deep dip per spin. In fact, what is different is that in the closer part of I27 within each 20 second cycle there is one of the two more field aligned directions (e.g. purple line left of the closest approach in in Figure \ref{fig:results1}, panel c) along which the depletion is now also approximately reproduced. Combined with the directions perpendicular to the field that were already depleted (previously referred to as deep dips \#1 and \#2) this depleted and field aligned direction forms the single deep dip we observe during one spin. The other field aligned direction (e.g. green line left of the closest approach, Figure \ref{fig:results1} left column, panel c) still underpredicts the loss. In I24, or at the extremities of I27 (e.g. purple and green line right of the closest approach in Figure \ref{fig:results1}, panel c), both of the field aligned directions produce the losses about equally poorly. Corresponding particle trajectories are shown in Figure \ref{fig:I27_traj}.
The field aligned direction which is depleted in the closer part of I27 corresponds to the direction southward of the trajectory. Considering that the I27 flyby is located north of Io's equator, particles going in these directions appear to have a higher chance of charge exchanging, even when they are more field aligned and have smaller gyroradii. We argue that this is because they traverse a larger section of the low altitude and therefore denser atmosphere, which results in a higher chance of charge exchange.

The underprediction in the northward field aligned directions appears to hold even in the case with the inhomogeneous fields (Figure \ref{fig:results1}, panel d). In the northward directions (e.g. closest to the green line on the left of panel c) the effect of the field configuration seems negligible. However, the southward field aligned directions (e.g. closest to the pink line on the left of panel c) is fully depleted in panel d. This means that we likely cannot derive any direct information about the atmospheric losses in those directions, since they are already fully saturated even without atmospheric charge exchange.

\begin{figure}
    \centering
    \makebox[\textwidth][c]{
    \includegraphics[width=0.8\textwidth]{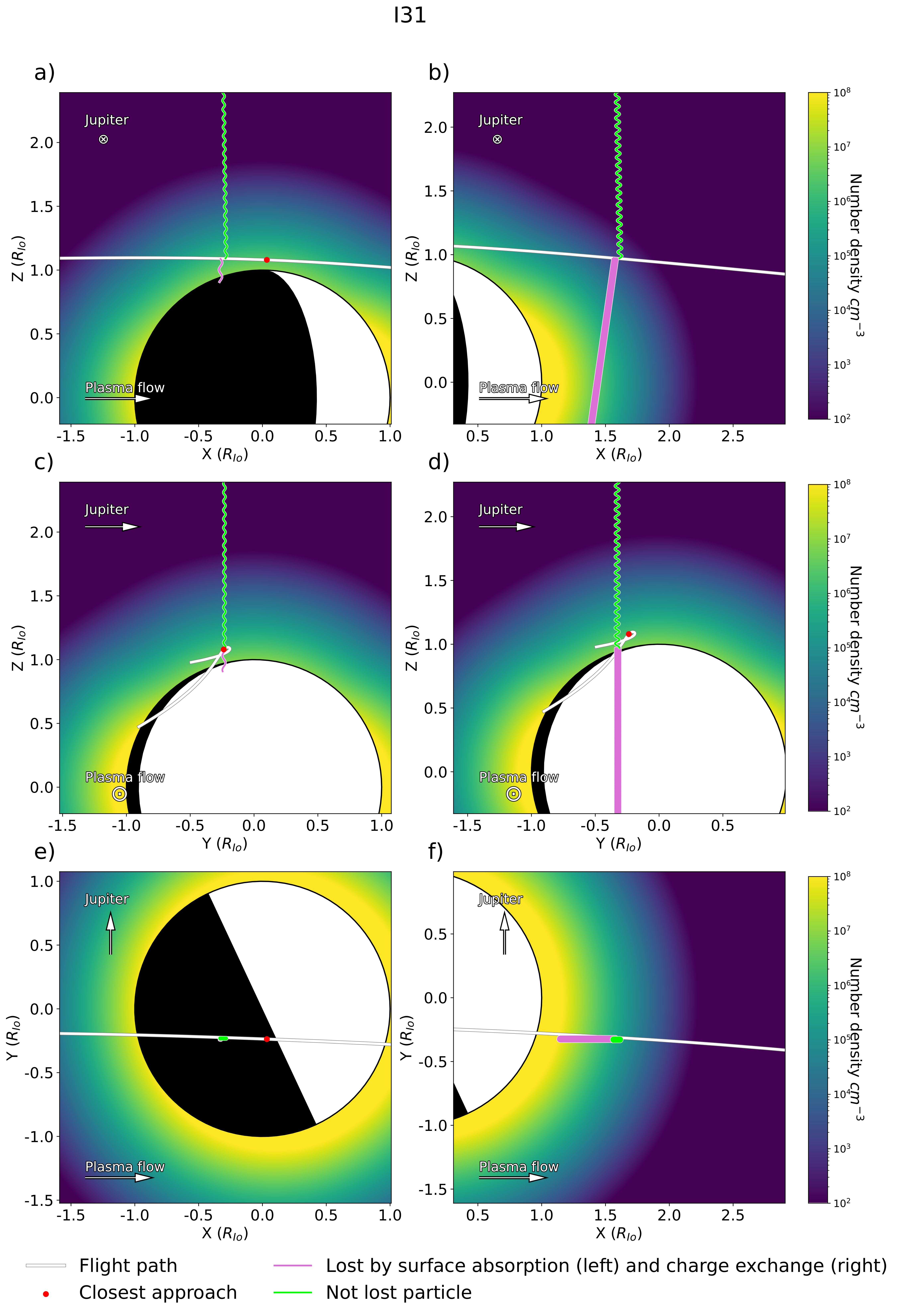}}
    \caption{Particle trajectories during flyby I31. The left column shows a particle that is lost by impact on Io's surface (pink) and a particle that is not lost (green). The right column shows a particle that is lost by charge exchange with the neutral atmosphere (pink) and a particle that is not lost (green). Note also that the gyromotion of the particle has been plotted. Furthermore, the pink particles correspond to points along the trajectory at which the depth of depletion is approximately reproduced, while the green ones correspond to points where the depletion is underpredicted. The points along the trajectory corresponding to these particles are indicated in Figure \ref{fig:results1} panel 'e', left column on the left side of the closest approach, right column on the right side.}
    \label{fig:I31_traj}
\end{figure}

Lastly, we discuss the periodic discrepancy for flyby I31. In Figure \ref{fig:I31_traj} we show particle trajectories for four points along the I31 trajectory where the depletion reaches approximately the depth of the data (pink trajectory), and one where it does not (green trajectory). First considering the two trajectories in the left column of Figure \ref{fig:I31_traj}. The direction in which the simulation matches the depth of depletion (pink) corresponds to a direction that intersects with the surface. As discussed in the previous section such a particle is considered lost.
The green trajectory is essentially moving in the opposite direction and does not intersect the surface, implying that the loss process associated with this direction has to be different from surface impact.
In the right panel the pink trajectory (depth of depletion reproduced) does not intersect the surface, rather the cause of the depletion is charge exchange. As for I24 and I27 it appears that both atmospheric models underestimate the losses in the more field aligned directions that do not encounter the denser part of the atmosphere. The conclusions for the 'green' particles hold even when including perturbed fields, at those points the depletion seems unaffected by the perturbed fields (black line).

However, we also observe that during the two minutes around the closest approach the cyclical mismatch disappears in the case with the inhomogeneous fields. This suggest that at least part of the mismatch could be attributed to a mismatch between the simulated field and the real configuration. Nonetheless, the simulations suggest that the role of the electromagnetic field in producing the losses only matters in a segment along the trajectory that is more confined than where atmospheric charge exchange plays a role. Suggesting that both charge exchange and field perturbations are required to reproduce the observed loss.

In conclusion, we find that the periodic discrepancy can be attributed to the properties of the atmospheric model. Directions in which the depletion is underpredicted correspond to ions that have small gyroradii and are travelling away from Io or the denser part of its atmosphere. Losses along these directions thus contain more information about possible charge exchange at larger distances from Io. We will discuss the implications of this finding further in Section \ref{ss_extended}.

\section{Discussion}\label{sec:discussion}

\subsection{Atmospheric charge exchange is a major or dominant loss process during flybys within $\sim2 $R$_{Io}$}

In this section we argue that atmospheric charge exchange is a necessary factor in order to explain the energetic proton depletion seen in the TP1 data during the Galileo flybys of Io. In fact, we suggest that near Io atmospheric charge exchange is the dominant loss process for flybys of 600 km altitude, while in closer flybys of 200km the electromagnetic field configuration and surface absorption also play a role in the measured proton depletion. In this sense the environment closest to Io is different than the environment along Io's orbit, but not close to Io, wherein charge exchange with the low energy plasma is thought to be the dominant loss process \cite{Mauk2022}. Table \ref{tab_summary} summarises the effects contributing to the proton loss and the derived atmospheric properties.

\begin{table}
\caption{Overview of the effects contributing to the proton losses and derived atmospheric properties, derived from the data-model comparison for the TP1 channel (155-224 keV).}
\centering
\begin{tabular}{l c c c}
\hline
 \textbf{Relevant loss processes} &
 \textbf{I24} &
 \textbf{I27} &
 \textbf{I31}  \\
 \hline
  Surface absorption  & No  & Yes & Yes \\
  Charge exchange  & Yes  & Yes & Yes \\
  Forbidden fields  & No  & No & Yes \\
  \hline
  \textbf{Atmospheric property} & & & \\
  \hline
  Day/downstream extension  & No  & Yes & Yes \\
  Full collapse on nightside & No & No & No \\
  Global extended component  & Yes  & Yes & Yes \\
\hline
\label{tab_summary}
\end{tabular}
\end{table}

As discussed in Section \ref{sec:epd_data}, at higher energies (540-1250 keV, TP3) the depletion region decreases in size along the flyby (I27 and I31) or disappears altogether (I24). This energy dependent size of the depletion region itself is already a strong indication for a role of charge exchange with the atmosphere as a loss process.
Furthermore, the simulation results discussed in the previous section indicate that the inclusion of atmospheric charge exchange leads to deeper energetic proton depletion for all flybys in all cases, including those with homogeneous electromagnetic fields, with perturbed fields and for both atmospheric models, in the energy range of the TP1 channel of the EPD (155-224 keV), thereby further supporting our hypothesis regarding the role of charge exchange.

The effect of charge exchange is most apparent for flyby I24, for which we argue that charge exchange is the dominant loss process of energetic protons. Our simulations can reproduce the extent of the depletion region, as well as attain the depth of observed depletion, while neither perturbed fields or impact of the surface play a significant role during this flyby. Furthermore, the lack of any depletion in the TP3 channel is fully consistent with the energy dependent charge exchange cross section. 

For flyby I27 the picture is more complex, as both electromagnetic fields which guide particles to the surface and charge exchange play a role in the formation of the depletion features, according to the simulations. The role of electromagnetic fields is confined to the region closest to Io, while the effect of charge exchange extends further. Furthermore, the depletion region in TP1 extends well beyond the depletion region in TP3, which is fully consistent with the energy dependence of charge exchange. We conclude that during I27 charge exchange is the major loss process during most of the flyby segment.

For flyby I31 the picture is different due to the polar geometry of the flyby. Here surface impact, perturbed fields and charge exchange play a role in the formation of the depletion features. However, also here the depletion region in TP1 extends well beyond the depletion region in TP3, which is fully consistent with the energy dependence of charge exchange. We conclude that during I31 charge exchange is a major loss process, becoming the dominant process on the outbound part of the trajectory which is on Io's dayside/downstream side.

Nearest Io during I27 and I31 we observe that situations occur where half of the spin cycle or the entire cycle is fully depleted due to perturbed fields. The corresponding directions are difficult to interpret in terms of charge exchange, considering that any atmospheric model would fit them. Furthermore, the directions along the trajectory nearest Io where fields have no effect are aimed away from Io, thus this limits the sensitivity of the analysis in interpreting the density of Io's atmosphere, nearest to Io.

Previously \citeA{Huybrighs2020AnDepletions,Huybrighs2021ReplyDepletions} have established that field perturbations and atmospheric charge exchange are required to explain particle dropouts measured during the E26 flyby of Europa (closest approach 350 km), in addition to losses by particles impacting on Europa's surface. Whereas the same processes take place at Io, we point out that the relative importance of atmospheric charge exchange is much larger at Io, due to its substantially denser neutral environment. This is true for all flybys considered in this study, from the lower altitude flybys I27 (closest approach at 198 km) and I31 (closest approach at 193 km) as well as for the higher altitude flyby I24 (closest approach at 611 km).

\subsection{Proton losses are sensitive to the electromagnetic field configuration within $\sim0.5$ R$_{Io}$}

Here we argue that an appropriate choice of electromagnetic field is also necessary to explain the  proton depletion seen at 155-224 keV (TP1), together with charge exchange and surface impact, and at 540-1250 keV (TP3) together with surface impact, in flybys closer to the surface such as I27 (198 km) and I31 (193 km). At higher altitudes, such as flyby I24 (611 km), the choice of field configuration does not appear to cause losses. Losses due to the field configuration are caused when the field guides protons to the surface and makes the impact. Perturbations in the electromagnetic field can also form 'forbidden regions' to which the access of the energetic proton is restricted (I31). 

During flyby I27 (closest approach at 198 km), the field configuration causes losses, up to 100\%, near the point of closest approach of Io in both TP1 and TP3 (Figures \ref{fig:results1} and \ref{sec:results} in panel d, black curve). This is in contrast to flyby I24 (closest approach 611 km), which has a qualitatively similar geometry with respect to the upstream flow and Io as I27 (see Figure \ref{fig:flight_overview}), however the choice of fields has no major effect on this flyby. Thus, due to the higher altitude the electromagnetic fields are unable to make the back-traced particles intersect with Io's surface.

It should be noted that for flyby I31, the perturbed fields also reduce the range along the trajectory within which the depletion occurs. This is due to the fact that the fields bend the trajectories of the charged particles around Io, allowing a greater number of particles to pass from the south to north side of the moon than if homogeneous fields are assumed. 

Even though the fields on their own do not cause proton losses during flyby I24, they do have a (minor) effect on the proton losses due to charge exchange. 
The simulated depletion pattern for I24 with atmospheric charge exchange and homogeneous fields (Figure \ref{fig:results1}, panel a) differs from the depletion pattern calculated with atmospheric charge exchange and perturbed fields (panel b in the same figure) in both shape and depth. With the inclusion of the perturbed fields points of maximum depletion are deeper by about 10\%. This implies that even if the field configuration does not lead to charged particle depletion directly, it does affect the depletion caused by the atmospheric charge exchange. Hence, we argue that perturbed fields are also an influential, though minor, factor in the charged particle depletion seen in TP1 during flyby I24. 

\subsection{Evidence for an extended atmosphere on the downstream/dayside, lack of evidence for full atmospheric collapse on upstream/nightside}

Here we argue for the existence of an extended downstream/dayside atmosphere during flybys I27 and I31, as well as the lack of evidence for complete atmospheric collapse on the upstream/nightside, based on the energetic proton measurements.

Figure \ref{fig:epd_data} reveals some evidence for an asymmetry during flyby I27. By visual inspection it can be determined that the trend of the loss in TP1 (155-224 keV) strongly correlates with the radial distance on the outbound (right side) part of the trajectory, while the trend of the loss on the inbound (left side) of the figure is lagging the trend of the radial distance. Considering the evidence for atmospheric charge exchange is the main loss process, during most of the segment, this suggests that the atmosphere is less expanded on the nightside than on the dayside.

Flyby I31 crosses from the night side into the dayside over Io's pole (see Figure \ref{fig:flight_overview}), as well as from the upstream to the downstream plasma environment. Figure \ref{fig:results1} shows that the measured depletion region during I31 extends further on the dayside/downstream region than on the nightside region/upstream. Thus, the energetic protons measurements are indicative of a dayside/upstream asymmetry in the proton losses. The simulation with the asymmetrical atmosphere (see Section \ref{sec:method_atmospheres}), which incorporates a day-night asymmetry as well as an upstream-downstream asymmetry, demonstrates that the extended depletion from approximately 05:04:30, corresponding to the dayside, is caused by atmospheric charge exchange alone. Whereas the radially symmetrical atmosphere is unable to reproduce the extended loss on the dayside. Therefore we interpret the extended loss region on the dayside as an indication of an extended atmosphere on the downstream/dayside.

The atmosphere is expected to be denser on the dayside due to enhanced sublimation (e.g. \citeA{Tsang2016TheEclipse}). It is also thought to have a higher scale height on the downstream side (e.g. \citeA{Saur2002InterpretationPasses}).
Because the downstream and dayside coincide during I24 and I27 we cannot distinguish between these two possible physical explanations for the extension on the downstream/dayside. 

For flyby I24 the trend of the loss follows the radial distance closely, we conclude that there is no clear sign of an atmospheric asymmetry in this flyby. The lack of a strong asymmetry does not necessarily imply that no asymmetry exists. We propose that it is the specific geometry of the I31 flyby, polar from the upstream/nightside to the downstream/dayside, that is most favourable to reveal the presence of asymmetries between the upstream/nightside and the downstream/dayside as compared to the other two. However, since I24 and I27 have similar geometries with the downstream/dayside, but different altitudes, it appears that the visibility of the asymmetry between the upstream/nightside and the downstream/dayside could be limited to lower altitudes.

We note that even though the atmosphere on the upstream/nightside extends less, there is still some loss on the upstream/nightside. From the proton measurements there is thus no evidence that the atmosphere is fully collapsed on the upstream/nightside.

Additional measurements, in particular those on the dayside when the dayside and the upstream side are not coinciding, could provide clarity on the role of enhanced sublimation in the formation of Io's extended atmosphere on the dayside, additional flybys on the Jovian side could reveal the Jovian/anti-Jovian asymmetry.

\subsection{Evidence for a global extended atmospheric component}
\label{ss_extended}

In Section \ref{s_discrepancy} we argued that for each of the three flybys there are viewing directions for which we underpredict the proton losses, that is where the modelled loss returns to unperturbed level whereas the data does not. We observe for the three flybys that the directions in which we underpredict the depth of the depletion corresponds to more field aligned particles that have a higher parallel velocity and thus smaller gyroradius. (An exception occurs when those particles pass through the denser part of the atmosphere closer to Io or the extended atmosphere on the upstream/dayside where they are depleted.) In contrast, the directions where the depth of depletion matches the data approximately correspond to particles that are less field aligned and have larger gyroradii. We suggest that a more extended atmosphere could compensate for the lack of depletion in the field aligned directions corresponding to smaller gyroradii, while retaining the relative agreement in other directions. 

The protons with the larger gyroradius spend more time in the high density part of the atmosphere (due to a lower velocity parallel to the magnetic field) and thus have a higher chance of charge-exchanging in this region than a particle with the same total energy that has a smaller gyroradius and higher parallel velocity. If an additional extended atmospheric component were to be included, the larger gyroradius particles would still charge exchange at a similar position along their trajectory in back-tracing sense (since they are most sensitive to the denser part). The smaller gyroradius particles, which would have had a low chance of charge exchange in the dense part of the atmosphere, would get a higher chance of charge exchanging if there were an extended atmosphere that is of a larger physical scale. In this sense an extended atmosphere could increase the depletion for smaller gyroradius particles while having less effect on the larger gyroradius particles. Indeed, during I27 we observe that field aligned directions could be depleted when the densities are increased. At the pink marker during I27 (Figure \ref{fig:results1}, left side, panel c) the direction corresponds to particles that are closer to field aligned (see Figure \ref{fig:I27_traj}), indeed the depletion at this point is reasonably well reproduced. On the other hand, at the green marker (left side) in the same figure, the field depletion is underpredicted. The difference between the two is that the pink marker travels through the densest part of the atmosphere, that is located at lower altitudes.

Additional simulations for a wide range of atmospheric properties are needed to constrain the properties of the extended atmosphere, we leave such a comprehensive enterprise as a subject for a future study. However, we do consider it likely that the missing atmospheric component extends beyond 2 Io radii, based on the fact that we observe the periodic discrepancy between the data and simulations throughout all flyby segments we considered.

The presence of such an extended atmosphere is supported by remote and in-situ observations of Io. Using the Hubble Space Telescope imaging spectrograph, \cite{Wolven2001EmissionCorona} detected Far Ultraviolet (FUV) emissions of O and S neutral atoms extending to at least 10 R$_{Io}$ around Io. Based on the direction of the instrument aperture, it can be inferred that these emissions are probably spherically distributed.  Assuming nominal torus values for the electron density and temperature, Wolven et al. infer an Oxygen density $\sim$ 1-2 $10^5$ cm$^{-3}$ at 2 R$_{Io}$. The charge exchange cross sections of protons on O follow the same trend as H on O$_2$, but have somewhat lower values \cite{Basu1987LinearObservations}. We therefore consider charge exchange with the S and O corona as a possible explanation. Follow-up studies should consider a multi-species atmospheric model that includes the O and S corona and appropriate cross-sections. 

Along most flybys of Io, Galileo observed Electro-Magnetic Ion Cyclotron (EMIC) waves at frequencies close to the SO$_2^+$ and SO$^+$ gyrofrequencies \cite{Huddleston1998,Russell1998,Russell2001,BlancoCano2001}. Such EMIC emissions are produced when neutrals are freshly ionized, forming a typical PickUp ring-distribution in phase space.
These EMIC emissions were observed mainly downstream of Io and sometimes far from the moon: along the J0 flyby in Io’s wake, EMIC waves were observed extending to $\sim$ 6-20 R$_{Io}$ in the inbound-outbound directions of the Galileo trajectory. These EMIC detections suggest the presence of a tenuous corona of SO$_2$ and SO extending far from Io, mainly in the downstream direction. Under simplifying assumptions, \citeA{Huddleston1997} infer a lower limit of the SO$_2$ density $\sim$ 10 cm$^{-3}$ at 4 R$_{Io}$ but this estimate is very model-dependent.  Although EMIC waves were also observed, at the S$^+$ gyrofrequencies,  such waves are usually damped by the torus thermal species and more difficult to observe.
Based on Galileo PLS measurements of the plasma density and temperature along the I32 flyby, \citeA{Dols2012AsymmetryFlybys} propose the presence of SO$_2$ and SO coronae extending far downstream of Io and that such downstream coronae could be produced by SO$_2$ resonant charge exchange in  Io’s atmosphere. On the contrary, our proton absorption data indicates an extended atmosphere both on the inbound (upstream) and outbound (downstream) parts of the spacecraft trajectories. In summary, there is observational evidence that Io’s atmosphere is very extended, which might explain some of the discrepancies between the observations and the simulations discussed above. 

Above we argue that the more field aligned particles are more sensitive to an extended atmospheric component, unless they pass through a denser part of the lower altitude atmosphere and are already depleted there. However, above a few 1000 km altitude the density of the atmosphere no longer exceeds that of the plasma torus ($\sim 2000$ cm$^{-3}$) and charge exchange losses with the torus become more important. We don't consider charge exchange with the low energy torus ions as a likely cause of the depletion features that we have attributed to a possible extended atmospheric component. The features that we focus on in this work only occur within 2 R$_{Io}$ and are therefore likely an aspect of the interaction with Io's local environment.

\subsection{Energetic particle data as an unexploited constraint on the electromagnetic field and atmosphere configuration at Io}

Our results show that the apparent proton losses during the flybys of Io are sensitive to both the electromagnetic field configuration and the atmospheric properties, because of neutral-ion charge exchange. These data thus provide an additional unexploited avenue to constrain the electromagnetic field and atmosphere structure around Io. A comprehensive parameter study to investigate the effect of the atmospheric and field properties on the proton flux forms the next logical step. Such studies should eventually take into account constraints provided by all Galileo datasets, including magnetic field (MAG), low energy plasma (PLS), plasma waves (PWS) and energetic ions (EPD). A multi-species atmospheric model or induced dipole, such as in \citeA{Dols2012AsymmetryFlybys}, should also be considered. However such a comprehensive endeavour is computationally intensive and we therefore consider it outside of the scope of this study. Instead this study aims to provide hypotheses and directions to guide such future modelling efforts.

Thanks to the fact that the TP3 channel of the EPD is sensitive to the field configuration but not sensitive to depletion through atmospheric charge exchange, data from this channel may be used as an initial constraint on the perturbed field model around Io. Once constraints are placed on what type of electromagnetic fields are present around Io, these may then be used in combination with atmospheric models to place constraints on the atmospheric structure of Io by looking at the depletion in the TP1 channel of the EPD. 

When comparing atmospheric models, it is of great use to have different flight geometries. Flybys that have a point of closest approach similar to that of I24 (611 km) allow us to fit atmospheric models with less influence on the depletion by the electromagnetic fields. Furthermore, when combining these observations with lower flybys at the same latitude (such as I27), information can be gained on the vertical structure of the atmosphere allowing us to fit surface density and scale height accordingly. Polar flybys, such as I31, could be used for the testing of different longitudinally dependent models. 

The Galileo data is of high value in this context, and will retain this value in the absence of a dedicated Io mission. The next missions destined for the Jupiter system, JUICE and Europa Clipper, are not scheduled to encounter Io from close by. However, the energetic particle data set has been expanded by two flybys by Juno with their closest approach at approximately 1500 km. Juno caries the energetic particle detector JEDI \cite{Mauk2017TheMission} which would be able to measure the proton flux.

Interpretation of the ion measurements is hampered by a lack of appropriate charge exchange cross sections. Here we used the cross section of H$^+$ on O$_2$ as a proxy for SO$_2$. This assumption leads to an underestimation of the charge exchange cross section, which makes it more difficult to accurately estimate the absolute values of atmospheric densities. However, our conclusions on the atmospheric configuration depend on relative differences between inbound and outbound segments of the trajectory, as well as relative differences between viewing directions that already represent the depletion well and those that don't, we therefore consider that these conclusions will hold, even for somewhat different charge exchange cross sections. Further laboratory studies could confirm our assumptions about the cross sections at higher energies as well as provide species specific cross sections, in particular cross sections for S and O ions are lacking.

\subsection{Io's role as a proton radiation belt sink in Jupiter's magnetosphere}

\citeA{Nenon2018AOrbit} has modelled Jovian trapped protons with kinetic energies higher than 1 MeV. In the \citeA{Nenon2018AOrbit} model Io is considered as an inert absorber of protons. This work shows that the electromagnetic field configuration is important in determining the exact size of Io as a sink. For example, we have shown that from approximately equatorial flybys of 200 km (e.g. I27) protons can be guided to the surface and impact. On the other hand, 
the particle tracing simulations presented in this work show that near Io the energetic protons of energies above $\sim 100$ keV can also be partly deflected by the perturbed fields (e.g. flyby I31). Thus when treating Io as an inert body the losses at these energies could also be overestimated. Furthermore, \citeA{Nenon2018AOrbit} does not take into account charge exchange with Io's atmosphere. Indeed, our simulations show that at energies of 1 MeV the effect of charge exchange will be minor. However, at lower energies where charge exchange is more efficient (e.g. below 224 keV) charge exchange will increase the size of Io as a proton sink in Jupiter's magnetosphere, causing charge exchange losses up to at least 1 Io radius away.

The extended loss region around Io could be measured along the field lines connected to the loss region. Indeed losses of energetic protons with energies in the 10-800 keV range over a region larger than Io have been measured by Juno along field lines connected with Io \cite{Paranicas2019IosData}. Whereas \citeA{Paranicas2019IosData} focuses on wave particle interactions as a source of loss, this work illustrates that charge exchange (at energies below 224 keV) and the electromagnetic field configuration near Io will also play a role in the proton depletion. In fact, we suggest that the field configuration is a more important effect than losses due to wave particle interaction because our simulations at 540-1250 keV (TP3), have a good qualitative agreement with the measured loss regions, without the need of invoking wave-particle interactions. 

In conclusion, we propose that the losses due to charge exchange near Io and apparent losses due to perturbed fields will also help explain losses observed along field lines connected to Io and its immediate environment.

\subsection{Io's atmosphere is a source of 150 $>$ keV ENAs}
Previously \citeA{Futaana2015Low-energyTori} have demonstrated that low-energy S and O energetic neutral atoms (referred to as LENA by the authors), of 460eV and 920eV respectively, resulting from the charge exchange of $<$1 keV plasma from the plasma torus with Io's neutral particle environment are detectable by the Jovian Neutrals Analyzer (JNA) instrument on the JUpiter ICy moon Explorer (JUICE).

High energy ($>$10 keV) ENA emissions were first reported by \citeA{Kirsch1981} based on Voyager 1 LECP measurements. However, the mass or charge state could not be determined. Later, \citeA{Krimigis2002} reported ENA emissions, possibly 50-80 keV hydrogen atoms, from Jupiter, based on Cassini INCA measurements. Subsequently, \citeA{Mauk2003} attributed the origin of the ENA emissions to Jupiter's atmosphere and a possible neutral cloud slightly outside of the orbit of Europa. The emissions were not attributed to Io's torus or Io itself.

This study suggests that energetic protons with energies exceeding 155 keV will charge exchange with Io's atmosphere, and produce ENAs of $\sim150$ keV energies. When the  energetic protons become Energetic Neutral Atoms (ENA) they are no longer bound by the electromagnetic field and will escape at high energies that are similar to their energies before neutralisation \cite<e.g.>[]{Rees1989PhysicsAtmosphere}. While the proton flux at $\sim$ 100 keV is about two orders of magnitude less at Io than at Europa, the total mass of Io's atmosphere is 100 times that of Europa \cite{Bagenal2020TheEuropa}. Therefore, we recommend that future studies on the interpretation of ENA imaging of Jupiter's magnetosphere also consider Io as a possible source of $\sim$100 keV ENAs. 

These $\sim$100 keV ENAs could be detected by the Jovian Energetic Neutral and Ions (JENI) detector part of the Particle Environment Package (PEP) on the JUICE mission \cite{Barabash2013ParticleMission}. Variations of keV ENA flux at Io could provide information about variations in the energetic proton flux and variations in the neutral environment of Io. While the LENAs will reach JUICE within a few hours \cite{Futaana2015Low-energyTori}, the ENAs could reach JUICE in a Ganymede orbit within just a few minutes, thereby providing a more instantaneous assessment of the neutral-ion interaction at Io.

\subsection{Energetic ion detectors on future missions to Io}
Proposals are put forward for new dedicated missions to Io \cite<e.g.>[]{McEwen2021FutureIo}. This study demonstrates the added value of including an energetic ion detector to investigate the electromagnetic field configuration and atmospheric structure around Io. While a magnetometer or neutral mass spectrometer only measures the field and atmospheric density along the flyby, the energetic ions are sensitive to the wider environment of Io. Their flux dropouts will thus provide information about the field configuration outside of the flyby trajectory. Thus, we recommend to include an energetic particle detector on future missions to Io.

A combination of lower and higher altitude flybys would be beneficial for isolating the relative importance of electromagnetic field perturbations and atmospheric charge exchange on the resulting proton loss. Flybys at different longitudes, latitudes and local times could help disentangle the atmospheric asymmetries. 

\section{Conclusion}\label{sec:conclusion}

In this work we used particle back-tracing in order to simulate the energetic proton losses measured in channels TP1 (155-224 keV) and TP3 (540-1250 keV) of the EPD during Galileo flybys of Io I24, I27 and I31. The three sources of depletion taken into account in the simulations are: absorption by the surface, atmospheric charge exchange, and perturbed fields creating `forbidden regions'. From these simulations we conclude the following:

\begin{itemize}
    \item Atmospheric charge exchange is a major loss process of energetic protons in the TP1 channel of the EPD during Galileo flybys I27, and I31 (closest approach near 200km); and the dominant process during I24 (closest approach at 611 km).
    \item The electromagnetic fields can also play a role in the proton losses in the energy range of TP1 and TP3, in particular within $\sim$0.5 R$_{Io}$ during flybys I27 and I31. This region is more confined than the region over which we expect charge exchange losses ($\sim$2 R$_{Io}$).
    \item A combination of atmospheric charge exchange and an appropriate description of the electromagnetic field is necessary to fully describe the energetic proton depletion seen in the TP1 channel of the EPD during all Galileo flybys I24, I27, and I31.
    \item Proton losses are not only sensitive to atmospheric densities or field perturbations along the trajectory, but to the entire region traversed by the energetic ions and thus offer more global information than other in-situ measurements.
\end{itemize}

None of our models are able to fully reproduce all properties of the proton loss observed in the data. However, from partial agreements and discrepancies we find that the losses in the TP1 channel (155-224 keV) suggest several properties of the structure of Io's atmosphere:
\begin{itemize}
    \item A longitudinal asymmetry with an extended atmosphere on the day/downstream side, which is apparent for lower flybys near 200 km, but not at higher altitudes at 600km on the night/upstream side.
    \item There is no evidence for full atmospheric collapse on the night/upstream side
    \item Underpredictions of the proton losses in more field aligned directions hint at the existence of a global extended atmosphere $> 1$ Io radii.
\end{itemize}

These findings open up a new set of data to be used in order to constrain atmospheric models of Io and the electromagnetic fields around Io. Future work should involve fitting a range of self-consistent atmospheric and electromagnetic field models to the EPD data set using this back-tracing method to establish confidence in our interpretation and to constrain the properties of the extended atmospheric component hinted at in our results. Improved understanding of the charge exchange cross section, including those for other energetic ions species (S, O), would greatly benefit the interpretation of the data. Lastly, we emphasize the importance of energetic ion detectors on future missions to Io. Having energetic particle data sets for a variety of flyby geometries would further increase the reliability of the constraints on the fields and atmosphere.

\appendix

\section{Charge exchange cross section}

\begin{figure}
\begin{center}
\noindent\includegraphics[width=0.9\textwidth]{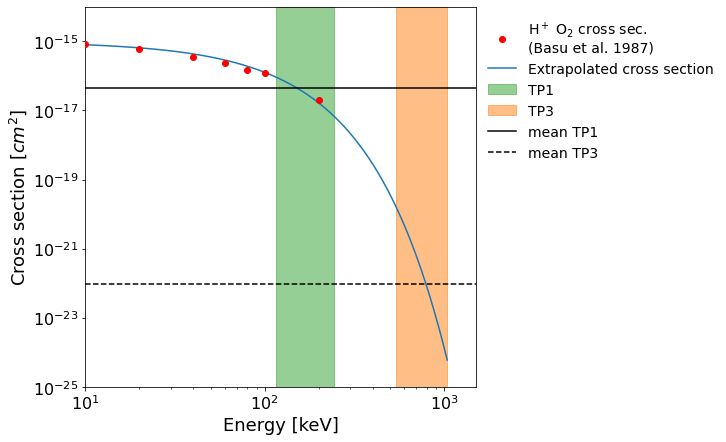}
\end{center}
\caption{Charge exchange cross section of H$^+$ on O$_2$ as a function of energy from \protect\citeA{Basu1987LinearObservations} (red dots) and a fit to the data (blue line).  TP1 (green band) covers an energy range from 115 to 244 keV and TP3 (orange band) covers a range from 540 to 12500 keV. The mean cross section values of the two channels are indicated using the solid black line for TP1 and the dashed black line for TP3.}
\label{fig_cross}
\end{figure}

The charge exchange cross section of H$^+$ on O$_2$ that is used in the simulations is shown in Figure  \ref{fig_cross}, together with the energy range of the TP1 and TP3 channels of the EPD instrument.

\section{Proton gyroradius}

\begin{figure}
\center
\noindent\includegraphics[width=1.0\textwidth]{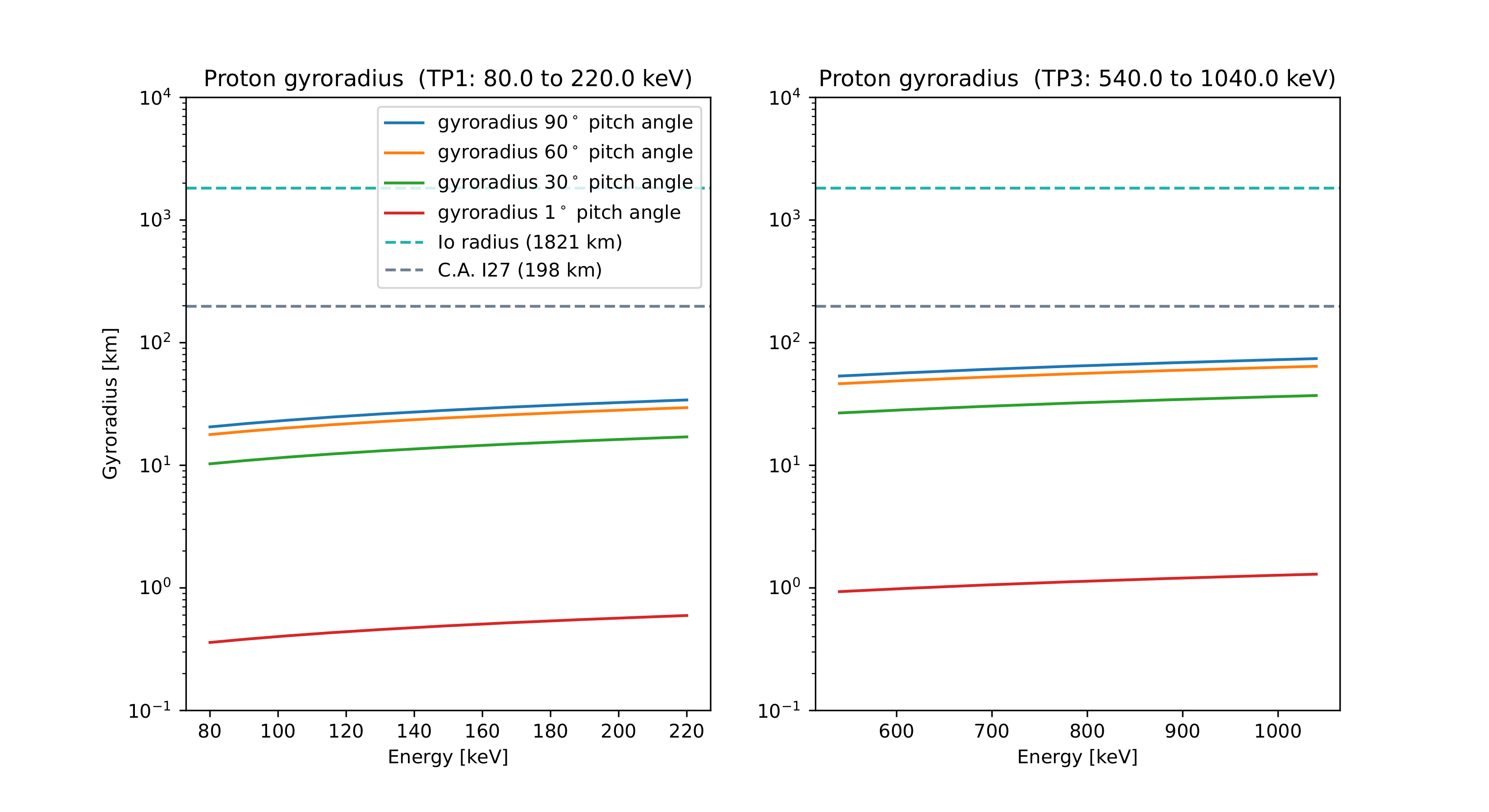}
\caption{Gyroradius as a function of pitch angle and the energy range of the TP1 channel (left) and TP3 (right). The magnetic field is the background field assumed in the simulations for I27 in \protect\citeA{Blocker2018MHDAtmosphere}. Also indicated are the radius of Io and the altitude of the closest approach.}
\label{fig_gyro}
\end{figure}

The proton gyroradius as a function of pitch angle and energy range of the TP1 and TP3 channels of EPD are shown in Figure \ref{fig_gyro}.

\section{Proton bounce period}
\label{appendix_bounce_period}

\begin{figure}
\begin{center}
\noindent\includegraphics[width=1.0\textwidth]{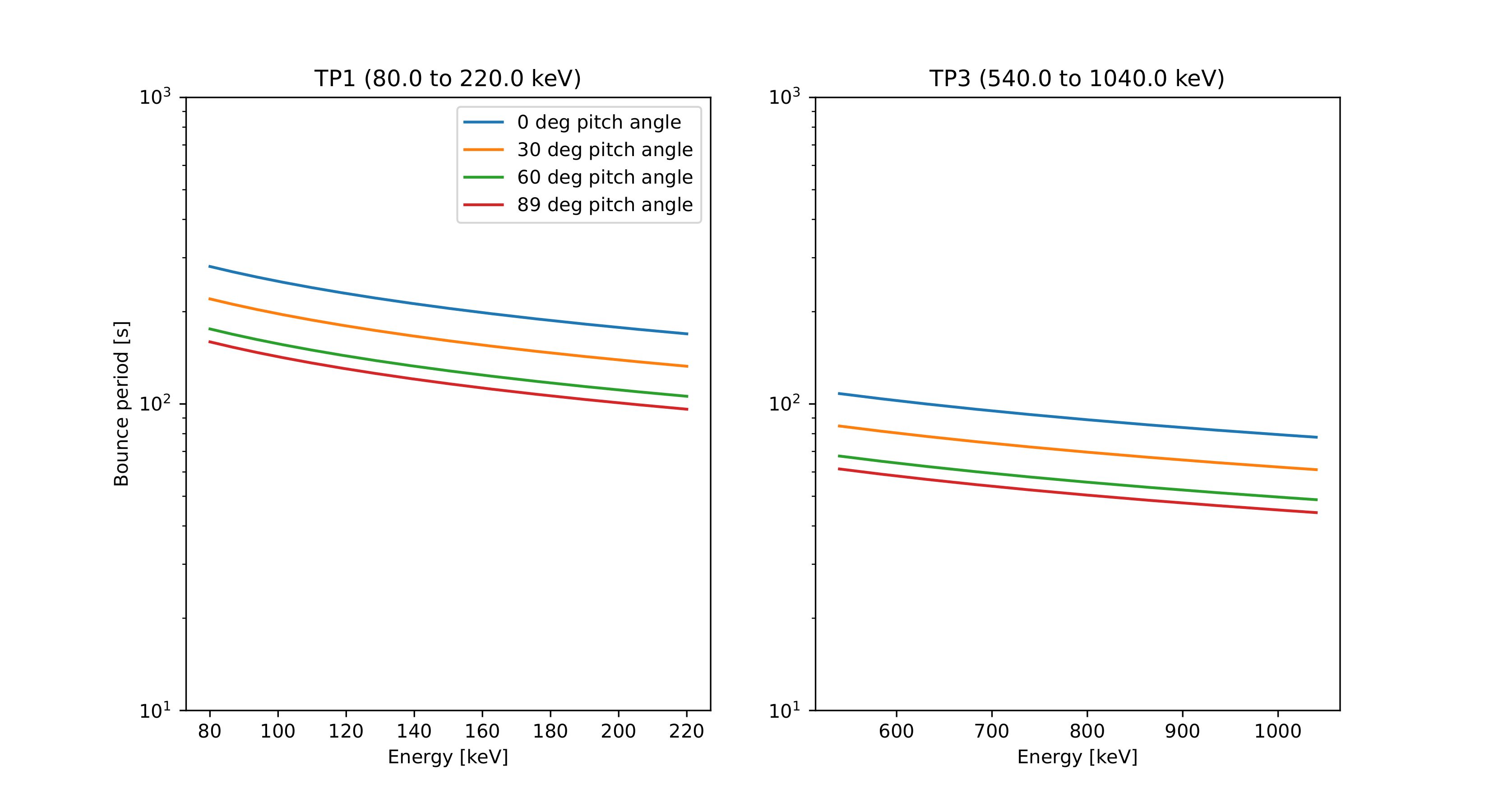}
\end{center}
\caption{Bounce period as a function of  pitch angle and energy range of the TP1 channel (left) and TP3 (right). The bounce period is calculated using Equation 2.2 from \protect\citeA{Baumjohann1997}.
}
\label{fig_bounce}
\end{figure}

The proton bounce period as a function of pitch angle and the energy range of the TP1 and TP3 channels of EPD is shown in Figure \ref{fig_bounce}.
Considering a corotation velocity of 57 km/s, and an Io radius of 1821 km it would take about 128 seconds for the corotation to cover the entire length of the simulation box.
Particles with a bounce period shorter than this occurring at the upstream edge of the simulation could in principle re-enter at the downstream edge. Meaning that any losses occurring at the upstream side could affect the ion distribution on the downstream edge, which is not accounted for in the simulation. We consider this not a major concern for the TP1 simulation since the bounce period of particles generally exceeds this limit. For the TP3 channel the bounce period becomes smaller, however the apparent region of proton losses associated with Io also becomes smaller than 1 Io radius. A lower limit of 63 seconds is applicable, which is still exceeded by the bounce periods. It is thus not likely in TP3 either that the distribution of the protons is affected by losses occurring along the flyby in such a way that it would cause the depletion features.

\section{MHD simulation for flyby I31}
\label{app_I31}
\begin{figure}
\begin{center}
\noindent\includegraphics[width=1.0\textwidth]{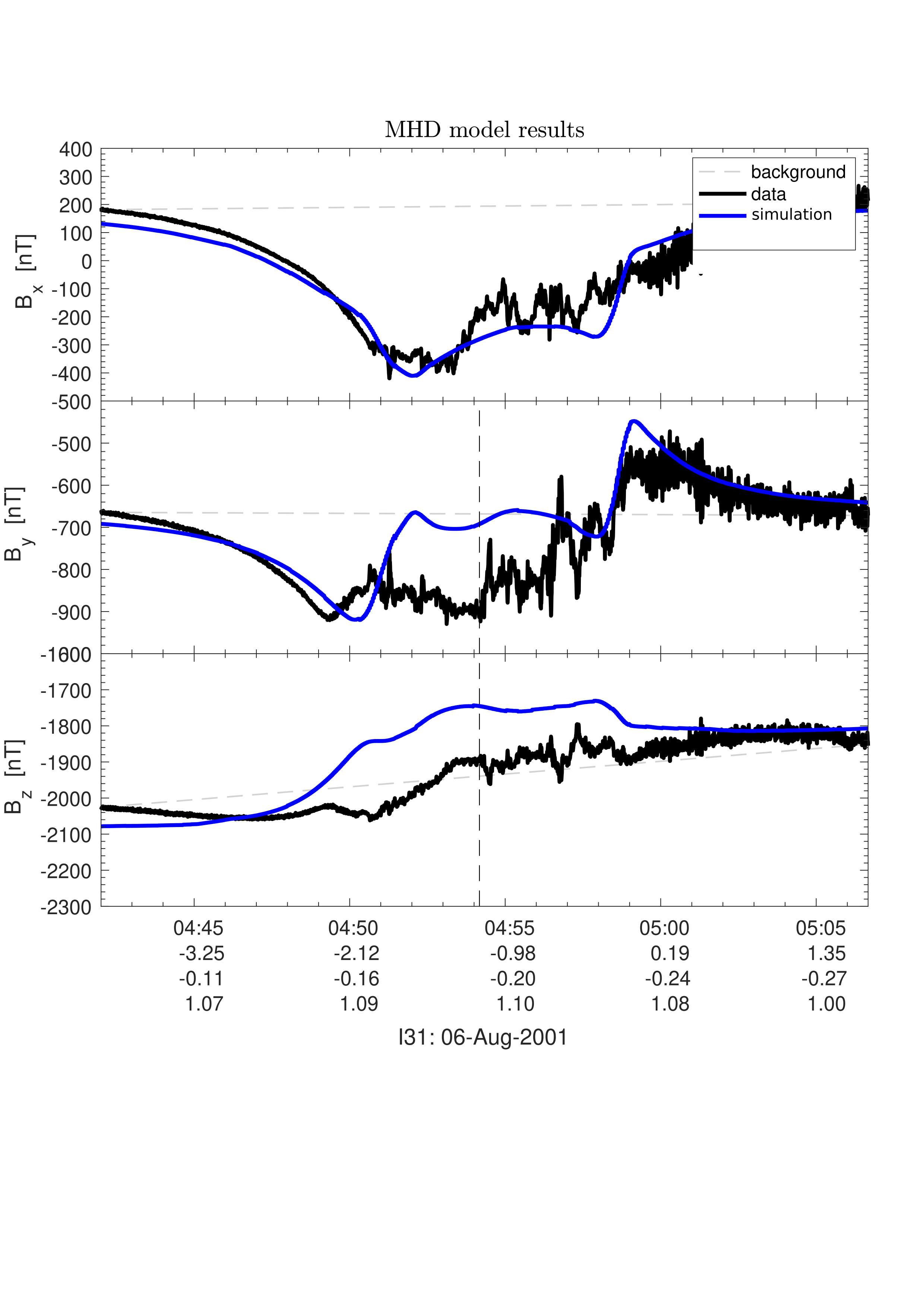}
\end{center}
\caption{Magnetic field components along the I31 flyby trajectory in the IPHIO coordinate system. Comparison between the Galileo MAG data (black lines) and the results obtained from the MHD model (blue lines). The dashed vertical line displays the closest approach at 04:59 UT. The dashed nearly horizontal line shows the background magnetic field. Simulations were obtained using the model from \protect\citeA{Blocker2018MHDAtmosphere}. Horizontal axis labels from top to bottom: time (UT), x (R$_{Io}$), y (R$_{Io}$), z (R$_{Io}$)}.
\label{fig_mhd_I31}
\end{figure}

Figure \ref{fig_mhd_I31} provides a comparison between the measured (black lines) and simulated magnetic field (blue lines) for flyby I31.
The I31 Galileo flyby was a close encounter above the north polar region, reaching its nearest point to Io at a distance of 194 km above its surface.
The trajectory intersected the northern Alfv\'en wing, evident in the significant magnetic field disturbance, especially in B$_x$  resulting from the bending of the wing. Given the spacecraft's close proximity to Io, variations in the magnetic field attributable to the ionospheric current system may also be evident in the magnetic field data.
The MHD model  reproduces the general structure of the magnetic field perturbation. 
However, B$_z$ is overestimated by the MHD model and the disturbance is displaced further towards the upstream side of the trajectory. 
The modeled B$_y$ component exhibits only weak perturbations inside the Alfv\'en wing (beetween 04:54 UT and 05:04 UT).
Measured small-scale perturbations during the wing crossing are obvious in the MAG data, implying that the atmosphere is probably not as smooth as it is described in our model.
The magnetic field could also be affected by intense wave emissions generating small-scale fluctuations.
Further investigations on the effect of different local atmospheric  asymmetries (e.g., volcanic plumes) in the atmosphere on the plasma interaction during I31 could provide a better fit to the variations of the magnetic field during the wing crossing.

\newpage

\acknowledgments
HH is supported by a DIAS Research Fellowship in Astrophysics. CPA van Buchem was supported by the ESA-Leiden University MSc program. HH was supported by an ESA research fellowship. HH gratefully acknowledges financial support from Khalifa University’s Space and Planetary Science Center (Abu Dhabi, UAE) under Grant no. KU-SPSC-8474000336. We also acknowledge: Andreas Lagg for the EPD data software, the 2019 and 2020 workshops on 'Outer planet moon-magnetosphere interactions' in fostering collaborations that contributed greatly to this work and the International Space Science Institute (ISSI) visiting scientist program. This study has been partially performed in the context of the activities of the ISSI International Team 515. CMJ's work at DIAS is supported by Science Foundation Ireland Grant 18/FRL/6199.
V. Dols is supported by the NASA System Workings 2020-0139 Grant 80NSSC22K0109 and also supported at the University of Colorado as a part of NASA's Juno mission supported by NASA through contract 699050X with the Southwest Research Institute

\section*{Open Research}

EPD data are available through NASA's Planetary Data System (PDS) \cite{EPD}. Simulation results are archived in \citeA{Huybrighs2024_data}.

\bibliography{references_hh.bib}

\begin{thebibliography}{}

\bibitem [\protect \citeauthoryear {%
Addison%
, Liuzzo%
, Arnold%
\BCBL {}\ \BBA {} Simon%
}{%
Addison%
\ \protect \BOthers {.}}{%
{\protect \APACyear {2021}}%
}]{%
Addison2021InfluenceWeathering}
\APACinsertmetastar {%
Addison2021InfluenceWeathering}%
\begin{APACrefauthors}%
Addison, P.%
, Liuzzo, L.%
, Arnold, H.%
\BCBL {}\ \BBA {} Simon, S.%
\end{APACrefauthors}%
\unskip\
\newblock
\APACrefYearMonthDay{2021}{5}{}.
\newblock
{\BBOQ}\APACrefatitle {{Influence of Europa’s Time-Varying Electromagnetic Environment on Magnetospheric Ion Precipitation and Surface Weathering}} {{Influence of Europa’s Time-Varying Electromagnetic Environment on Magnetospheric Ion Precipitation and Surface Weathering}}.{\BBCQ}
\newblock
\APACjournalVolNumPages{Journal of Geophysical Research: Space Physics}{126}{5}{e2020JA029087}.
\newblock
\begin{APACrefURL} \url{https://onlinelibrary-wiley-com.vu-nl.idm.oclc.org/doi/full/10.1029/2020JA029087 https://onlinelibrary-wiley-com.vu-nl.idm.oclc.org/doi/abs/10.1029/2020JA029087 https://agupubs-onlinelibrary-wiley-com.vu-nl.idm.oclc.org/doi/10.1029/2020JA029087} \end{APACrefURL}
\newblock
\begin{APACrefDOI} \doi{10.1029/2020JA029087} \end{APACrefDOI}
\PrintBackRefs{\CurrentBib}

\bibitem [\protect \citeauthoryear {%
Addison%
, Liuzzo%
\BCBL {}\ \BBA {} Simon%
}{%
Addison%
\ \protect \BOthers {.}}{%
{\protect \APACyear {2022}}%
}]{%
Addison2022EffectIce}
\APACinsertmetastar {%
Addison2022EffectIce}%
\begin{APACrefauthors}%
Addison, P.%
, Liuzzo, L.%
\BCBL {}\ \BBA {} Simon, S.%
\end{APACrefauthors}%
\unskip\
\newblock
\APACrefYearMonthDay{2022}{2}{}.
\newblock
{\BBOQ}\APACrefatitle {{Effect of the Magnetospheric Plasma Interaction and Solar Illumination on Ion Sputtering of Europa’s Surface Ice}} {{Effect of the Magnetospheric Plasma Interaction and Solar Illumination on Ion Sputtering of Europa’s Surface Ice}}.{\BBCQ}
\newblock
\APACjournalVolNumPages{Journal of Geophysical Research: Space Physics}{127}{2}{e2021JA030136}.
\newblock
\begin{APACrefURL} \url{https://onlinelibrary-wiley-com.vu-nl.idm.oclc.org/doi/full/10.1029/2021JA030136 https://onlinelibrary-wiley-com.vu-nl.idm.oclc.org/doi/abs/10.1029/2021JA030136 https://agupubs-onlinelibrary-wiley-com.vu-nl.idm.oclc.org/doi/10.1029/2021JA030136} \end{APACrefURL}
\newblock
\begin{APACrefDOI} \doi{10.1029/2021JA030136} \end{APACrefDOI}
\PrintBackRefs{\CurrentBib}

\bibitem [\protect \citeauthoryear {%
Armstrong%
, Paonessa%
, Brandon%
, Krimigis%
\BCBL {}\ \BBA {} Lanzerotti%
}{%
Armstrong%
\ \protect \BOthers {.}}{%
{\protect \APACyear {1981}}%
}]{%
Armstrong1981Low-energyMagnetosphere}
\APACinsertmetastar {%
Armstrong1981Low-energyMagnetosphere}%
\begin{APACrefauthors}%
Armstrong, T\BPBI P.%
, Paonessa, M\BPBI T.%
, Brandon, S\BPBI T.%
, Krimigis, S\BPBI M.%
\BCBL {}\ \BBA {} Lanzerotti, L\BPBI J.%
\end{APACrefauthors}%
\unskip\
\newblock
\APACrefYearMonthDay{1981}{9}{}.
\newblock
{\BBOQ}\APACrefatitle {{Low-energy charged particle observations in the 5–20 RJ region of the Jovian magnetosphere}} {{Low-energy charged particle observations in the 5–20 RJ region of the Jovian magnetosphere}}.{\BBCQ}
\newblock
\APACjournalVolNumPages{Journal of Geophysical Research: Space Physics}{86}{A10}{8343--8355}.
\newblock
\begin{APACrefURL} \url{https://onlinelibrary-wiley-com.vu-nl.idm.oclc.org/doi/full/10.1029/JA086iA10p08343 https://onlinelibrary-wiley-com.vu-nl.idm.oclc.org/doi/abs/10.1029/JA086iA10p08343 https://agupubs-onlinelibrary-wiley-com.vu-nl.idm.oclc.org/doi/10.1029/JA086iA10p08343} \end{APACrefURL}
\newblock
\begin{APACrefDOI} \doi{10.1029/JA086IA10P08343} \end{APACrefDOI}
\PrintBackRefs{\CurrentBib}

\bibitem [\protect \citeauthoryear {%
Bagenal%
}{%
Bagenal%
}{%
{\protect \APACyear {1989}}%
}]{%
Bagenal1989Torus-MagnetosphereCoupling}
\APACinsertmetastar {%
Bagenal1989Torus-MagnetosphereCoupling}%
\begin{APACrefauthors}%
Bagenal, F.%
\end{APACrefauthors}%
\unskip\
\newblock
\APACrefYearMonthDay{1989}{}{}.
\newblock
{\BBOQ}\APACrefatitle {{Torus-Magnetosphere Coupling}} {{Torus-Magnetosphere Coupling}}.{\BBCQ}
\newblock
\BIn{} \APACrefbtitle {NASA Special Publication} {Nasa special publication}\ (\BVOL~494, \BPGS\ 196--210).
\PrintBackRefs{\CurrentBib}

\bibitem [\protect \citeauthoryear {%
Bagenal%
\ \BBA {} Dols%
}{%
Bagenal%
\ \BBA {} Dols%
}{%
{\protect \APACyear {2020}}%
}]{%
Bagenal2020TheEuropa}
\APACinsertmetastar {%
Bagenal2020TheEuropa}%
\begin{APACrefauthors}%
Bagenal, F.%
\BCBT {}\ \BBA {} Dols, V.%
\end{APACrefauthors}%
\unskip\
\newblock
\APACrefYearMonthDay{2020}{5}{}.
\newblock
{\BBOQ}\APACrefatitle {{The Space Environment of Io and Europa}} {{The Space Environment of Io and Europa}}.{\BBCQ}
\newblock
\APACjournalVolNumPages{Journal of Geophysical Research: Space Physics}{125}{5}{}.
\newblock
\begin{APACrefDOI} \doi{10.1029/2019ja027485} \end{APACrefDOI}
\PrintBackRefs{\CurrentBib}

\bibitem [\protect \citeauthoryear {%
Barabash%
, Wurz%
\BCBL {}\ \BBA {} Team%
}{%
Barabash%
\ \protect \BOthers {.}}{%
{\protect \APACyear {2013}}%
}]{%
Barabash2013ParticleMission}
\APACinsertmetastar {%
Barabash2013ParticleMission}%
\begin{APACrefauthors}%
Barabash, S.%
, Wurz, P.%
\BCBL {}\ \BBA {} Team, P.%
\end{APACrefauthors}%
\unskip\
\newblock
\APACrefYearMonthDay{2013}{}{}.
\newblock
{\BBOQ}\APACrefatitle {{Particle Environment Package (PEP) for the ESA JUICE mission}} {{Particle Environment Package (PEP) for the ESA JUICE mission}}.{\BBCQ}
\newblock
\APACjournalVolNumPages{EGUGA}{15}{}{2013--9745}.
\newblock
\begin{APACrefURL} \url{https://ui.adsabs.harvard.edu/abs/2013EGUGA..15.9745B/abstract} \end{APACrefURL}
\PrintBackRefs{\CurrentBib}

\bibitem [\protect \citeauthoryear {%
Basu%
, Jasperse%
, Robinson%
, Vondrak%
\BCBL {}\ \BBA {} Evans%
}{%
Basu%
\ \protect \BOthers {.}}{%
{\protect \APACyear {1987}}%
}]{%
Basu1987LinearObservations}
\APACinsertmetastar {%
Basu1987LinearObservations}%
\begin{APACrefauthors}%
Basu, B.%
, Jasperse, J\BPBI R.%
, Robinson, R\BPBI M.%
, Vondrak, R\BPBI R.%
\BCBL {}\ \BBA {} Evans, D\BPBI S.%
\end{APACrefauthors}%
\unskip\
\newblock
\APACrefYearMonthDay{1987}{6}{}.
\newblock
{\BBOQ}\APACrefatitle {{Linear transport theory of auroral proton precipitation: A comparison with observations}} {{Linear transport theory of auroral proton precipitation: A comparison with observations}}.{\BBCQ}
\newblock
\APACjournalVolNumPages{Journal of Geophysical Research}{92}{A6}{5920}.
\newblock
\begin{APACrefDOI} \doi{10.1029/ja092ia06p05920} \end{APACrefDOI}
\PrintBackRefs{\CurrentBib}

\bibitem [\protect \citeauthoryear {%
Baumjohann%
\ \BBA {} Treumann%
}{%
Baumjohann%
\ \BBA {} Treumann%
}{%
{\protect \APACyear {1997}}%
}]{%
Baumjohann1997}
\APACinsertmetastar {%
Baumjohann1997}%
\begin{APACrefauthors}%
Baumjohann, W.%
\BCBT {}\ \BBA {} Treumann, R.%
\end{APACrefauthors}%
\unskip\
\newblock
\APACrefYear{1997}.
\newblock
\APACrefbtitle {Basic space plasma physics} {Basic space plasma physics}.
\newblock
\APACaddressPublisher{}{Imperial College Press}.
\PrintBackRefs{\CurrentBib}

\bibitem [\protect \citeauthoryear {%
{Birdsall}%
}{%
{Birdsall}%
}{%
{\protect \APACyear {1991}}%
}]{%
Birdsall1991}
\APACinsertmetastar {%
Birdsall1991}%
\begin{APACrefauthors}%
{Birdsall}, C\BPBI K.%
\end{APACrefauthors}%
\unskip\
\newblock
\APACrefYearMonthDay{1991}{April}{}.
\newblock
{\BBOQ}\APACrefatitle {Particle-in-cell charged-particle simulations, plus Monte Carlo collisions with neutral atoms, PIC-MCC} {Particle-in-cell charged-particle simulations, plus monte carlo collisions with neutral atoms, pic-mcc}.{\BBCQ}
\newblock
\APACjournalVolNumPages{IEEE Transactions on Plasma Science}{19}{2}{65-85}.
\newblock
\begin{APACrefDOI} \doi{10.1109/27.106800} \end{APACrefDOI}
\PrintBackRefs{\CurrentBib}

\bibitem [\protect \citeauthoryear {%
Blanco-Cano%
, Russell%
\BCBL {}\ \BBA {} Strangeway%
}{%
Blanco-Cano%
\ \protect \BOthers {.}}{%
{\protect \APACyear {2001}}%
}]{%
BlancoCano2001}
\APACinsertmetastar {%
BlancoCano2001}%
\begin{APACrefauthors}%
Blanco-Cano, X.%
, Russell, C\BPBI T.%
\BCBL {}\ \BBA {} Strangeway, R\BPBI J.%
\end{APACrefauthors}%
\unskip\
\newblock
\APACrefYearMonthDay{2001}{}{}.
\newblock
{\BBOQ}\APACrefatitle {The Io mass-loading disk: Wave dispersion analysis} {The io mass-loading disk: Wave dispersion analysis}.{\BBCQ}
\newblock
\APACjournalVolNumPages{Journal of Geophysical Research: Space Physics}{106}{A11}{26261-26275}.
\newblock
\begin{APACrefURL} \url{https://agupubs.onlinelibrary.wiley.com/doi/abs/10.1029/2001JA900090} \end{APACrefURL}
\newblock
\begin{APACrefDOI} \doi{https://doi.org/10.1029/2001JA900090} \end{APACrefDOI}
\PrintBackRefs{\CurrentBib}

\bibitem [\protect \citeauthoryear {%
Bl{\"{o}}cker%
, Saur%
, Roth%
\BCBL {}\ \BBA {} Strobel%
}{%
Bl{\"{o}}cker%
\ \protect \BOthers {.}}{%
{\protect \APACyear {2018}}%
}]{%
Blocker2018MHDAtmosphere}
\APACinsertmetastar {%
Blocker2018MHDAtmosphere}%
\begin{APACrefauthors}%
Bl{\"{o}}cker, A.%
, Saur, J.%
, Roth, L.%
\BCBL {}\ \BBA {} Strobel, D\BPBI F.%
\end{APACrefauthors}%
\unskip\
\newblock
\APACrefYearMonthDay{2018}{11}{}.
\newblock
{\BBOQ}\APACrefatitle {{MHD Modeling of the Plasma Interaction With Io's Asymmetric Atmosphere}} {{MHD Modeling of the Plasma Interaction With Io's Asymmetric Atmosphere}}.{\BBCQ}
\newblock
\APACjournalVolNumPages{Journal of Geophysical Research: Space Physics}{123}{11}{9286--9311}.
\newblock
\begin{APACrefDOI} \doi{10.1029/2018JA025747} \end{APACrefDOI}
\PrintBackRefs{\CurrentBib}

\bibitem [\protect \citeauthoryear {%
Breer%
, Liuzzo%
, Arnold%
, Andersson%
\BCBL {}\ \BBA {} Simon%
}{%
Breer%
\ \protect \BOthers {.}}{%
{\protect \APACyear {2019}}%
}]{%
Breer2019EnergeticEuropa}
\APACinsertmetastar {%
Breer2019EnergeticEuropa}%
\begin{APACrefauthors}%
Breer, B\BPBI R.%
, Liuzzo, L.%
, Arnold, H.%
, Andersson, P\BPBI N.%
\BCBL {}\ \BBA {} Simon, S.%
\end{APACrefauthors}%
\unskip\
\newblock
\APACrefYearMonthDay{2019}{9}{}.
\newblock
{\BBOQ}\APACrefatitle {{Energetic Ion Dynamics in the Perturbed Electromagnetic Fields Near Europa}} {{Energetic Ion Dynamics in the Perturbed Electromagnetic Fields Near Europa}}.{\BBCQ}
\newblock
\APACjournalVolNumPages{Journal of Geophysical Research: Space Physics}{124}{9}{7592--7613}.
\newblock
\begin{APACrefURL} \url{https://onlinelibrary-wiley-com.vu-nl.idm.oclc.org/doi/full/10.1029/2019JA027147 https://onlinelibrary-wiley-com.vu-nl.idm.oclc.org/doi/abs/10.1029/2019JA027147 https://agupubs-onlinelibrary-wiley-com.vu-nl.idm.oclc.org/doi/10.1029/2019JA027147} \end{APACrefURL}
\newblock
\begin{APACrefDOI} \doi{10.1029/2019JA027147} \end{APACrefDOI}
\PrintBackRefs{\CurrentBib}

\bibitem [\protect \citeauthoryear {%
Brieda%
}{%
Brieda%
}{%
{\protect \APACyear {2011}}%
}]{%
Brieda2011ChargeCEX}
\APACinsertmetastar {%
Brieda2011ChargeCEX}%
\begin{APACrefauthors}%
Brieda, L.%
\end{APACrefauthors}%
\unskip\
\newblock
\APACrefYearMonthDay{2011}{}{}.
\newblock
\APACrefbtitle {{Charge Exchange Collisions (CEX)}.} {{Charge Exchange Collisions (CEX)}.}
\newblock
\begin{APACrefURL} \url{https://www.particleincell.com/2011/charge-exchange/} \end{APACrefURL}
\PrintBackRefs{\CurrentBib}

\bibitem [\protect \citeauthoryear {%
Carnielli%
\ \protect \BOthers {.}}{%
Carnielli%
\ \protect \BOthers {.}}{%
{\protect \APACyear {2019}}%
}]{%
Carnielli2019FirstIonosphere}
\APACinsertmetastar {%
Carnielli2019FirstIonosphere}%
\begin{APACrefauthors}%
Carnielli, G.%
, Galand, M.%
, Leblanc, F.%
, Leclercq, L.%
, Modolo, R.%
, Beth, A.%
\BDBL {}Jia, X.%
\end{APACrefauthors}%
\unskip\
\newblock
\APACrefYearMonthDay{2019}{9}{}.
\newblock
{\BBOQ}\APACrefatitle {{First 3D test particle model of Ganymede's ionosphere}} {{First 3D test particle model of Ganymede's ionosphere}}.{\BBCQ}
\newblock
\APACjournalVolNumPages{Icarus}{330}{}{42--59}.
\newblock
\begin{APACrefDOI} \doi{10.1016/J.ICARUS.2019.04.016} \end{APACrefDOI}
\PrintBackRefs{\CurrentBib}

\bibitem [\protect \citeauthoryear {%
Cheng%
, Maclennan%
, Lanzerotti%
, Paonessa%
\BCBL {}\ \BBA {} Armstrong%
}{%
Cheng%
\ \protect \BOthers {.}}{%
{\protect \APACyear {1983}}%
{\protect \APACexlab {{\protect \BCnt {1}}}}}]{%
Cheng1983EnergeticOrbit}
\APACinsertmetastar {%
Cheng1983EnergeticOrbit}%
\begin{APACrefauthors}%
Cheng, A\BPBI F.%
, Maclennan, C\BPBI G.%
, Lanzerotti, L\BPBI J.%
, Paonessa, M.%
\BCBL {}\ \BBA {} Armstrong, T\BPBI P.%
\end{APACrefauthors}%
\unskip\
\newblock
\APACrefYearMonthDay{1983{\protect \BCnt {1}}}{}{}.
\newblock
{\BBOQ}\APACrefatitle {{Energetic Ion Losses Near Io's Orbit}} {{Energetic Ion Losses Near Io's Orbit}}.{\BBCQ}
\newblock
\APACjournalVolNumPages{Journal of Geophysical Research}{88}{}{3936--3944}.
\PrintBackRefs{\CurrentBib}

\bibitem [\protect \citeauthoryear {%
Cheng%
, Maclennan%
, Lanzerotti%
, Paonessa%
\BCBL {}\ \BBA {} Armstrong%
}{%
Cheng%
\ \protect \BOthers {.}}{%
{\protect \APACyear {1983}}%
{\protect \APACexlab {{\protect \BCnt {2}}}}}]{%
Cheng1983EnergeticOrbitb}
\APACinsertmetastar {%
Cheng1983EnergeticOrbitb}%
\begin{APACrefauthors}%
Cheng, A\BPBI F.%
, Maclennan, C\BPBI G.%
, Lanzerotti, L\BPBI J.%
, Paonessa, M\BPBI T.%
\BCBL {}\ \BBA {} Armstrong, T\BPBI P.%
\end{APACrefauthors}%
\unskip\
\newblock
\APACrefYearMonthDay{1983{\protect \BCnt {2}}}{5}{}.
\newblock
{\BBOQ}\APACrefatitle {{Energetic ion losses near Io's orbit}} {{Energetic ion losses near Io's orbit}}.{\BBCQ}
\newblock
\APACjournalVolNumPages{Journal of Geophysical Research}{88}{A5}{3936}.
\newblock
\begin{APACrefURL} \url{http://doi.wiley.com/10.1029/JA088iA05p03936} \end{APACrefURL}
\newblock
\begin{APACrefDOI} \doi{10.1029/JA088iA05p03936} \end{APACrefDOI}
\PrintBackRefs{\CurrentBib}

\bibitem [\protect \citeauthoryear {%
Clark%
, Mauk%
, Paranicas%
, Kollmann%
\BCBL {}\ \BBA {} Smith%
}{%
Clark%
\ \protect \BOthers {.}}{%
{\protect \APACyear {2016}}%
}]{%
Clark2016ChargeMagnetosphere}
\APACinsertmetastar {%
Clark2016ChargeMagnetosphere}%
\begin{APACrefauthors}%
Clark, G.%
, Mauk, B\BPBI H.%
, Paranicas, C.%
, Kollmann, P.%
\BCBL {}\ \BBA {} Smith, H\BPBI T.%
\end{APACrefauthors}%
\unskip\
\newblock
\APACrefYearMonthDay{2016}{3}{}.
\newblock
{\BBOQ}\APACrefatitle {{Charge states of energetic oxygen and sulfur ions in Jupiter's magnetosphere}} {{Charge states of energetic oxygen and sulfur ions in Jupiter's magnetosphere}}.{\BBCQ}
\newblock
\APACjournalVolNumPages{Journal of Geophysical Research: Space Physics}{121}{3}{2264--2273}.
\newblock
\begin{APACrefURL} \url{https://onlinelibrary-wiley-com.vu-nl.idm.oclc.org/doi/full/10.1002/2015JA022257 https://onlinelibrary-wiley-com.vu-nl.idm.oclc.org/doi/abs/10.1002/2015JA022257 https://agupubs-onlinelibrary-wiley-com.vu-nl.idm.oclc.org/doi/10.1002/2015JA022257} \end{APACrefURL}
\newblock
\begin{APACrefDOI} \doi{10.1002/2015JA022257} \end{APACrefDOI}
\PrintBackRefs{\CurrentBib}

\bibitem [\protect \citeauthoryear {%
Dols%
, Delamere%
\BCBL {}\ \BBA {} Bagenal%
}{%
Dols%
\ \protect \BOthers {.}}{%
{\protect \APACyear {2008}}%
}]{%
Dols2008ATorus}
\APACinsertmetastar {%
Dols2008ATorus}%
\begin{APACrefauthors}%
Dols, V.%
, Delamere, P\BPBI A.%
\BCBL {}\ \BBA {} Bagenal, F.%
\end{APACrefauthors}%
\unskip\
\newblock
\APACrefYearMonthDay{2008}{9}{}.
\newblock
{\BBOQ}\APACrefatitle {{A multispecies chemistry model of Io's local interaction with the Plasma Torus}} {{A multispecies chemistry model of Io's local interaction with the Plasma Torus}}.{\BBCQ}
\newblock
\APACjournalVolNumPages{Journal of Geophysical Research: Space Physics}{113}{9}{}.
\newblock
\begin{APACrefDOI} \doi{10.1029/2007JA012805} \end{APACrefDOI}
\PrintBackRefs{\CurrentBib}

\bibitem [\protect \citeauthoryear {%
Dols%
, Delamere%
, Bagenal%
, Kurth%
\BCBL {}\ \BBA {} Paterson%
}{%
Dols%
\ \protect \BOthers {.}}{%
{\protect \APACyear {2012}}%
}]{%
Dols2012AsymmetryFlybys}
\APACinsertmetastar {%
Dols2012AsymmetryFlybys}%
\begin{APACrefauthors}%
Dols, V.%
, Delamere, P\BPBI A.%
, Bagenal, F.%
, Kurth, W\BPBI S.%
\BCBL {}\ \BBA {} Paterson, W\BPBI R.%
\end{APACrefauthors}%
\unskip\
\newblock
\APACrefYearMonthDay{2012}{10}{}.
\newblock
{\BBOQ}\APACrefatitle {{Asymmetry of Io's outer atmosphere: Constraints from five Galileo flybys}} {{Asymmetry of Io's outer atmosphere: Constraints from five Galileo flybys}}.{\BBCQ}
\newblock
\APACjournalVolNumPages{Journal of Geophysical Research: Planets}{117}{E10}{}.
\newblock
\begin{APACrefURL} \url{http://doi.wiley.com/10.1029/2012JE004076} \end{APACrefURL}
\newblock
\begin{APACrefDOI} \doi{10.1029/2012JE004076} \end{APACrefDOI}
\PrintBackRefs{\CurrentBib}

\bibitem [\protect \citeauthoryear {%
Douté%
\ \protect \BOthers {.}}{%
Douté%
\ \protect \BOthers {.}}{%
{\protect \APACyear {2001}}%
}]{%
Doute2001}
\APACinsertmetastar {%
Doute2001}%
\begin{APACrefauthors}%
Douté, S.%
, Schmitt, B.%
, Lopes-Gautier, R.%
, Carlson, R.%
, Soderblom, L.%
, Shirley, J.%
\BCBL {}\ \BBA {} {the Galileo NIMS Team}.%
\end{APACrefauthors}%
\unskip\
\newblock
\APACrefYearMonthDay{2001}{}{}.
\newblock
{\BBOQ}\APACrefatitle {Mapping SO2 Frost on Io by the Modeling of NIMS Hyperspectral Images} {Mapping so2 frost on io by the modeling of nims hyperspectral images}.{\BBCQ}
\newblock
\APACjournalVolNumPages{Icarus}{149}{1}{107-132}.
\newblock
\begin{APACrefURL} \url{https://www.sciencedirect.com/science/article/pii/S0019103500965138} \end{APACrefURL}
\newblock
\begin{APACrefDOI} \doi{https://doi.org/10.1006/icar.2000.6513} \end{APACrefDOI}
\PrintBackRefs{\CurrentBib}

\bibitem [\protect \citeauthoryear {%
Duling%
, Saur%
\BCBL {}\ \BBA {} Wicht%
}{%
Duling%
\ \protect \BOthers {.}}{%
{\protect \APACyear {2014}}%
}]{%
Duling2014}
\APACinsertmetastar {%
Duling2014}%
\begin{APACrefauthors}%
Duling, S.%
, Saur, J.%
\BCBL {}\ \BBA {} Wicht, J.%
\end{APACrefauthors}%
\unskip\
\newblock
\APACrefYearMonthDay{2014}{}{}.
\newblock
{\BBOQ}\APACrefatitle {Consistent boundary conditions at nonconducting surfaces of planetary bodies: Applications in a new Ganymede MHD model} {Consistent boundary conditions at nonconducting surfaces of planetary bodies: Applications in a new ganymede mhd model}.{\BBCQ}
\newblock
\APACjournalVolNumPages{Journal of Geophysical Research: Space Physics}{119}{6}{4412-4440}.
\newblock
\begin{APACrefURL} \url{https://agupubs.onlinelibrary.wiley.com/doi/abs/10.1002/2013JA019554} \end{APACrefURL}
\newblock
\begin{APACrefDOI} \doi{https://doi.org/10.1002/2013JA019554} \end{APACrefDOI}
\PrintBackRefs{\CurrentBib}

\bibitem [\protect \citeauthoryear {%
Feaga%
, McGrath%
\BCBL {}\ \BBA {} Feldman%
}{%
Feaga%
\ \protect \BOthers {.}}{%
{\protect \APACyear {2009}}%
}]{%
Feaga2009IosAtmosphere}
\APACinsertmetastar {%
Feaga2009IosAtmosphere}%
\begin{APACrefauthors}%
Feaga, L\BPBI M.%
, McGrath, M.%
\BCBL {}\ \BBA {} Feldman, P\BPBI D.%
\end{APACrefauthors}%
\unskip\
\newblock
\APACrefYearMonthDay{2009}{6}{}.
\newblock
{\BBOQ}\APACrefatitle {{Io's dayside SO2 atmosphere}} {{Io's dayside SO2 atmosphere}}.{\BBCQ}
\newblock
\APACjournalVolNumPages{Icarus}{201}{2}{570--584}.
\newblock
\begin{APACrefDOI} \doi{10.1016/j.icarus.2009.01.029} \end{APACrefDOI}
\PrintBackRefs{\CurrentBib}

\bibitem [\protect \citeauthoryear {%
Frank%
\ \BBA {} Paterson%
}{%
Frank%
\ \BBA {} Paterson%
}{%
{\protect \APACyear {2000}}%
}]{%
Frank2000}
\APACinsertmetastar {%
Frank2000}%
\begin{APACrefauthors}%
Frank, L\BPBI A.%
\BCBT {}\ \BBA {} Paterson, W\BPBI R.%
\end{APACrefauthors}%
\unskip\
\newblock
\APACrefYearMonthDay{2000}{}{}.
\newblock
{\BBOQ}\APACrefatitle {Return to Io by the Galileo spacecraft: Plasma observations} {Return to io by the galileo spacecraft: Plasma observations}.{\BBCQ}
\newblock
\APACjournalVolNumPages{Journal of Geophysical Research: Space Physics}{105}{A11}{25363-25378}.
\newblock
\begin{APACrefURL} \url{https://agupubs.onlinelibrary.wiley.com/doi/abs/10.1029/1999JA000460} \end{APACrefURL}
\newblock
\begin{APACrefDOI} \doi{https://doi.org/10.1029/1999JA000460} \end{APACrefDOI}
\PrintBackRefs{\CurrentBib}

\bibitem [\protect \citeauthoryear {%
Frank%
\ \BBA {} Paterson%
}{%
Frank%
\ \BBA {} Paterson%
}{%
{\protect \APACyear {2001}}%
}]{%
Frank2001}
\APACinsertmetastar {%
Frank2001}%
\begin{APACrefauthors}%
Frank, L\BPBI A.%
\BCBT {}\ \BBA {} Paterson, W\BPBI R.%
\end{APACrefauthors}%
\unskip\
\newblock
\APACrefYearMonthDay{2001}{}{}.
\newblock
{\BBOQ}\APACrefatitle {Passage through lo's ionospheric plasmas by the Galileo spacecraft} {Passage through lo's ionospheric plasmas by the galileo spacecraft}.{\BBCQ}
\newblock
\APACjournalVolNumPages{Journal of Geophysical Research: Space Physics}{106}{A11}{26209-26224}.
\newblock
\begin{APACrefURL} \url{https://agupubs.onlinelibrary.wiley.com/doi/abs/10.1029/2000JA002503} \end{APACrefURL}
\newblock
\begin{APACrefDOI} \doi{https://doi.org/10.1029/2000JA002503} \end{APACrefDOI}
\PrintBackRefs{\CurrentBib}

\bibitem [\protect \citeauthoryear {%
Futaana%
\ \protect \BOthers {.}}{%
Futaana%
\ \protect \BOthers {.}}{%
{\protect \APACyear {2015}}%
}]{%
Futaana2015Low-energyTori}
\APACinsertmetastar {%
Futaana2015Low-energyTori}%
\begin{APACrefauthors}%
Futaana, Y.%
, Barabash, S.%
, Wang, X\BPBI D.%
, Wieser, M.%
, Wieser, G\BPBI S.%
, Wurz, P.%
\BDBL {}Brandt, P\BPBI C.%
\end{APACrefauthors}%
\unskip\
\newblock
\APACrefYearMonthDay{2015}{4}{}.
\newblock
{\BBOQ}\APACrefatitle {{Low-energy energetic neutral atom imaging of Io plasma and neutral tori}} {{Low-energy energetic neutral atom imaging of Io plasma and neutral tori}}.{\BBCQ}
\newblock
\APACjournalVolNumPages{Planetary and Space Science}{108}{}{41--53}.
\newblock
\begin{APACrefDOI} \doi{10.1016/J.PSS.2014.12.022} \end{APACrefDOI}
\PrintBackRefs{\CurrentBib}

\bibitem [\protect \citeauthoryear {%
Huddleston%
, Strangeway%
, Warnecke%
, Russell%
\BCBL {}\ \BBA {} Kivelson%
}{%
Huddleston%
\ \protect \BOthers {.}}{%
{\protect \APACyear {1998}}%
}]{%
Huddleston1998}
\APACinsertmetastar {%
Huddleston1998}%
\begin{APACrefauthors}%
Huddleston, D\BPBI E.%
, Strangeway, R\BPBI J.%
, Warnecke, J.%
, Russell, C\BPBI T.%
\BCBL {}\ \BBA {} Kivelson, M\BPBI G.%
\end{APACrefauthors}%
\unskip\
\newblock
\APACrefYearMonthDay{1998}{}{}.
\newblock
{\BBOQ}\APACrefatitle {Ion cyclotron waves in the Io torus: Wave dispersion, free energy analysis, and SO2 + source rate estimates} {Ion cyclotron waves in the io torus: Wave dispersion, free energy analysis, and so2 + source rate estimates}.{\BBCQ}
\newblock
\APACjournalVolNumPages{Journal of Geophysical Research: Planets}{103}{E9}{19887-19899}.
\newblock
\begin{APACrefURL} \url{https://agupubs.onlinelibrary.wiley.com/doi/abs/10.1029/97JE03557} \end{APACrefURL}
\newblock
\begin{APACrefDOI} \doi{https://doi.org/10.1029/97JE03557} \end{APACrefDOI}
\PrintBackRefs{\CurrentBib}

\bibitem [\protect \citeauthoryear {%
Huddleston%
\ \protect \BOthers {.}}{%
Huddleston%
\ \protect \BOthers {.}}{%
{\protect \APACyear {1997}}%
}]{%
Huddleston1997}
\APACinsertmetastar {%
Huddleston1997}%
\begin{APACrefauthors}%
Huddleston, D\BPBI E.%
, Strangeway, R\BPBI J.%
, Warnecke, J.%
, Russell, C\BPBI T.%
, Kivelson, M\BPBI G.%
\BCBL {}\ \BBA {} Bagenal, F.%
\end{APACrefauthors}%
\unskip\
\newblock
\APACrefYearMonthDay{1997}{}{}.
\newblock
{\BBOQ}\APACrefatitle {Ion cyclotron waves in the Io torus during the Galileo encounter: Warm plasma dispersion analysis} {Ion cyclotron waves in the io torus during the galileo encounter: Warm plasma dispersion analysis}.{\BBCQ}
\newblock
\APACjournalVolNumPages{Geophysical Research Letters}{24}{17}{2143-2146}.
\newblock
\begin{APACrefURL} \url{https://agupubs.onlinelibrary.wiley.com/doi/abs/10.1029/97GL01203} \end{APACrefURL}
\newblock
\begin{APACrefDOI} \doi{https://doi.org/10.1029/97GL01203} \end{APACrefDOI}
\PrintBackRefs{\CurrentBib}

\bibitem [\protect \citeauthoryear {%
Huybrighs%
\ \protect \BOthers {.}}{%
Huybrighs%
\ \protect \BOthers {.}}{%
{\protect \APACyear {2023}}%
}]{%
Huybrighs2023}
\APACinsertmetastar {%
Huybrighs2023}%
\begin{APACrefauthors}%
Huybrighs, H\BPBI L\BPBI F.%
, Blöcker, A.%
, Roussos, E.%
, van Buchem, C.%
, Futaana, Y.%
, Holmberg, M\BPBI K\BPBI G.%
\BCBL {}\ \BBA {} Witasse, O.%
\end{APACrefauthors}%
\unskip\
\newblock
\APACrefYearMonthDay{2023}{}{}.
\newblock
{\BBOQ}\APACrefatitle {Europa’s perturbed fields and induced dipole affect energetic proton depletions during distant Alfven wing flybys} {Europa’s perturbed fields and induced dipole affect energetic proton depletions during distant alfven wing flybys}.{\BBCQ}
\newblock
\APACjournalVolNumPages{Geophysical Research Letters}{}{}{}.
\PrintBackRefs{\CurrentBib}

\bibitem [\protect \citeauthoryear {%
Huybrighs%
\ \protect \BOthers {.}}{%
Huybrighs%
\ \protect \BOthers {.}}{%
{\protect \APACyear {2017}}%
}]{%
Huybrighs2017OnMission}
\APACinsertmetastar {%
Huybrighs2017OnMission}%
\begin{APACrefauthors}%
Huybrighs, H\BPBI L\BPBI F.%
, Futaana, Y.%
, Barabash, S.%
, Wieser, M.%
, Wurz, P.%
, Krupp, N.%
\BDBL {}Vermeersen, B.%
\end{APACrefauthors}%
\unskip\
\newblock
\APACrefYearMonthDay{2017}{}{}.
\newblock
{\BBOQ}\APACrefatitle {On the in-situ detectability of Europa's water vapour plumes from a flyby mission} {On the in-situ detectability of europa's water vapour plumes from a flyby mission}.{\BBCQ}
\newblock
\APACjournalVolNumPages{Icarus}{289}{}{270-280}.
\newblock
\begin{APACrefURL} \url{https://www.sciencedirect.com/science/article/pii/S0019103516301968} \end{APACrefURL}
\newblock
\begin{APACrefDOI} \doi{https://doi.org/10.1016/j.icarus.2016.10.026} \end{APACrefDOI}
\PrintBackRefs{\CurrentBib}

\bibitem [\protect \citeauthoryear {%
Huybrighs%
\ \protect \BOthers {.}}{%
Huybrighs%
\ \protect \BOthers {.}}{%
{\protect \APACyear {2020}}%
}]{%
Huybrighs2020AnDepletions}
\APACinsertmetastar {%
Huybrighs2020AnDepletions}%
\begin{APACrefauthors}%
Huybrighs, H\BPBI L\BPBI F.%
, Roussos, E.%
, Blöcker, A.%
, Krupp, N.%
, Futaana, Y.%
, Barabash, S.%
\BDBL {}Witasse, O.%
\end{APACrefauthors}%
\unskip\
\newblock
\APACrefYearMonthDay{2020}{}{}.
\newblock
{\BBOQ}\APACrefatitle {An Active Plume Eruption on Europa During Galileo Flyby E26 as Indicated by Energetic Proton Depletions} {An active plume eruption on europa during galileo flyby e26 as indicated by energetic proton depletions}.{\BBCQ}
\newblock
\APACjournalVolNumPages{Geophysical Research Letters}{47}{10}{e2020GL087806}.
\newblock
\begin{APACrefURL} \url{https://agupubs.onlinelibrary.wiley.com/doi/abs/10.1029/2020GL087806} \end{APACrefURL}
\newblock
\APACrefnote{e2020GL087806 10.1029/2020GL087806}
\newblock
\begin{APACrefDOI} \doi{https://doi.org/10.1029/2020GL087806} \end{APACrefDOI}
\PrintBackRefs{\CurrentBib}

\bibitem [\protect \citeauthoryear {%
Huybrighs%
\ \protect \BOthers {.}}{%
Huybrighs%
\ \protect \BOthers {.}}{%
{\protect \APACyear {2021}}%
}]{%
Huybrighs2021ReplyDepletions}
\APACinsertmetastar {%
Huybrighs2021ReplyDepletions}%
\begin{APACrefauthors}%
Huybrighs, H\BPBI L\BPBI F.%
, Roussos, E.%
, Blöcker, A.%
, Krupp, N.%
, Futaana, Y.%
, Barabash, S.%
\BDBL {}Witasse, O.%
\end{APACrefauthors}%
\unskip\
\newblock
\APACrefYearMonthDay{2021}{}{}.
\newblock
{\BBOQ}\APACrefatitle {Reply to Comment on “An Active Plume Eruption on Europa During Galileo Flyby E26 as Indicated by Energetic Proton Depletions”} {Reply to comment on “an active plume eruption on europa during galileo flyby e26 as indicated by energetic proton depletions”}.{\BBCQ}
\newblock
\APACjournalVolNumPages{Geophysical Research Letters}{48}{18}{e2021GL095240}.
\newblock
\begin{APACrefURL} \url{https://agupubs.onlinelibrary.wiley.com/doi/abs/10.1029/2021GL095240} \end{APACrefURL}
\newblock
\APACrefnote{e2021GL095240 2021GL095240}
\newblock
\begin{APACrefDOI} \doi{https://doi.org/10.1029/2021GL095240} \end{APACrefDOI}
\PrintBackRefs{\CurrentBib}

\bibitem [\protect \citeauthoryear {%
Huybrighs%
\ \protect \BOthers {.}}{%
Huybrighs%
\ \protect \BOthers {.}}{%
{\protect \APACyear {2024}}%
}]{%
Huybrighs2024_data}
\APACinsertmetastar {%
Huybrighs2024_data}%
\begin{APACrefauthors}%
Huybrighs, H\BPBI L\BPBI F.%
, van Buchem, C\BPBI P\BPBI A.%
, Bl\"{o}cker, A.%
, Dols, V.%
, Bowers, C\BPBI F.%
\BCBL {}\ \BBA {} Jackman, C\BPBI M.%
\end{APACrefauthors}%
\unskip\
\newblock
\APACrefYearMonthDay{2024}{}{}.
\newblock
\APACrefbtitle {Simulations pertaining to this publication [Dataset].} {Simulations pertaining to this publication [dataset].}
\newblock
\APAChowpublished {https://doi.org/10.5281/zenodo.11653182}.
\PrintBackRefs{\CurrentBib}

\bibitem [\protect \citeauthoryear {%
Ingersoll%
, Summers%
\BCBL {}\ \BBA {} Schlipf%
}{%
Ingersoll%
\ \protect \BOthers {.}}{%
{\protect \APACyear {1985}}%
}]{%
Ingersoll1985}
\APACinsertmetastar {%
Ingersoll1985}%
\begin{APACrefauthors}%
Ingersoll, A\BPBI P.%
, Summers, M\BPBI E.%
\BCBL {}\ \BBA {} Schlipf, S\BPBI G.%
\end{APACrefauthors}%
\unskip\
\newblock
\APACrefYearMonthDay{1985}{}{}.
\newblock
{\BBOQ}\APACrefatitle {Supersonic meteorology of Io: Sublimation-driven flow of SO2} {Supersonic meteorology of io: Sublimation-driven flow of so2}.{\BBCQ}
\newblock
\APACjournalVolNumPages{Icarus}{64}{3}{375-390}.
\newblock
\begin{APACrefURL} \url{https://www.sciencedirect.com/science/article/pii/0019103585900624} \end{APACrefURL}
\newblock
\begin{APACrefDOI} \doi{https://doi.org/10.1016/0019-1035(85)90062-4} \end{APACrefDOI}
\PrintBackRefs{\CurrentBib}

\bibitem [\protect \citeauthoryear {%
Jia%
, Kivelson%
\BCBL {}\ \BBA {} Paranicas%
}{%
Jia%
\ \protect \BOthers {.}}{%
{\protect \APACyear {2021}}%
}]{%
Jia2021CommentEtal.}
\APACinsertmetastar {%
Jia2021CommentEtal.}%
\begin{APACrefauthors}%
Jia, X.%
, Kivelson, M\BPBI G.%
\BCBL {}\ \BBA {} Paranicas, C.%
\end{APACrefauthors}%
\unskip\
\newblock
\APACrefYearMonthDay{2021}{3}{}.
\newblock
{\BBOQ}\APACrefatitle {{Comment on “An Active Plume Eruption on Europa During Galileo Flyby E26 as Indicated by Energetic Proton Depletions” by Huybrighs et al.}} {{Comment on “An Active Plume Eruption on Europa During Galileo Flyby E26 as Indicated by Energetic Proton Depletions” by Huybrighs et al.}}{\BBCQ}
\newblock
\APACjournalVolNumPages{Geophysical Research Letters}{48}{6}{e2020GL091550}.
\newblock
\begin{APACrefURL} \url{https://onlinelibrary.wiley.com/doi/10.1029/2020GL091550} \end{APACrefURL}
\newblock
\begin{APACrefDOI} \doi{10.1029/2020GL091550} \end{APACrefDOI}
\PrintBackRefs{\CurrentBib}

\bibitem [\protect \citeauthoryear {%
Khurana%
\ \protect \BOthers {.}}{%
Khurana%
\ \protect \BOthers {.}}{%
{\protect \APACyear {2011}}%
}]{%
Khurana2011EvidenceInterior}
\APACinsertmetastar {%
Khurana2011EvidenceInterior}%
\begin{APACrefauthors}%
Khurana, K\BPBI K.%
, Jia, X.%
, Kivelson, M\BPBI G.%
, Nimmo, F.%
, Schubert, G.%
\BCBL {}\ \BBA {} Russell, C\BPBI T.%
\end{APACrefauthors}%
\unskip\
\newblock
\APACrefYearMonthDay{2011}{6}{}.
\newblock
{\BBOQ}\APACrefatitle {{Evidence of a global magma ocean in Io's interior}} {{Evidence of a global magma ocean in Io's interior}}.{\BBCQ}
\newblock
\APACjournalVolNumPages{Science}{332}{6034}{1186--1189}.
\newblock
\begin{APACrefDOI} \doi{10.1126/science.1201425} \end{APACrefDOI}
\PrintBackRefs{\CurrentBib}

\bibitem [\protect \citeauthoryear {%
Khurana%
\ \protect \BOthers {.}}{%
Khurana%
\ \protect \BOthers {.}}{%
{\protect \APACyear {2004}}%
}]{%
Khurana2004TheMagnetosphere}
\APACinsertmetastar {%
Khurana2004TheMagnetosphere}%
\begin{APACrefauthors}%
Khurana, K\BPBI K.%
, Kivelson, M\BPBI G.%
, Vasyliunas, V\BPBI M.%
, Krupp, N.%
, Woch, J.%
, Lagg, A.%
\BDBL {}Kurth, W\BPBI S.%
\end{APACrefauthors}%
\unskip\
\newblock
\APACrefYearMonthDay{2004}{}{}.
\newblock
{\BBOQ}\APACrefatitle {{The Configuration of Jupiter's Magnetosphere}} {{The Configuration of Jupiter's Magnetosphere}}.{\BBCQ}
\newblock
\BIn{} F.~Bagenal, T\BPBI E.~Dowling\BCBL {}\ \BBA {} W\BPBI B.~McKinnon\ (\BEDS), \APACrefbtitle {Jupiter: The Planet, Satellites and Magnetosphere} {Jupiter: The planet, satellites and magnetosphere}\ (\BCHAP~24).
\newblock
\APACaddressPublisher{}{Cambridge University Press}.
\PrintBackRefs{\CurrentBib}

\bibitem [\protect \citeauthoryear {%
Kirsch%
, Krimigis%
, Kohl%
\BCBL {}\ \BBA {} Keath%
}{%
Kirsch%
\ \protect \BOthers {.}}{%
{\protect \APACyear {1981}}%
}]{%
Kirsch1981}
\APACinsertmetastar {%
Kirsch1981}%
\begin{APACrefauthors}%
Kirsch, E.%
, Krimigis, S\BPBI M.%
, Kohl, J\BPBI W.%
\BCBL {}\ \BBA {} Keath, E\BPBI P.%
\end{APACrefauthors}%
\unskip\
\newblock
\APACrefYearMonthDay{1981}{}{}.
\newblock
{\BBOQ}\APACrefatitle {Upper limits for X - ray and energetic neutral particle emission from Jupiter: Voyager-1 results} {Upper limits for x - ray and energetic neutral particle emission from jupiter: Voyager-1 results}.{\BBCQ}
\newblock
\APACjournalVolNumPages{Geophysical Research Letters}{8}{2}{169-172}.
\newblock
\begin{APACrefURL} \url{https://agupubs.onlinelibrary.wiley.com/doi/abs/10.1029/GL008i002p00169} \end{APACrefURL}
\newblock
\begin{APACrefDOI} \doi{https://doi.org/10.1029/GL008i002p00169} \end{APACrefDOI}
\PrintBackRefs{\CurrentBib}

\bibitem [\protect \citeauthoryear {%
Kivelson%
\ \protect \BOthers {.}}{%
Kivelson%
\ \protect \BOthers {.}}{%
{\protect \APACyear {2004}}%
}]{%
Kivelson2004MagnetosphericSatellites}
\APACinsertmetastar {%
Kivelson2004MagnetosphericSatellites}%
\begin{APACrefauthors}%
Kivelson, M\BPBI G.%
, Bagenal, F.%
, Kurth, W\BPBI S.%
, Neubauer, F\BPBI M.%
, Paranicas, C.%
\BCBL {}\ \BBA {} Saur, J.%
\end{APACrefauthors}%
\unskip\
\newblock
\APACrefYearMonthDay{2004}{}{}.
\newblock
{\BBOQ}\APACrefatitle {{Magnetospheric Interactions with Satellites}} {{Magnetospheric Interactions with Satellites}}.{\BBCQ}
\newblock
\BIn{} F.~Bagenal, T\BPBI E.~Dowling\BCBL {}\ \BBA {} W\BPBI B.~McKinnon\ (\BEDS), \APACrefbtitle {Jupiter: The Planet, Satellites and Magnetosphere} {Jupiter: The planet, satellites and magnetosphere}\ (\BCHAP~21).
\newblock
\APACaddressPublisher{}{Cambridge University Press}.
\PrintBackRefs{\CurrentBib}

\bibitem [\protect \citeauthoryear {%
Kollmann%
\ \protect \BOthers {.}}{%
Kollmann%
\ \protect \BOthers {.}}{%
{\protect \APACyear {2022}}%
}]{%
Kollmann2022}
\APACinsertmetastar {%
Kollmann2022}%
\begin{APACrefauthors}%
Kollmann, P.%
, Paranicas, C.%
, Lagg, A.%
, Roussos, E.%
, Z.H., L\BHBI P.%
, Kusterer, M.%
\BDBL {}J., V.%
\end{APACrefauthors}%
\unskip\
\newblock
\APACrefYearMonthDay{2022}{}{}.
\newblock
\APACrefbtitle {Galileo/EPD user guide} {Galileo/epd user guide}\ \APACbVolEdTR{}{\BTR{}}.
\newblock
\begin{APACrefDOI} \doi{10.1002/essoar.10503620.3} \end{APACrefDOI}
\PrintBackRefs{\CurrentBib}

\bibitem [\protect \citeauthoryear {%
Kotova%
, Roussos%
, Krupp%
\BCBL {}\ \BBA {} Dandouras%
}{%
Kotova%
\ \protect \BOthers {.}}{%
{\protect \APACyear {2015}}%
}]{%
Kotova2015ModelingDione}
\APACinsertmetastar {%
Kotova2015ModelingDione}%
\begin{APACrefauthors}%
Kotova, A.%
, Roussos, E.%
, Krupp, N.%
\BCBL {}\ \BBA {} Dandouras, I.%
\end{APACrefauthors}%
\unskip\
\newblock
\APACrefYearMonthDay{2015}{9}{}.
\newblock
{\BBOQ}\APACrefatitle {{Modeling of the energetic ion observations in the vicinity of Rhea and Dione}} {{Modeling of the energetic ion observations in the vicinity of Rhea and Dione}}.{\BBCQ}
\newblock
\APACjournalVolNumPages{Icarus}{258}{}{402--417}.
\newblock
\begin{APACrefDOI} \doi{10.1016/J.ICARUS.2015.06.031} \end{APACrefDOI}
\PrintBackRefs{\CurrentBib}

\bibitem [\protect \citeauthoryear {%
Krimigis%
\ \protect \BOthers {.}}{%
Krimigis%
\ \protect \BOthers {.}}{%
{\protect \APACyear {2002}}%
}]{%
Krimigis2002}
\APACinsertmetastar {%
Krimigis2002}%
\begin{APACrefauthors}%
Krimigis, S\BPBI M.%
, Mitchell, D\BPBI G.%
, Hamilton, D\BPBI C.%
, Dandouras, J.%
, Armstrong, T\BPBI P.%
, Bolton, S\BPBI J.%
\BDBL {}Williams, D\BPBI J.%
\end{APACrefauthors}%
\unskip\
\newblock
\APACrefYearMonthDay{2002}{Feb}{01}.
\newblock
{\BBOQ}\APACrefatitle {A nebula of gases from Io surrounding Jupiter} {A nebula of gases from io surrounding jupiter}.{\BBCQ}
\newblock
\APACjournalVolNumPages{Nature}{415}{6875}{994-996}.
\newblock
\begin{APACrefURL} \url{https://doi.org/10.1038/415994a} \end{APACrefURL}
\newblock
\begin{APACrefDOI} \doi{10.1038/415994a} \end{APACrefDOI}
\PrintBackRefs{\CurrentBib}

\bibitem [\protect \citeauthoryear {%
Krupp%
\ \protect \BOthers {.}}{%
Krupp%
\ \protect \BOthers {.}}{%
{\protect \APACyear {2020}}%
}]{%
Krupp2020Magnetospheric20052015}
\APACinsertmetastar {%
Krupp2020Magnetospheric20052015}%
\begin{APACrefauthors}%
Krupp, N.%
, Kotova, A.%
, Roussos, E.%
, Simon, S.%
, Liuzzo, L.%
, Paranicas, C.%
\BDBL {}Jones, G\BPBI H.%
\end{APACrefauthors}%
\unskip\
\newblock
\APACrefYearMonthDay{2020}{6}{}.
\newblock
{\BBOQ}\APACrefatitle {{Magnetospheric Interactions of Saturn's Moon Dione (2005–2015)}} {{Magnetospheric Interactions of Saturn's Moon Dione (2005–2015)}}.{\BBCQ}
\newblock
\APACjournalVolNumPages{Journal of Geophysical Research: Space Physics}{125}{6}{e2019JA027688}.
\newblock
\begin{APACrefURL} \url{https://onlinelibrary-wiley-com.vu-nl.idm.oclc.org/doi/full/10.1029/2019JA027688 https://onlinelibrary-wiley-com.vu-nl.idm.oclc.org/doi/abs/10.1029/2019JA027688 https://agupubs-onlinelibrary-wiley-com.vu-nl.idm.oclc.org/doi/10.1029/2019JA027688} \end{APACrefURL}
\newblock
\begin{APACrefDOI} \doi{10.1029/2019JA027688} \end{APACrefDOI}
\PrintBackRefs{\CurrentBib}

\bibitem [\protect \citeauthoryear {%
Lagg%
\ \protect \BOthers {.}}{%
Lagg%
\ \protect \BOthers {.}}{%
{\protect \APACyear {1998}}%
}]{%
Lagg1998DeterminationMeasurements}
\APACinsertmetastar {%
Lagg1998DeterminationMeasurements}%
\begin{APACrefauthors}%
Lagg, A.%
, Krupp, N.%
, Woch, J.%
, Livi, S.%
, Wilken, B.%
\BCBL {}\ \BBA {} Williams, D\BPBI J.%
\end{APACrefauthors}%
\unskip\
\newblock
\APACrefYearMonthDay{1998}{11}{}.
\newblock
{\BBOQ}\APACrefatitle {{Determination of the neutral number density in the Io torus from Galileo-EPD measurements}} {{Determination of the neutral number density in the Io torus from Galileo-EPD measurements}}.{\BBCQ}
\newblock
\APACjournalVolNumPages{Geophysical Research Letters}{25}{21}{4039--4042}.
\newblock
\begin{APACrefURL} \url{http://doi.wiley.com/10.1029/1998GL900070} \end{APACrefURL}
\newblock
\begin{APACrefDOI} \doi{10.1029/1998GL900070} \end{APACrefDOI}
\PrintBackRefs{\CurrentBib}

\bibitem [\protect \citeauthoryear {%
Lee-Payne%
, Kollmann%
, Grande%
\BCBL {}\ \BBA {} Knight%
}{%
Lee-Payne%
\ \protect \BOthers {.}}{%
{\protect \APACyear {2020}}%
}]{%
Lee-Payne2020CorrectionCorrection}
\APACinsertmetastar {%
Lee-Payne2020CorrectionCorrection}%
\begin{APACrefauthors}%
Lee-Payne, Z.%
, Kollmann, P.%
, Grande, M.%
\BCBL {}\ \BBA {} Knight, T.%
\end{APACrefauthors}%
\unskip\
\newblock
\APACrefYearMonthDay{2020}{2}{}.
\newblock
{\BBOQ}\APACrefatitle {{Correction of Galileo Energetic Particle Detector, Composition Measurement System High Rate Data: Semiconductor Dead Layer Correction}} {{Correction of Galileo Energetic Particle Detector, Composition Measurement System High Rate Data: Semiconductor Dead Layer Correction}}.{\BBCQ}
\newblock
\APACjournalVolNumPages{Space Science Reviews}{216}{1}{1--14}.
\newblock
\begin{APACrefURL} \url{https://doi.org/10.1007/s11214-019-0621-y} \end{APACrefURL}
\newblock
\begin{APACrefDOI} \doi{10.1007/s11214-019-0621-y} \end{APACrefDOI}
\PrintBackRefs{\CurrentBib}

\bibitem [\protect \citeauthoryear {%
Lellouch%
, Paubert%
, Moses%
, Schneider%
\BCBL {}\ \BBA {} Strobel%
}{%
Lellouch%
\ \protect \BOthers {.}}{%
{\protect \APACyear {2003}}%
}]{%
Lellouch2003VolcanicallyTorus}
\APACinsertmetastar {%
Lellouch2003VolcanicallyTorus}%
\begin{APACrefauthors}%
Lellouch, E.%
, Paubert, G.%
, Moses, J\BPBI I.%
, Schneider, N\BPBI M.%
\BCBL {}\ \BBA {} Strobel, D\BPBI F.%
\end{APACrefauthors}%
\unskip\
\newblock
\APACrefYearMonthDay{2003}{1}{}.
\newblock
{\BBOQ}\APACrefatitle {{Volcanically emitted sodium chloride as a source for Io's neutral clouds and plasma torus}} {{Volcanically emitted sodium chloride as a source for Io's neutral clouds and plasma torus}}.{\BBCQ}
\newblock
\APACjournalVolNumPages{Nature}{421}{6918}{45--47}.
\newblock
\begin{APACrefDOI} \doi{10.1038/nature01292} \end{APACrefDOI}
\PrintBackRefs{\CurrentBib}

\bibitem [\protect \citeauthoryear {%
Lellouch%
\ \protect \BOthers {.}}{%
Lellouch%
\ \protect \BOthers {.}}{%
{\protect \APACyear {1996}}%
}]{%
Lellouch1996DetectionAtmosphere}
\APACinsertmetastar {%
Lellouch1996DetectionAtmosphere}%
\begin{APACrefauthors}%
Lellouch, E.%
, Strobel, D\BPBI F.%
, Belton, M\BPBI J\BPBI S.%
, Summers, M\BPBI E.%
, Paubert, G.%
\BCBL {}\ \BBA {} Moreno, R.%
\end{APACrefauthors}%
\unskip\
\newblock
\APACrefYearMonthDay{1996}{3}{}.
\newblock
{\BBOQ}\APACrefatitle {{Detection of Sulfur Monoxide in Io's Atmosphere}} {{Detection of Sulfur Monoxide in Io's Atmosphere}}.{\BBCQ}
\newblock
\APACjournalVolNumPages{The Astrophysical Journal}{459}{2}{L107}.
\newblock
\begin{APACrefURL} \url{https://ui.adsabs.harvard.edu/abs/1996ApJ...459L.107L/abstract} \end{APACrefURL}
\newblock
\begin{APACrefDOI} \doi{10.1086/309956} \end{APACrefDOI}
\PrintBackRefs{\CurrentBib}

\bibitem [\protect \citeauthoryear {%
Liuzzo%
, Simon%
\BCBL {}\ \BBA {} Regoli%
}{%
Liuzzo%
\ \protect \BOthers {.}}{%
{\protect \APACyear {2019}}%
}]{%
Liuzzo2019EnergeticCallistob}
\APACinsertmetastar {%
Liuzzo2019EnergeticCallistob}%
\begin{APACrefauthors}%
Liuzzo, L.%
, Simon, S.%
\BCBL {}\ \BBA {} Regoli, L.%
\end{APACrefauthors}%
\unskip\
\newblock
\APACrefYearMonthDay{2019}{2}{}.
\newblock
{\BBOQ}\APACrefatitle {{Energetic ion dynamics near Callisto}} {{Energetic ion dynamics near Callisto}}.{\BBCQ}
\newblock
\APACjournalVolNumPages{Planetary and Space Science}{166}{}{23--53}.
\newblock
\begin{APACrefDOI} \doi{10.1016/J.PSS.2018.07.014} \end{APACrefDOI}
\PrintBackRefs{\CurrentBib}

\bibitem [\protect \citeauthoryear {%
Lo%
, Kurzweg%
, Brackman%
\BCBL {}\ \BBA {} Fite%
}{%
Lo%
\ \protect \BOthers {.}}{%
{\protect \APACyear {1971}}%
}]{%
Lo1971}
\APACinsertmetastar {%
Lo1971}%
\begin{APACrefauthors}%
Lo, H\BPBI H.%
, Kurzweg, L.%
, Brackman, R\BPBI T.%
\BCBL {}\ \BBA {} Fite, W\BPBI L.%
\end{APACrefauthors}%
\unskip\
\newblock
\APACrefYearMonthDay{1971}{Oct}{}.
\newblock
{\BBOQ}\APACrefatitle {Electron Capture and Loss in Collisions of Heavy Ions with Atomic Oxygen} {Electron capture and loss in collisions of heavy ions with atomic oxygen}.{\BBCQ}
\newblock
\APACjournalVolNumPages{Phys. Rev. A}{4}{}{1462--1476}.
\newblock
\begin{APACrefURL} \url{https://link.aps.org/doi/10.1103/PhysRevA.4.1462} \end{APACrefURL}
\newblock
\begin{APACrefDOI} \doi{10.1103/PhysRevA.4.1462} \end{APACrefDOI}
\PrintBackRefs{\CurrentBib}

\bibitem [\protect \citeauthoryear {%
Mauk%
\ \protect \BOthers {.}}{%
Mauk%
\ \protect \BOthers {.}}{%
{\protect \APACyear {2022}}%
}]{%
Mauk2022}
\APACinsertmetastar {%
Mauk2022}%
\begin{APACrefauthors}%
Mauk, B\BPBI H.%
, Allegrini, F.%
, Bagenal, F.%
, Bolton, S\BPBI J.%
, Clark, G.%
, Connerney, J\BPBI E\BPBI P.%
\BDBL {}Sulaiman, A\BPBI H.%
\end{APACrefauthors}%
\unskip\
\newblock
\APACrefYearMonthDay{2022}{}{}.
\newblock
{\BBOQ}\APACrefatitle {Loss of Energetic Ions Comprising the Ring Current Populations of Jupiter's Middle and Inner Magnetosphere} {Loss of energetic ions comprising the ring current populations of jupiter's middle and inner magnetosphere}.{\BBCQ}
\newblock
\APACjournalVolNumPages{Journal of Geophysical Research: Space Physics}{127}{5}{e2022JA030293}.
\newblock
\begin{APACrefURL} \url{https://agupubs.onlinelibrary.wiley.com/doi/abs/10.1029/2022JA030293} \end{APACrefURL}
\newblock
\APACrefnote{e2022JA030293 2022JA030293}
\newblock
\begin{APACrefDOI} \doi{https://doi.org/10.1029/2022JA030293} \end{APACrefDOI}
\PrintBackRefs{\CurrentBib}

\bibitem [\protect \citeauthoryear {%
Mauk%
\ \protect \BOthers {.}}{%
Mauk%
\ \protect \BOthers {.}}{%
{\protect \APACyear {2017}}%
}]{%
Mauk2017TheMission}
\APACinsertmetastar {%
Mauk2017TheMission}%
\begin{APACrefauthors}%
Mauk, B\BPBI H.%
, Haggerty, D\BPBI K.%
, Jaskulek, S\BPBI E.%
, Schlemm, C\BPBI E.%
, Brown, L\BPBI E.%
, Cooper, S\BPBI A.%
\BDBL {}Stokes, M\BPBI R.%
\end{APACrefauthors}%
\unskip\
\newblock
\APACrefYearMonthDay{2017}{11}{}.
\newblock
{\BBOQ}\APACrefatitle {{The Jupiter Energetic Particle Detector Instrument (JEDI) Investigation for the Juno Mission}} {{The Jupiter Energetic Particle Detector Instrument (JEDI) Investigation for the Juno Mission}}.{\BBCQ}
\newblock
\APACjournalVolNumPages{Space Science Reviews}{213}{1-4}{289--346}.
\newblock
\begin{APACrefURL} \url{https://link-springer-com.vu-nl.idm.oclc.org/article/10.1007/s11214-013-0025-3} \end{APACrefURL}
\newblock
\begin{APACrefDOI} \doi{10.1007/S11214-013-0025-3/TABLES/21} \end{APACrefDOI}
\PrintBackRefs{\CurrentBib}

\bibitem [\protect \citeauthoryear {%
Mauk%
\ \protect \BOthers {.}}{%
Mauk%
\ \protect \BOthers {.}}{%
{\protect \APACyear {1998}}%
}]{%
Mauk1998Galileo-measuredEpoch}
\APACinsertmetastar {%
Mauk1998Galileo-measuredEpoch}%
\begin{APACrefauthors}%
Mauk, B\BPBI H.%
, McEntire, R\BPBI W.%
, Williams, D\BPBI J.%
, Lagg, A.%
, Roelof, E\BPBI C.%
, Krimigis, S\BPBI M.%
\BDBL {}Wilken, B.%
\end{APACrefauthors}%
\unskip\
\newblock
\APACrefYearMonthDay{1998}{3}{}.
\newblock
{\BBOQ}\APACrefatitle {{Galileo-measured depletion of near-Io hot ring current plasmas since the Voyager epoch}} {{Galileo-measured depletion of near-Io hot ring current plasmas since the Voyager epoch}}.{\BBCQ}
\newblock
\APACjournalVolNumPages{Journal of Geophysical Research: Space Physics}{103}{A3}{4715--4722}.
\newblock
\begin{APACrefURL} \url{https://agupubs.onlinelibrary.wiley.com/doi/10.1029/97JA02343} \end{APACrefURL}
\newblock
\begin{APACrefDOI} \doi{10.1029/97ja02343} \end{APACrefDOI}
\PrintBackRefs{\CurrentBib}

\bibitem [\protect \citeauthoryear {%
Mauk%
, Mitchell%
, Krimigis%
, Roelof%
\BCBL {}\ \BBA {} Paranicas%
}{%
Mauk%
\ \protect \BOthers {.}}{%
{\protect \APACyear {2003}}%
}]{%
Mauk2003}
\APACinsertmetastar {%
Mauk2003}%
\begin{APACrefauthors}%
Mauk, B\BPBI H.%
, Mitchell, D\BPBI G.%
, Krimigis, S\BPBI M.%
, Roelof, E\BPBI C.%
\BCBL {}\ \BBA {} Paranicas, C\BPBI P.%
\end{APACrefauthors}%
\unskip\
\newblock
\APACrefYearMonthDay{2003}{Feb}{01}.
\newblock
{\BBOQ}\APACrefatitle {Energetic neutral atoms from a trans-Europa gas torus at Jupiter} {Energetic neutral atoms from a trans-europa gas torus at jupiter}.{\BBCQ}
\newblock
\APACjournalVolNumPages{Nature}{421}{6926}{920-922}.
\newblock
\begin{APACrefURL} \url{https://doi.org/10.1038/nature01431} \end{APACrefURL}
\newblock
\begin{APACrefDOI} \doi{10.1038/nature01431} \end{APACrefDOI}
\PrintBackRefs{\CurrentBib}

\bibitem [\protect \citeauthoryear {%
McEwen%
, Haapala%
, Keszthelyi%
\BCBL {}\ \BBA {} Mandt%
}{%
McEwen%
\ \protect \BOthers {.}}{%
{\protect \APACyear {2021}}%
}]{%
McEwen2021FutureIo}
\APACinsertmetastar {%
McEwen2021FutureIo}%
\begin{APACrefauthors}%
McEwen, A.%
, Haapala, A.%
, Keszthelyi, L.%
\BCBL {}\ \BBA {} Mandt, K.%
\end{APACrefauthors}%
\unskip\
\newblock
\APACrefYearMonthDay{2021}{12}{}.
\newblock
{\BBOQ}\APACrefatitle {{Future Exploration of Io}} {{Future Exploration of Io}}.{\BBCQ}
\newblock
\APACjournalVolNumPages{AGU Fall Meeting 2021, held in New Orleans, LA, 13-17 December 2021, id. P15A-02.}{2021}{}{P15A-02}.
\newblock
\begin{APACrefURL} \url{https://ui.adsabs.harvard.edu/abs/2021AGUFM.P15A..02M/abstract} \end{APACrefURL}
\PrintBackRefs{\CurrentBib}

\bibitem [\protect \citeauthoryear {%
McGrath%
\ \BBA {} Johnson%
}{%
McGrath%
\ \BBA {} Johnson%
}{%
{\protect \APACyear {1987}}%
}]{%
McGrath1987MagnetosphericAtmosphere}
\APACinsertmetastar {%
McGrath1987MagnetosphericAtmosphere}%
\begin{APACrefauthors}%
McGrath, M\BPBI A.%
\BCBT {}\ \BBA {} Johnson, R\BPBI E.%
\end{APACrefauthors}%
\unskip\
\newblock
\APACrefYearMonthDay{1987}{3}{}.
\newblock
{\BBOQ}\APACrefatitle {{Magnetospheric plasma sputtering of Io's atmosphere}} {{Magnetospheric plasma sputtering of Io's atmosphere}}.{\BBCQ}
\newblock
\APACjournalVolNumPages{Icarus}{69}{3}{519--531}.
\newblock
\begin{APACrefDOI} \doi{10.1016/0019-1035(87)90021-2} \end{APACrefDOI}
\PrintBackRefs{\CurrentBib}

\bibitem [\protect \citeauthoryear {%
Mitchell%
\ \protect \BOthers {.}}{%
Mitchell%
\ \protect \BOthers {.}}{%
{\protect \APACyear {2005}}%
}]{%
Mitchell2005EnergeticMagnetosphere}
\APACinsertmetastar {%
Mitchell2005EnergeticMagnetosphere}%
\begin{APACrefauthors}%
Mitchell, D\BPBI G.%
, Brandt, P\BPBI C.%
, Roelof, E\BPBI C.%
, Dandouras, J.%
, Krimigis, S\BPBI M.%
\BCBL {}\ \BBA {} Mauk, B\BPBI H.%
\end{APACrefauthors}%
\unskip\
\newblock
\APACrefYearMonthDay{2005}{5}{}.
\newblock
{\BBOQ}\APACrefatitle {{Energetic neutral atom emissions from Titan interaction with Saturn's magnetosphere}} {{Energetic neutral atom emissions from Titan interaction with Saturn's magnetosphere}}.{\BBCQ}
\newblock
\APACjournalVolNumPages{Science}{308}{5724}{989--992}.
\newblock
\begin{APACrefURL} \url{https://www-science-org.vu-nl.idm.oclc.org/doi/10.1126/science.1109805} \end{APACrefURL}
\newblock
\begin{APACrefDOI} \doi{10.1126/SCIENCE.1109805/ASSET/F5666609-8BFB-4F95-84AB-660B0F1B1A40/ASSETS/GRAPHIC/308{\_}989{\_}F4.JPEG} \end{APACrefDOI}
\PrintBackRefs{\CurrentBib}

\bibitem [\protect \citeauthoryear {%
Morabito%
, Synnott%
, Kupferman%
\BCBL {}\ \BBA {} Collins%
}{%
Morabito%
\ \protect \BOthers {.}}{%
{\protect \APACyear {1979}}%
}]{%
Morabito1979}
\APACinsertmetastar {%
Morabito1979}%
\begin{APACrefauthors}%
Morabito, L\BPBI A.%
, Synnott, S\BPBI P.%
, Kupferman, P\BPBI N.%
\BCBL {}\ \BBA {} Collins, S\BPBI A.%
\end{APACrefauthors}%
\unskip\
\newblock
\APACrefYearMonthDay{1979}{}{}.
\newblock
{\BBOQ}\APACrefatitle {Discovery of Currently Active Extraterrestrial Volcanism} {Discovery of currently active extraterrestrial volcanism}.{\BBCQ}
\newblock
\APACjournalVolNumPages{Science}{204}{4396}{972-972}.
\newblock
\begin{APACrefURL} \url{https://www.science.org/doi/abs/10.1126/science.204.4396.972.a} \end{APACrefURL}
\newblock
\begin{APACrefDOI} \doi{10.1126/science.204.4396.972.a} \end{APACrefDOI}
\PrintBackRefs{\CurrentBib}

\bibitem [\protect \citeauthoryear {%
Moreno%
, Schubert%
, Baumgardner%
, Kivelson%
\BCBL {}\ \BBA {} Paige%
}{%
Moreno%
\ \protect \BOthers {.}}{%
{\protect \APACyear {1991}}%
}]{%
Moreno1991}
\APACinsertmetastar {%
Moreno1991}%
\begin{APACrefauthors}%
Moreno, M\BPBI A.%
, Schubert, G.%
, Baumgardner, J.%
, Kivelson, M\BPBI G.%
\BCBL {}\ \BBA {} Paige, D\BPBI A.%
\end{APACrefauthors}%
\unskip\
\newblock
\APACrefYearMonthDay{1991}{}{}.
\newblock
{\BBOQ}\APACrefatitle {Io's volcanic and sublimation atmospheres} {Io's volcanic and sublimation atmospheres}.{\BBCQ}
\newblock
\APACjournalVolNumPages{Icarus}{93}{1}{63-81}.
\newblock
\begin{APACrefURL} \url{https://www.sciencedirect.com/science/article/pii/001910359190164O} \end{APACrefURL}
\newblock
\begin{APACrefDOI} \doi{https://doi.org/10.1016/0019-1035(91)90164-O} \end{APACrefDOI}
\PrintBackRefs{\CurrentBib}

\bibitem [\protect \citeauthoryear {%
Moullet%
, Gurwell%
, Lellouch%
\BCBL {}\ \BBA {} Moreno%
}{%
Moullet%
\ \protect \BOthers {.}}{%
{\protect \APACyear {2010}}%
}]{%
Moullet2010SimultaneousArray}
\APACinsertmetastar {%
Moullet2010SimultaneousArray}%
\begin{APACrefauthors}%
Moullet, A.%
, Gurwell, M\BPBI A.%
, Lellouch, E.%
\BCBL {}\ \BBA {} Moreno, R.%
\end{APACrefauthors}%
\unskip\
\newblock
\APACrefYearMonthDay{2010}{7}{}.
\newblock
{\BBOQ}\APACrefatitle {{Simultaneous mapping of SO2, SO, NaCl in Io's atmosphere with the Submillimeter Array}} {{Simultaneous mapping of SO2, SO, NaCl in Io's atmosphere with the Submillimeter Array}}.{\BBCQ}
\newblock
\APACjournalVolNumPages{Icarus}{208}{1}{353--365}.
\newblock
\begin{APACrefDOI} \doi{10.1016/j.icarus.2010.02.009} \end{APACrefDOI}
\PrintBackRefs{\CurrentBib}

\bibitem [\protect \citeauthoryear {%
N{\'{e}}non%
\ \BBA {} Andr{\'{e}}%
}{%
N{\'{e}}non%
\ \BBA {} Andr{\'{e}}%
}{%
{\protect \APACyear {2019}}%
}]{%
Nenon2019EvidenceMeasurements}
\APACinsertmetastar {%
Nenon2019EvidenceMeasurements}%
\begin{APACrefauthors}%
N{\'{e}}non, Q.%
\BCBT {}\ \BBA {} Andr{\'{e}}, N.%
\end{APACrefauthors}%
\unskip\
\newblock
\APACrefYearMonthDay{2019}{4}{}.
\newblock
{\BBOQ}\APACrefatitle {{Evidence of Europa Neutral Gas Torii From Energetic Sulfur Ion Measurements}} {{Evidence of Europa Neutral Gas Torii From Energetic Sulfur Ion Measurements}}.{\BBCQ}
\newblock
\APACjournalVolNumPages{Geophysical Research Letters}{46}{7}{3599--3606}.
\newblock
\begin{APACrefURL} \url{https://onlinelibrary-wiley-com.vu-nl.idm.oclc.org/doi/full/10.1029/2019GL082200 https://onlinelibrary-wiley-com.vu-nl.idm.oclc.org/doi/abs/10.1029/2019GL082200 https://agupubs-onlinelibrary-wiley-com.vu-nl.idm.oclc.org/doi/10.1029/2019GL082200} \end{APACrefURL}
\newblock
\begin{APACrefDOI} \doi{10.1029/2019GL082200} \end{APACrefDOI}
\PrintBackRefs{\CurrentBib}

\bibitem [\protect \citeauthoryear {%
N{\'{e}}non%
\ \protect \BOthers {.}}{%
N{\'{e}}non%
\ \protect \BOthers {.}}{%
{\protect \APACyear {2018}}%
}]{%
Nenon2018AOrbit}
\APACinsertmetastar {%
Nenon2018AOrbit}%
\begin{APACrefauthors}%
N{\'{e}}non, Q.%
, Sicard, A.%
, Kollmann, P.%
, Garrett, H\BPBI B.%
, Sauer, S\BPBI P.%
\BCBL {}\ \BBA {} Paranicas, C.%
\end{APACrefauthors}%
\unskip\
\newblock
\APACrefYearMonthDay{2018}{5}{}.
\newblock
{\BBOQ}\APACrefatitle {{A Physical Model of the Proton Radiation Belts of Jupiter inside Europa's Orbit}} {{A Physical Model of the Proton Radiation Belts of Jupiter inside Europa's Orbit}}.{\BBCQ}
\newblock
\APACjournalVolNumPages{Journal of Geophysical Research: Space Physics}{123}{5}{3512--3532}.
\newblock
\begin{APACrefURL} \url{https://onlinelibrary-wiley-com.vu-nl.idm.oclc.org/doi/full/10.1029/2018JA025216 https://onlinelibrary-wiley-com.vu-nl.idm.oclc.org/doi/abs/10.1029/2018JA025216 https://agupubs-onlinelibrary-wiley-com.vu-nl.idm.oclc.org/doi/10.1029/2018JA025216} \end{APACrefURL}
\newblock
\begin{APACrefDOI} \doi{10.1029/2018JA025216} \end{APACrefDOI}
\PrintBackRefs{\CurrentBib}

\bibitem [\protect \citeauthoryear {%
Nordheim%
\ \protect \BOthers {.}}{%
Nordheim%
\ \protect \BOthers {.}}{%
{\protect \APACyear {2022}}%
}]{%
Nordheim2022MagnetosphericSurface}
\APACinsertmetastar {%
Nordheim2022MagnetosphericSurface}%
\begin{APACrefauthors}%
Nordheim, T\BPBI A.%
, Regoli, L\BPBI H.%
, Harris, C\BPBI D.%
, Paranicas, C.%
, Hand, K\BPBI P.%
\BCBL {}\ \BBA {} Jia, X.%
\end{APACrefauthors}%
\unskip\
\newblock
\APACrefYearMonthDay{2022}{1}{}.
\newblock
{\BBOQ}\APACrefatitle {{Magnetospheric Ion Bombardment of Europa’s Surface}} {{Magnetospheric Ion Bombardment of Europa’s Surface}}.{\BBCQ}
\newblock
\APACjournalVolNumPages{The Planetary Science Journal}{3}{1}{5}.
\newblock
\begin{APACrefURL} \url{https://iopscience-iop-org.vu-nl.idm.oclc.org/article/10.3847/PSJ/ac382a https://iopscience-iop-org.vu-nl.idm.oclc.org/article/10.3847/PSJ/ac382a/meta} \end{APACrefURL}
\newblock
\begin{APACrefDOI} \doi{10.3847/PSJ/AC382A} \end{APACrefDOI}
\PrintBackRefs{\CurrentBib}

\bibitem [\protect \citeauthoryear {%
Paranicas%
\ \protect \BOthers {.}}{%
Paranicas%
\ \protect \BOthers {.}}{%
{\protect \APACyear {2019}}%
}]{%
Paranicas2019IosData}
\APACinsertmetastar {%
Paranicas2019IosData}%
\begin{APACrefauthors}%
Paranicas, C.%
, Mauk, B.%
, Haggerty, D.%
, Clark, G.%
, Kollmann, P.%
, Rymer, A.%
\BDBL {}Bolton, S.%
\end{APACrefauthors}%
\unskip\
\newblock
\APACrefYearMonthDay{2019}{11}{}.
\newblock
{\BBOQ}\APACrefatitle {{Io's effect on energetic charged particles as seen in Juno data}} {{Io's effect on energetic charged particles as seen in Juno data}}.{\BBCQ}
\newblock
\APACjournalVolNumPages{Geophysical Research Letters}{}{}{2019GL085393}.
\newblock
\begin{APACrefURL} \url{https://onlinelibrary.wiley.com/doi/abs/10.1029/2019GL085393} \end{APACrefURL}
\newblock
\begin{APACrefDOI} \doi{10.1029/2019GL085393} \end{APACrefDOI}
\PrintBackRefs{\CurrentBib}

\bibitem [\protect \citeauthoryear {%
Paranicas%
, Mauk%
, McEntire%
\BCBL {}\ \BBA {} Armstrong%
}{%
Paranicas%
\ \protect \BOthers {.}}{%
{\protect \APACyear {2003}}%
}]{%
Paranicas2003}
\APACinsertmetastar {%
Paranicas2003}%
\begin{APACrefauthors}%
Paranicas, C.%
, Mauk, B\BPBI H.%
, McEntire, R\BPBI W.%
\BCBL {}\ \BBA {} Armstrong, T\BPBI P.%
\end{APACrefauthors}%
\unskip\
\newblock
\APACrefYearMonthDay{2003}{}{}.
\newblock
{\BBOQ}\APACrefatitle {The radiation environment near Io} {The radiation environment near io}.{\BBCQ}
\newblock
\APACjournalVolNumPages{Geophysical Research Letters}{30}{18}{}.
\newblock
\begin{APACrefURL} \url{https://agupubs.onlinelibrary.wiley.com/doi/abs/10.1029/2003GL017682} \end{APACrefURL}
\newblock
\begin{APACrefDOI} \doi{https://doi.org/10.1029/2003GL017682} \end{APACrefDOI}
\PrintBackRefs{\CurrentBib}

\bibitem [\protect \citeauthoryear {%
Paranicas%
, McEntire%
, Cheng%
, Lagg%
\BCBL {}\ \BBA {} Williams%
}{%
Paranicas%
\ \protect \BOthers {.}}{%
{\protect \APACyear {2000}}%
}]{%
Paranicas2000EnergeticEuropa}
\APACinsertmetastar {%
Paranicas2000EnergeticEuropa}%
\begin{APACrefauthors}%
Paranicas, C.%
, McEntire, R\BPBI W.%
, Cheng, A\BPBI F.%
, Lagg, A.%
\BCBL {}\ \BBA {} Williams, D\BPBI J.%
\end{APACrefauthors}%
\unskip\
\newblock
\APACrefYearMonthDay{2000}{7}{}.
\newblock
{\BBOQ}\APACrefatitle {{Energetic charged particles near Europa}} {{Energetic charged particles near Europa}}.{\BBCQ}
\newblock
\APACjournalVolNumPages{Journal of Geophysical Research: Space Physics}{105}{A7}{16005--16015}.
\newblock
\begin{APACrefURL} \url{https://agupubs.onlinelibrary.wiley.com/doi/full/10.1029/1999JA000350 https://agupubs.onlinelibrary.wiley.com/doi/abs/10.1029/1999JA000350 https://agupubs.onlinelibrary.wiley.com/doi/10.1029/1999JA000350} \end{APACrefURL}
\newblock
\begin{APACrefDOI} \doi{10.1029/1999ja000350} \end{APACrefDOI}
\PrintBackRefs{\CurrentBib}

\bibitem [\protect \citeauthoryear {%
PDS%
}{%
PDS%
}{%
{\protect \APACyear {2022}}%
}]{%
EPD}
\APACinsertmetastar {%
EPD}%
\begin{APACrefauthors}%
PDS.%
\end{APACrefauthors}%
\unskip\
\newblock
\APACrefYearMonthDay{2022}{}{}.
\newblock
\APACrefbtitle {Galileo EPD data [Dataset].} {Galileo epd data [dataset].}
\newblock
\APAChowpublished {https://doi.org/10.17189/n0dm-0014}.
\PrintBackRefs{\CurrentBib}

\bibitem [\protect \citeauthoryear {%
Plainaki%
\ \protect \BOthers {.}}{%
Plainaki%
\ \protect \BOthers {.}}{%
{\protect \APACyear {2020}}%
}]{%
Plainaki2020}
\APACinsertmetastar {%
Plainaki2020}%
\begin{APACrefauthors}%
Plainaki, C.%
, Massetti, S.%
, Jia, X.%
, Mura, A.%
, Milillo, A.%
, Grassi, D.%
\BDBL {}Filacchione, G.%
\end{APACrefauthors}%
\unskip\
\newblock
\APACrefYearMonthDay{2020}{sep}{}.
\newblock
{\BBOQ}\APACrefatitle {Kinetic Simulations of the Jovian Energetic Ion Circulation around Ganymede} {Kinetic simulations of the jovian energetic ion circulation around ganymede}.{\BBCQ}
\newblock
\APACjournalVolNumPages{The Astrophysical Journal}{900}{1}{74}.
\newblock
\begin{APACrefURL} \url{https://doi.org/10.3847/1538-4357/aba94c} \end{APACrefURL}
\newblock
\begin{APACrefDOI} \doi{10.3847/1538-4357/aba94c} \end{APACrefDOI}
\PrintBackRefs{\CurrentBib}

\bibitem [\protect \citeauthoryear {%
Poppe%
, Fatemi%
\BCBL {}\ \BBA {} Khurana%
}{%
Poppe%
\ \protect \BOthers {.}}{%
{\protect \APACyear {2018}}%
}]{%
Poppe2018ThermalMagnetosphere}
\APACinsertmetastar {%
Poppe2018ThermalMagnetosphere}%
\begin{APACrefauthors}%
Poppe, A\BPBI R.%
, Fatemi, S.%
\BCBL {}\ \BBA {} Khurana, K\BPBI K.%
\end{APACrefauthors}%
\unskip\
\newblock
\APACrefYearMonthDay{2018}{6}{}.
\newblock
{\BBOQ}\APACrefatitle {{Thermal and Energetic Ion Dynamics in Ganymede's Magnetosphere}} {{Thermal and Energetic Ion Dynamics in Ganymede's Magnetosphere}}.{\BBCQ}
\newblock
\APACjournalVolNumPages{Journal of Geophysical Research: Space Physics}{123}{6}{4614--4637}.
\newblock
\begin{APACrefURL} \url{https://onlinelibrary-wiley-com.vu-nl.idm.oclc.org/doi/full/10.1029/2018JA025312 https://onlinelibrary-wiley-com.vu-nl.idm.oclc.org/doi/abs/10.1029/2018JA025312 https://agupubs-onlinelibrary-wiley-com.vu-nl.idm.oclc.org/doi/10.1029/2018JA025312} \end{APACrefURL}
\newblock
\begin{APACrefDOI} \doi{10.1029/2018JA025312} \end{APACrefDOI}
\PrintBackRefs{\CurrentBib}

\bibitem [\protect \citeauthoryear {%
Pospieszalska%
\ \BBA {} Johnson%
}{%
Pospieszalska%
\ \BBA {} Johnson%
}{%
{\protect \APACyear {1996}}%
}]{%
Pospieszalska1996}
\APACinsertmetastar {%
Pospieszalska1996}%
\begin{APACrefauthors}%
Pospieszalska, M\BPBI K.%
\BCBT {}\ \BBA {} Johnson, R\BPBI E.%
\end{APACrefauthors}%
\unskip\
\newblock
\APACrefYearMonthDay{1996}{}{}.
\newblock
{\BBOQ}\APACrefatitle {Monte Carlo calculations of plasma ion-induced sputtering of an atmosphere: SO2 ejected from Io} {Monte carlo calculations of plasma ion-induced sputtering of an atmosphere: So2 ejected from io}.{\BBCQ}
\newblock
\APACjournalVolNumPages{Journal of Geophysical Research: Planets}{101}{E3}{7565-7573}.
\newblock
\begin{APACrefURL} \url{https://agupubs.onlinelibrary.wiley.com/doi/abs/10.1029/95JE03650} \end{APACrefURL}
\newblock
\begin{APACrefDOI} \doi{https://doi.org/10.1029/95JE03650} \end{APACrefDOI}
\PrintBackRefs{\CurrentBib}

\bibitem [\protect \citeauthoryear {%
Rees%
}{%
Rees%
}{%
{\protect \APACyear {1989}}%
}]{%
Rees1989PhysicsAtmosphere}
\APACinsertmetastar {%
Rees1989PhysicsAtmosphere}%
\begin{APACrefauthors}%
Rees, M\BPBI H.%
\end{APACrefauthors}%
\unskip\
\newblock
\APACrefYear{1989}.
\newblock
\APACrefbtitle {{Physics and Chemistry of the Upper Atmosphere}} {{Physics and Chemistry of the Upper Atmosphere}}.
\newblock
\APACaddressPublisher{}{Cambridge University Press}.
\newblock
\begin{APACrefURL} \url{https://www-cambridge-org.vu-nl.idm.oclc.org/core/books/physics-and-chemistry-of-the-upper-atmosphere/B92A6E9E87492A19A407C4681E674EA6} \end{APACrefURL}
\newblock
\begin{APACrefDOI} \doi{10.1017/CBO9780511573118} \end{APACrefDOI}
\PrintBackRefs{\CurrentBib}

\bibitem [\protect \citeauthoryear {%
Roth%
\ \protect \BOthers {.}}{%
Roth%
\ \protect \BOthers {.}}{%
{\protect \APACyear {2024}}%
}]{%
Roth2024}
\APACinsertmetastar {%
Roth2024}%
\begin{APACrefauthors}%
Roth, L.%
, Blöcker, A.%
, de Kleer, K.%
, Goldstein, D.%
, Lellouch, E.%
, Saur, J.%
\BDBL {}Vorburger, A.%
\end{APACrefauthors}%
\unskip\
\newblock
\APACrefYearMonthDay{2024}{}{}.
\newblock
{\BBOQ}\APACrefatitle {Mass supply from Io to Jupiter's magnetosphere} {Mass supply from io to jupiter's magnetosphere}.{\BBCQ}
\newblock
\APACjournalVolNumPages{Space Science Reviews}{}{}{}.
\newblock
\begin{APACrefDOI} \doi{https://doi.org/10.48550/arXiv.2403.13970} \end{APACrefDOI}
\PrintBackRefs{\CurrentBib}

\bibitem [\protect \citeauthoryear {%
Roussos%
\ \protect \BOthers {.}}{%
Roussos%
\ \protect \BOthers {.}}{%
{\protect \APACyear {2022}}%
}]{%
Roussos2022ABelts}
\APACinsertmetastar {%
Roussos2022ABelts}%
\begin{APACrefauthors}%
Roussos, E.%
, Cohen, C.%
, Kollmann, P.%
, Pinto, M.%
, Krupp, N.%
, Gon{\c{c}}alves, P.%
\BCBL {}\ \BBA {} Dialynas, K.%
\end{APACrefauthors}%
\unskip\
\newblock
\APACrefYearMonthDay{2022}{1}{}.
\newblock
{\BBOQ}\APACrefatitle {{A source of very energetic oxygen located in Jupiter’s inner radiation belts}} {{A source of very energetic oxygen located in Jupiter’s inner radiation belts}}.{\BBCQ}
\newblock
\APACjournalVolNumPages{Science Advances}{8}{2}{4234}.
\newblock
\begin{APACrefURL} \url{https://www-science-org.vu-nl.idm.oclc.org/doi/10.1126/sciadv.abm4234} \end{APACrefURL}
\newblock
\begin{APACrefDOI} \doi{10.1126/SCIADV.ABM4234/SUPPL{\_}FILE/SCIADV.ABM4234{\_}SM.PDF} \end{APACrefDOI}
\PrintBackRefs{\CurrentBib}

\bibitem [\protect \citeauthoryear {%
Roussos%
\ \protect \BOthers {.}}{%
Roussos%
\ \protect \BOthers {.}}{%
{\protect \APACyear {2012}}%
}]{%
Roussos2012EnergeticInteraction}
\APACinsertmetastar {%
Roussos2012EnergeticInteraction}%
\begin{APACrefauthors}%
Roussos, E.%
, Kollmann, P.%
, Krupp, N.%
, Paranicas, C.%
, Krimigis, S\BPBI M.%
, Mitchell, D\BPBI G.%
\BDBL {}Holmberg, M\BPBI K.%
\end{APACrefauthors}%
\unskip\
\newblock
\APACrefYearMonthDay{2012}{9}{}.
\newblock
{\BBOQ}\APACrefatitle {{Energetic electron observations of Rhea’s magnetospheric interaction}} {{Energetic electron observations of Rhea’s magnetospheric interaction}}.{\BBCQ}
\newblock
\APACjournalVolNumPages{Icarus}{221}{1}{116--134}.
\newblock
\begin{APACrefDOI} \doi{10.1016/J.ICARUS.2012.07.006} \end{APACrefDOI}
\PrintBackRefs{\CurrentBib}

\bibitem [\protect \citeauthoryear {%
Russell%
\ \BBA {} Kivelson%
}{%
Russell%
\ \BBA {} Kivelson%
}{%
{\protect \APACyear {2000}}%
}]{%
Russell1998}
\APACinsertmetastar {%
Russell1998}%
\begin{APACrefauthors}%
Russell, C\BPBI T.%
\BCBT {}\ \BBA {} Kivelson, M\BPBI G.%
\end{APACrefauthors}%
\unskip\
\newblock
\APACrefYearMonthDay{2000}{}{}.
\newblock
{\BBOQ}\APACrefatitle {Detection of SO in Io's Exosphere} {Detection of so in io's exosphere}.{\BBCQ}
\newblock
\APACjournalVolNumPages{Science}{287}{5460}{1998-1999}.
\newblock
\begin{APACrefURL} \url{https://www.science.org/doi/abs/10.1126/science.287.5460.1998} \end{APACrefURL}
\newblock
\begin{APACrefDOI} \doi{10.1126/science.287.5460.1998} \end{APACrefDOI}
\PrintBackRefs{\CurrentBib}

\bibitem [\protect \citeauthoryear {%
Russell%
, Wang%
, Blanco-Cano%
\BCBL {}\ \BBA {} Strangeway%
}{%
Russell%
\ \protect \BOthers {.}}{%
{\protect \APACyear {2001}}%
}]{%
Russell2001}
\APACinsertmetastar {%
Russell2001}%
\begin{APACrefauthors}%
Russell, C\BPBI T.%
, Wang, Y\BPBI L.%
, Blanco-Cano, X.%
\BCBL {}\ \BBA {} Strangeway, R\BPBI J.%
\end{APACrefauthors}%
\unskip\
\newblock
\APACrefYearMonthDay{2001}{}{}.
\newblock
{\BBOQ}\APACrefatitle {The Io mass-loading disk: Constraints provided by ion cyclotron wave observations} {The io mass-loading disk: Constraints provided by ion cyclotron wave observations}.{\BBCQ}
\newblock
\APACjournalVolNumPages{Journal of Geophysical Research: Space Physics}{106}{A11}{26233-26242}.
\newblock
\begin{APACrefURL} \url{https://agupubs.onlinelibrary.wiley.com/doi/abs/10.1029/2001JA900029} \end{APACrefURL}
\newblock
\begin{APACrefDOI} \doi{https://doi.org/10.1029/2001JA900029} \end{APACrefDOI}
\PrintBackRefs{\CurrentBib}

\bibitem [\protect \citeauthoryear {%
Saur%
, Neubauer%
, Strobel%
\BCBL {}\ \BBA {} Summers%
}{%
Saur%
\ \protect \BOthers {.}}{%
{\protect \APACyear {2002}}%
}]{%
Saur2002InterpretationPasses}
\APACinsertmetastar {%
Saur2002InterpretationPasses}%
\begin{APACrefauthors}%
Saur, J.%
, Neubauer, F\BPBI M.%
, Strobel, D\BPBI F.%
\BCBL {}\ \BBA {} Summers, M\BPBI E.%
\end{APACrefauthors}%
\unskip\
\newblock
\APACrefYearMonthDay{2002}{12}{}.
\newblock
{\BBOQ}\APACrefatitle {{Interpretation of Galileo's Io plasma and field observations: I0, I24, and I27 flybys and close polar passes}} {{Interpretation of Galileo's Io plasma and field observations: I0, I24, and I27 flybys and close polar passes}}.{\BBCQ}
\newblock
\APACjournalVolNumPages{Journal of Geophysical Research: Space Physics}{107}{A12}{1422}.
\newblock
\begin{APACrefDOI} \doi{10.1029/2001JA005067} \end{APACrefDOI}
\PrintBackRefs{\CurrentBib}

\bibitem [\protect \citeauthoryear {%
Saur%
, Strobel%
, Neubauer%
\BCBL {}\ \BBA {} Summers%
}{%
Saur%
\ \protect \BOthers {.}}{%
{\protect \APACyear {2003}}%
}]{%
Saur2003TheIo}
\APACinsertmetastar {%
Saur2003TheIo}%
\begin{APACrefauthors}%
Saur, J.%
, Strobel, D\BPBI F.%
, Neubauer, F\BPBI M.%
\BCBL {}\ \BBA {} Summers, M\BPBI E.%
\end{APACrefauthors}%
\unskip\
\newblock
\APACrefYearMonthDay{2003}{6}{}.
\newblock
{\BBOQ}\APACrefatitle {{The ion mass loading rate at Io}} {{The ion mass loading rate at Io}}.{\BBCQ}
\newblock
\APACjournalVolNumPages{Icarus}{163}{2}{456--468}.
\newblock
\begin{APACrefDOI} \doi{10.1016/S0019-1035(03)00085-X} \end{APACrefDOI}
\PrintBackRefs{\CurrentBib}

\bibitem [\protect \citeauthoryear {%
Selesnick%
\ \BBA {} Cohen%
}{%
Selesnick%
\ \BBA {} Cohen%
}{%
{\protect \APACyear {2009}}%
}]{%
Selesnick2009ChargeIo}
\APACinsertmetastar {%
Selesnick2009ChargeIo}%
\begin{APACrefauthors}%
Selesnick, R\BPBI S.%
\BCBT {}\ \BBA {} Cohen, C\BPBI M.%
\end{APACrefauthors}%
\unskip\
\newblock
\APACrefYearMonthDay{2009}{1}{}.
\newblock
{\BBOQ}\APACrefatitle {{Charge states of energetic ions in Jupiter's radiation belt inferred from absorption microsignatures of Io}} {{Charge states of energetic ions in Jupiter's radiation belt inferred from absorption microsignatures of Io}}.{\BBCQ}
\newblock
\APACjournalVolNumPages{Journal of Geophysical Research: Space Physics}{114}{A1}{1207}.
\newblock
\begin{APACrefURL} \url{https://onlinelibrary-wiley-com.vu-nl.idm.oclc.org/doi/full/10.1029/2008JA013722 https://onlinelibrary-wiley-com.vu-nl.idm.oclc.org/doi/abs/10.1029/2008JA013722 https://agupubs-onlinelibrary-wiley-com.vu-nl.idm.oclc.org/doi/10.1029/2008JA013722} \end{APACrefURL}
\newblock
\begin{APACrefDOI} \doi{10.1029/2008JA013722} \end{APACrefDOI}
\PrintBackRefs{\CurrentBib}

\bibitem [\protect \citeauthoryear {%
Smith%
, Mitchell%
, Johnson%
, Mauk%
\BCBL {}\ \BBA {} Smith%
}{%
Smith%
\ \protect \BOthers {.}}{%
{\protect \APACyear {2019}}%
}]{%
Smith2019}
\APACinsertmetastar {%
Smith2019}%
\begin{APACrefauthors}%
Smith, H\BPBI T.%
, Mitchell, D\BPBI G.%
, Johnson, R\BPBI E.%
, Mauk, B\BPBI H.%
\BCBL {}\ \BBA {} Smith, J\BPBI E.%
\end{APACrefauthors}%
\unskip\
\newblock
\APACrefYearMonthDay{2019}{jan}{}.
\newblock
{\BBOQ}\APACrefatitle {Europa Neutral Torus Confirmation and Characterization Based on Observations and Modeling} {Europa neutral torus confirmation and characterization based on observations and modeling}.{\BBCQ}
\newblock
\APACjournalVolNumPages{The Astrophysical Journal}{871}{1}{69}.
\newblock
\begin{APACrefURL} \url{https://dx.doi.org/10.3847/1538-4357/aaed38} \end{APACrefURL}
\newblock
\begin{APACrefDOI} \doi{10.3847/1538-4357/aaed38} \end{APACrefDOI}
\PrintBackRefs{\CurrentBib}

\bibitem [\protect \citeauthoryear {%
Spencer%
, Jessup%
, McGrath%
, Ballester%
\BCBL {}\ \BBA {} Yelle%
}{%
Spencer%
\ \protect \BOthers {.}}{%
{\protect \APACyear {2000}}%
}]{%
Spencer2000}
\APACinsertmetastar {%
Spencer2000}%
\begin{APACrefauthors}%
Spencer, J\BPBI R.%
, Jessup, K\BPBI L.%
, McGrath, M\BPBI A.%
, Ballester, G\BPBI E.%
\BCBL {}\ \BBA {} Yelle, R.%
\end{APACrefauthors}%
\unskip\
\newblock
\APACrefYearMonthDay{2000}{}{}.
\newblock
{\BBOQ}\APACrefatitle {Discovery of Gaseous S<sub>2</sub> in Io's Pele Plume} {Discovery of gaseous s<sub>2</sub> in io's pele plume}.{\BBCQ}
\newblock
\APACjournalVolNumPages{Science}{288}{5469}{1208-1210}.
\newblock
\begin{APACrefURL} \url{https://www.science.org/doi/abs/10.1126/science.288.5469.1208} \end{APACrefURL}
\newblock
\begin{APACrefDOI} \doi{10.1126/science.288.5469.1208} \end{APACrefDOI}
\PrintBackRefs{\CurrentBib}

\bibitem [\protect \citeauthoryear {%
Spencer%
\ \protect \BOthers {.}}{%
Spencer%
\ \protect \BOthers {.}}{%
{\protect \APACyear {2005}}%
}]{%
Spencer2005Mid-infraredAtmosphere}
\APACinsertmetastar {%
Spencer2005Mid-infraredAtmosphere}%
\begin{APACrefauthors}%
Spencer, J\BPBI R.%
, Lellouch, E.%
, Richter, M\BPBI J.%
, L{\'{o}}pez-Valverde, M\BPBI A.%
, Lea~Jessup, K.%
, Greathouse, T\BPBI K.%
\BCBL {}\ \BBA {} Flaud, J\BPBI M.%
\end{APACrefauthors}%
\unskip\
\newblock
\APACrefYearMonthDay{2005}{8}{}.
\newblock
{\BBOQ}\APACrefatitle {{Mid-infrared detection of large longitudinal asymmetries in Io's SO2 atmosphere}} {{Mid-infrared detection of large longitudinal asymmetries in Io's SO2 atmosphere}}.{\BBCQ}
\newblock
\APACjournalVolNumPages{Icarus}{176}{2}{283--304}.
\newblock
\begin{APACrefDOI} \doi{10.1016/j.icarus.2005.01.019} \end{APACrefDOI}
\PrintBackRefs{\CurrentBib}

\bibitem [\protect \citeauthoryear {%
Strobel%
\ \BBA {} Wolven%
}{%
Strobel%
\ \BBA {} Wolven%
}{%
{\protect \APACyear {2001}}%
}]{%
Strobel2001TheHydrogen}
\APACinsertmetastar {%
Strobel2001TheHydrogen}%
\begin{APACrefauthors}%
Strobel, D\BPBI F.%
\BCBT {}\ \BBA {} Wolven, B\BPBI C.%
\end{APACrefauthors}%
\unskip\
\newblock
\APACrefYearMonthDay{2001}{Jun}{01}.
\newblock
{\BBOQ}\APACrefatitle {The Atmosphere of Io: Abundances and Sources of Sulfur Dioxide and Atomic Hydrogen} {The atmosphere of io: Abundances and sources of sulfur dioxide and atomic hydrogen}.{\BBCQ}
\newblock
\APACjournalVolNumPages{Astrophysics and Space Science}{277}{1}{271-287}.
\newblock
\begin{APACrefURL} \url{https://doi.org/10.1023/A:1012261209678} \end{APACrefURL}
\newblock
\begin{APACrefDOI} \doi{10.1023/A:1012261209678} \end{APACrefDOI}
\PrintBackRefs{\CurrentBib}

\bibitem [\protect \citeauthoryear {%
Tsang%
, Spencer%
, Lellouch%
, Lopez-Valverde%
\BCBL {}\ \BBA {} Richter%
}{%
Tsang%
\ \protect \BOthers {.}}{%
{\protect \APACyear {2016}}%
}]{%
Tsang2016TheEclipse}
\APACinsertmetastar {%
Tsang2016TheEclipse}%
\begin{APACrefauthors}%
Tsang, C\BPBI C\BPBI C.%
, Spencer, J\BPBI R.%
, Lellouch, E.%
, Lopez-Valverde, M\BPBI A.%
\BCBL {}\ \BBA {} Richter, M\BPBI J.%
\end{APACrefauthors}%
\unskip\
\newblock
\APACrefYearMonthDay{2016}{8}{}.
\newblock
{\BBOQ}\APACrefatitle {{The collapse of Io's primary atmosphere in Jupiter eclipse}} {{The collapse of Io's primary atmosphere in Jupiter eclipse}}.{\BBCQ}
\newblock
\APACjournalVolNumPages{Journal of Geophysical Research: Planets}{121}{8}{1400--1410}.
\newblock
\begin{APACrefURL} \url{http://doi.wiley.com/10.1002/2016JE005025} \end{APACrefURL}
\newblock
\begin{APACrefDOI} \doi{10.1002/2016JE005025} \end{APACrefDOI}
\PrintBackRefs{\CurrentBib}

\bibitem [\protect \citeauthoryear {%
Walker%
\ \protect \BOthers {.}}{%
Walker%
\ \protect \BOthers {.}}{%
{\protect \APACyear {2010}}%
}]{%
Walker2010}
\APACinsertmetastar {%
Walker2010}%
\begin{APACrefauthors}%
Walker, A\BPBI C.%
, Gratiy, S\BPBI L.%
, Goldstein, D\BPBI B.%
, Moore, C\BPBI H.%
, Varghese, P\BPBI L.%
, Trafton, L\BPBI M.%
\BDBL {}Stewart, B.%
\end{APACrefauthors}%
\unskip\
\newblock
\APACrefYearMonthDay{2010}{}{}.
\newblock
{\BBOQ}\APACrefatitle {A comprehensive numerical simulation of Io’s sublimation-driven atmosphere} {A comprehensive numerical simulation of io’s sublimation-driven atmosphere}.{\BBCQ}
\newblock
\APACjournalVolNumPages{Icarus}{207}{1}{409-432}.
\newblock
\begin{APACrefURL} \url{https://www.sciencedirect.com/science/article/pii/S0019103510000229} \end{APACrefURL}
\newblock
\begin{APACrefDOI} \doi{https://doi.org/10.1016/j.icarus.2010.01.012} \end{APACrefDOI}
\PrintBackRefs{\CurrentBib}

\bibitem [\protect \citeauthoryear {%
Williams%
\ \protect \BOthers {.}}{%
Williams%
\ \protect \BOthers {.}}{%
{\protect \APACyear {1996}}%
}]{%
Williams1996}
\APACinsertmetastar {%
Williams1996}%
\begin{APACrefauthors}%
Williams, D\BPBI J.%
, Mauk, B\BPBI H.%
, McEntire, R\BPBI E.%
, Roelof, E\BPBI C.%
, Armstrong, T\BPBI P.%
, Wilken, B.%
\BDBL {}Lanzerotti, L\BPBI J.%
\end{APACrefauthors}%
\unskip\
\newblock
\APACrefYearMonthDay{1996}{}{}.
\newblock
{\BBOQ}\APACrefatitle {Electron Beams and Ion Composition Measured at Io and in Its Torus} {Electron beams and ion composition measured at io and in its torus}.{\BBCQ}
\newblock
\APACjournalVolNumPages{Science}{274}{5286}{401-403}.
\newblock
\begin{APACrefURL} \url{https://www.science.org/doi/abs/10.1126/science.274.5286.401} \end{APACrefURL}
\newblock
\begin{APACrefDOI} \doi{10.1126/science.274.5286.401} \end{APACrefDOI}
\PrintBackRefs{\CurrentBib}

\bibitem [\protect \citeauthoryear {%
Williams%
, McEntire%
, Jaskulek%
\BCBL {}\ \BBA {} Wilken%
}{%
Williams%
\ \protect \BOthers {.}}{%
{\protect \APACyear {1992}}%
}]{%
Williams1992TheDetector}
\APACinsertmetastar {%
Williams1992TheDetector}%
\begin{APACrefauthors}%
Williams, D\BPBI J.%
, McEntire, R\BPBI W.%
, Jaskulek, S.%
\BCBL {}\ \BBA {} Wilken, B.%
\end{APACrefauthors}%
\unskip\
\newblock
\APACrefYearMonthDay{1992}{5}{}.
\newblock
{\BBOQ}\APACrefatitle {{The Galileo Energetic Particles Detector}} {{The Galileo Energetic Particles Detector}}.{\BBCQ}
\newblock
\APACjournalVolNumPages{Space Science Reviews}{60}{1-4}{385--412}.
\newblock
\begin{APACrefURL} \url{https://link.springer.com/article/10.1007/BF00216863} \end{APACrefURL}
\newblock
\begin{APACrefDOI} \doi{10.1007/BF00216863} \end{APACrefDOI}
\PrintBackRefs{\CurrentBib}

\bibitem [\protect \citeauthoryear {%
Wolven%
\ \protect \BOthers {.}}{%
Wolven%
\ \protect \BOthers {.}}{%
{\protect \APACyear {2001}}%
}]{%
Wolven2001EmissionCorona}
\APACinsertmetastar {%
Wolven2001EmissionCorona}%
\begin{APACrefauthors}%
Wolven, B\BPBI C.%
, Moos, H\BPBI W.%
, Retherford, K\BPBI D.%
, Feldman, P\BPBI D.%
, Strobel, D\BPBI F.%
, Smyth, W\BPBI H.%
\BCBL {}\ \BBA {} Roesler, F\BPBI L.%
\end{APACrefauthors}%
\unskip\
\newblock
\APACrefYearMonthDay{2001}{11}{}.
\newblock
{\BBOQ}\APACrefatitle {{Emission profiles of neutral oxygen and sulfur in Io's exospheric corona}} {{Emission profiles of neutral oxygen and sulfur in Io's exospheric corona}}.{\BBCQ}
\newblock
\APACjournalVolNumPages{Journal of Geophysical Research: Space Physics}{106}{A11}{26155--26182}.
\newblock
\begin{APACrefURL} \url{https://ui.adsabs.harvard.edu/abs/2001JGR...10626155W/abstract} \end{APACrefURL}
\newblock
\begin{APACrefDOI} \doi{10.1029/2000ja002506} \end{APACrefDOI}
\PrintBackRefs{\CurrentBib}

\bibitem [\protect \citeauthoryear {%
Wulms%
, Saur%
, Strobel%
, Simon%
\BCBL {}\ \BBA {} Mitchell%
}{%
Wulms%
\ \protect \BOthers {.}}{%
{\protect \APACyear {2010}}%
}]{%
Wulms2010EnergeticFields}
\APACinsertmetastar {%
Wulms2010EnergeticFields}%
\begin{APACrefauthors}%
Wulms, V.%
, Saur, J.%
, Strobel, D\BPBI F.%
, Simon, S.%
\BCBL {}\ \BBA {} Mitchell, D\BPBI G.%
\end{APACrefauthors}%
\unskip\
\newblock
\APACrefYearMonthDay{2010}{6}{}.
\newblock
{\BBOQ}\APACrefatitle {{Energetic neutral atoms from Titan: Particle simulations in draped magnetic and electric fields}} {{Energetic neutral atoms from Titan: Particle simulations in draped magnetic and electric fields}}.{\BBCQ}
\newblock
\APACjournalVolNumPages{Journal of Geophysical Research: Space Physics}{115}{A6}{6310}.
\newblock
\begin{APACrefURL} \url{https://onlinelibrary-wiley-com.vu-nl.idm.oclc.org/doi/full/10.1029/2009JA014893 https://onlinelibrary-wiley-com.vu-nl.idm.oclc.org/doi/abs/10.1029/2009JA014893 https://agupubs-onlinelibrary-wiley-com.vu-nl.idm.oclc.org/doi/10.1029/2009JA014893} \end{APACrefURL}
\newblock
\begin{APACrefDOI} \doi{10.1029/2009JA014893} \end{APACrefDOI}
\PrintBackRefs{\CurrentBib}

\end{thebibliography}

\end{document}